\documentclass[12pt,preprint]{aastex}
\newcommand{\msuns}{M$_{\odot}$}
\newcommand{\msun}{M$_{\odot}$~}

\newcommand{\mh}{H$_{2}$~}
\newcommand{\mhs}{H$_{2}$}
\newcommand{\um}{$\mu$m~}
\newcommand{\ums}{$\mu$m}

\def\cs33l{[S\kern.2em{\sc iii}] 33.5 $\mu$m}

\def\h0{{\rm H_0}}

\def\pm{^+_-}
\def\ciil{[C\kern.2em{\sc ii}] 158 $\mu$m }
\def\cii{[C\kern.2em{\sc ii}] 158 $\mu$m }

\def\oii{[O\kern.2em{\sc ii}] }

\def\oiii{[O\kern.2em{\sc iii}] }

\def\oiiil{[O\kern.2em{\sc iii}] 88.36 $\mu$m }

\def\oiiif{[O\kern.2em{\sc iii}] \kern.2em{3p2-3p1} }

\def\oiiie{[O\kern.2em{\sc iii}] \kern.2em{3p1-3p0} }

\def\oiv{[O\kern.2em{\sc iv}] }

\def\siiil{[S\kern.2em{\sc iii}] 18.71 $\mu$m }

\def\sivl{[S\kern.2em{\sc iv}]  10.51  $\mu$m }

\def\siil{[Si\kern.2em{\sc ii}]  34.8152  $\mu$m }

\def\neiil{[Ne\kern.2em{\sc ii}]  12.81  $\mu$m}
\def\neiiil{[Ne\kern.2em{\sc iii}]  15.56  $\mu$m}

\def\siii{[S\kern.2em{\sc iii}]}

\def\siv{[S\kern.2em{\sc iv}]}

\def\si2{[Si\kern.2em{\sc ii}]}

\def\neii{[Ne\kern.2em{\sc ii}]}
\def\neiii{[Ne\kern.2em{\sc iii}]}
\newcommand{\iraca}{$[3.6] $ }
\newcommand{\iracb}{$[4.5] $ }

\newcommand{\iracd}{$[8.0] $ }

\newcommand{\mipsa}{$[24.0] $ }

\shorttitle{SAINTS}
\shortauthors{Higdon et al.}

\begin{document}

\title{STAR FORMATION AND THE INTERSTELLAR MEDIUM IN NEARBY TIDAL STREAMS (SAINTS): SPITZER MID-INFRARED SPECTROSCOPY AND IMAGING OF INTERGALACTIC STAR-FORMING OBJECTS}

\author{S. J. U. Higdon \altaffilmark{1},
  J. L. Higdon\altaffilmark{1}, B. J. Smith\altaffilmark{2}, AND
  M. Hancock\altaffilmark{3}}

\altaffiltext{1}{Physics Department, Georgia Southern University, Statesboro, GA 30460, USA}
\altaffiltext{2}{Department of Physics and Astronomy, East Tennessee State University, Johnson City, TN 37614, USA}
\altaffiltext{3}{Bishop Gorman High School, 5959 Hualapai Way, Las Vegas, NV 89148, USA}

\begin{abstract}
A spectroscopic analysis of 10 intergalactic star forming objects
(ISFOs) and a photometric analysis of 67 ISFOs in a sample of 14
interacting systems is presented. The majority of the ISFOs have
relative polycyclic aromatic hydrocarbon (PAH) band strengths similar
to those of nearby spiral and starburst galaxies. In contrast to what
is observed in blue compact dwarfs (BCDs) and local giant HII regions
in the Milky Way (NGC 3603) and the Magellanic Clouds (30 Doradus and
N 66), the relative PAH band strengths in ISFOs correspond to models
with a significant PAH ion fraction ($<$ 50\%) and bright emission
from large PAHs ($\sim$ 100 carbon atoms).  The \neiii$/$\neii ~and
\siv$/$\siii ~line flux ratios indicate moderate levels of excitation
with an interstellar radiation field that is harder than the majority
of the Spitzer Infrared Nearby Galaxies Survey and starburst galaxies,
but softer than BCDs and local giant HII regions. The ISFO neon line
flux ratios are consistent with a burst of star formation $\la$ 6
million years ago.  Most of the ISFOs have $\sim 10^6$ \msun of warm
\mh with a likely origin in photo-dissociation regions
(PDRs). Infrared Array Camera photometry shows the ISFOs to be
bright at 8 \ums, with one third having \iracb - \iracd $>$ 3.7, i.e.,
enhanced non-stellar emission, most likely due to PAHs, relative to
normal spirals, dwarf irregulars and BCD galaxies. The relative
strength of the 8 \um emission compared to that at 3.6 \um or 24 \um
separates ISFOs from dwarf galaxies in Spitzer two color diagrams. The
infrared power in two thirds of the ISFOs is dominated by emission
from grains in a diffuse interstellar medium. One in six ISFOs have
significant emission from PDRs, contributing $\sim 30$ \% $- 60$\% of the
total power. ISFOs are young knots of intense star formation.

\end{abstract}


\section{INTRODUCTION}

In addition to triggering starbursts, mergers of dusty, gas
rich disk galaxies frequently lead to the formation of tidal tails
that can stretch many disk diameters from the site of the collision
(Toomre \& Toomre 1972; Schweizer 1978; Sanders \& Mirabel
1996). These structures are typically HI rich with blue UV/optical
color, reflecting both their origin in the outer spiral disk and
on-going star formation (van der Hulst 1979; Schombert et al. 1990;
Mirabel et al. 1991; Hibbard \& van Gorkom 1996, Smith et al. 2010).
Zwicky (1956) proposed that dwarf galaxies might form out of
self-gravitating clumps within tidal tails, and indeed, concentrations
of gas and star forming regions are commonly found there, with HI
masses and optical luminosities comparable with dwarf galaxies. A wide
range in HI masses have been derived for these objects, from $3 \times
10^{8}$ M$_{\odot}$ for the HI condensations in NGC 7252's plumes
(Hibbard et al. 1994) to $\approx 10^{10}$ M$_{\odot}$ in NGC 5291
(Duc \& Mirabel 1998), with $\approx 10^{9}$ M$_{\odot}$ being typical
(e.g., Duc \& Mirabel 1994; Hibbard \& van Gorkom 1996).

In this paper we use the broad term ``intergalactic star forming
object'' (ISFO), recognizing that some of them may formed as a result
of tidal interactions between the parent galaxies, others from
ram-sweeping of debris material, or inflow of material, or the
interplay of more than one of these physical processes.  These ISFOs
range in mass from super star clusters (10$^4-10^6$ \msuns) to
tidal dwarf galaxies (TDGs; $\ga10^9$ \msuns). The bound ISFOs will
become bona-fide TDGs or intergalactic clusters. Many ISFOs and their
associated tidal streams may be accreted back onto the parent galaxy,
possibly triggering further star formation in the disk. Leftover
material will enrich the intergalactic medium with metals and
dust (Morris \& van den Bergh 1994). The fraction of ISFOs that are
gravitationally bound and long-lived is still an open question (e.g.,
Hibbard \& Mihos 1995, Bournaud \& Duc 2006, Duc et al. 2014). Kaviraj
et al. (2012) analysed data from a sample of mergers observed as part
of the Sloan Digital Sky Survey, and concluded that $\sim$6\% of
dwarfs in nearby clusters have a tidal origin. In contrast, Hunsberger
et al. (1996) concluded that TDGs may make up 30\% - 50\% of the dwarf
galaxy population in compact groups.  Observations with high spatial
and velocity resolution are necessary to confirm that a ISFO is a
physical condensation, and not just a chance alignment of material
along our line of sight. Ideally a rotation curve will be constructed
to determine whether the ISFO is gravitationally bound, i.e., a
bona-fide tidal dwarf galaxy. These observations are generally beyond,
or at the limits of the current observatories.

Here we are studying ISFOs to look for insights into star formation
that is influenced by an external event, i.e., interacting
systems. Computer simulations by Elmegreen et al. (1993) showed that
gas properties play a key role in the formation of TDGs. The elevated
velocity dispersion in the tidal stream leads to an increase in the
Jeans mass resulting in massive clouds which may be more stable and
long-lived. TDGs are formed from gas stripped from the outer portions
of the parent galaxy disk. This material has a higher metallicity
(e.g., the two TDGs in NGC 5291 have log(O/H) $+ 12 = 8.4$; Longmore
et al. 1979; Pena et al. 1991; Duc \& Mirabel 1998) than the gas found
in isolated low-metallicity dwarfs hence the threshold HI column
density for star formation may be lower in TDGs. A typical dynamical
timescale is $\sim$10$^8$ yr for the formation of condensations at
the tips of tidal tails, indicating that any young stellar population
has been formed in situ, and has not simply been stripped from the
disk of the progenitor galaxy along with the older stars, gas and dust
(Elmegreen et al. 1993). A luminous ISFO at the base of a tidal
feature, referred to as a ``hinge clump'' (e.g., Hancock et al. 2007)
may be produced by orbit crowding in the tidal feature (e.g., Struck
\& Smith 2012). Additional ISFOs can be distributed in a seemingly
random pattern or distributed more regularly, in a
``beads-on-a-string'' structure, which is a hallmark of large scale
gravitational instability (e.g., Elmegreen \& Efremov 1996, Hancock et
al. 2007).

Earlier observations have shown that the mid-infrared spectra of ISFOs
are rich in atomic and molecular features from the ISM, including fine
structure lines such as \neiil, ~\neiiil, polycyclic aromatic
hydrocarbons (PAHs) and warm molecular hydrogen ( Higdon, Higdon \&
Marshal 2006b, hereafter HHM06). The mid-infrared spectra of
star-forming dwarf galaxies are known to differ substantially from
those of spirals, with weaker PAH emission and higher \neiii$/$\neii
~line ratios (e.g., Thuan et al. 1999; Madden 2000; Galliano et
al.\ 2003; Houck et al.\ 2004a; Madden et al.\ 2006; Hunt et
al. 2010).  The higher [Ne~III]/[Ne~II] ratio is a consequence of the
lower metallicity (stars are hotter and dust absorption is lower)
resulting in harder ultraviolet interstellar radiation fields (ISRFs).
The weaker PAH features in dwarfs may also be due to metallicity,
either directly as a result of the paucity of available enriched raw
material to build the PAHs or an indirect consequence of the harder UV
field. PAHs can be destroyed in hard UV fields (Plante \& Sauvage
2002; Gordon et al.\ 2008), while young systems may not have the low
mass asymptotic giant branch (AGB) stars to create PAH precursors in
the interstellar medium (e.g., Galliano et al.\ 2008).  Hard ISRFs may
preferentially destroy smaller PAHs, leading to larger PAHs on average
in dwarf galaxies, as indicated by the relative strengths of the
different mid-infrared PAH features in blue compact dwarfs (BCDs; Hunt et
al. 2010). Extreme low metallicity systems have a paucity of PAHs and
a hard ISRF, but other factors in the physical environment may drive
the relative PAH strengths in more enriched environments.

It is unknown whether the mid-infrared spectra of most tidal features
more closely resemble the global spectra of spiral galaxies or those
of dwarf galaxies, since only a handful have been studied in detail
with the Spitzer Infrared Spectrograph (IRS, Houck et al. 2004b; e.g.,
HHM06, Higdon \& Higdon 2008, Higdon et al. 2010).  Tidal features
tend to have lower metallicities than the average values for inner
disks of spirals, thus may be expected to have weaker PAH features.
However, tidal structures tend to have higher metallicities on average
than classical dwarfs with the same luminosity (e.g., Weilbacher et
al. 2003; de Mello et al. 2012). For example, the spectra of the TDGs
in NGC 5291 show prominent PAH features (HHM06).

Since the PAH features contribute significantly to the broadband 8
$\mu$m flux as measured by the Spitzer Infrared Array Camera (IRAC,
Fazio et al. 2004b), their emission directly affects the IRAC colors
of galaxies. Low metallicity dwarfs tend to be deficient in 8 $\mu$m
flux due to weak PAH emission relative to the underlying continuum
(e.g., Engelbracht et al. 2005, Rosenberg et al. 2006, 2008, Draine et
al.  2007, Madden et al. 2006, Wu et al. 2007). ISFOs in NGC 5291 tend
to have red Spitzer \iracb - \iracd colors (HHM06). This may be due to
higher mass-normalized star formation rates and lower proportions of
underlying older stars.

Here we investigate the Spitzer spectral properties of a set of 10
ISFOs and present a photometric analysis of a larger sample of 67
ISFOs. In Section 2 we describe our sample selection.  In Section 3 we
describe the observations and data analysis. The results are used in
Section 4 to address these key questions pertaining to the properties of
ISFOs:

{\em 1) Are the relative PAH strengths best matched to models with
  bright emission from small or large PAHs? 2) are the PAHs mainly
  neutral or charged? 3) Are the interstellar radiation fields
  (ISRF) hard? 4) How much warm molecular gas do the ISFOs contain? 5)
  Can ISFOs be identified by their mid-infrared colors? 6) Does the
  bulk of the infrared emission arise in dust illuminated by a
  diffuse ISRF or by dust embedded in intense fields associated with
  photo-dissociation regions (PDRs)?}

While addressing each question we compare the results for the ISFOs to
the global properties of different galaxy types (starburst, spirals
and BCDs) and also to individual sources in three local giant HII
regions, NGC 3603 in the Milky Way, 30 Doradus (hereafter 30 Dor.) in
the Large Magellanic Cloud (LMC), and N 66 in the Small Magellanic
Cloud (SMC) and knots of star formation in the disks of nearby
interacting galaxies. Our findings are summarized in Section 5.

\section{THE SAMPLE}

The ISFO sample was derived from 12 systems showing clear signs of
recent or on-going gravitational interactions as evidenced by highly
extended optical tails and/or bridges, plus two additional systems
with more complex interaction histories (NGC 5291/Seashell and
Stephan’s Quintet) with known ISFO populations.  These are shown in
Figure 1 and the 14 systems are briefly described in the figure
captions. We purposely chose relatively nearby systems in order to
distinguish individual ISFOs within a given tidal feature subject to
the angular resolution constraints of IRAC and Multiband Imaging
Photometer (MIPS).  Our intent is to identify a sufficiently large
sample of ISFOs in order to derive basic properties of the population,
rather than conduct a survey of every known system with
tails/plumes. The targets were originally observed as part of Spitzer
programs Tidal Dwarf Galaxies (Program name SJH-TDGS; HHM06, Higdon \&
Higdon 2008), Spiral Bridges and Tails (Program name SB\&T; Smith et
al. 2007a, 2008, 2010, Hancock et al. 2007) and Star Formation and the
ISM in Nearby Tidal Streams (Program name SAINTS; Higdon et al. 2010).
Candidate selection was made using the $3.6 \mu$m and $8 \mu$m
images. This is discussed in Section 3.1. Our sample consists of 67
ISFOs, which we list in Table 1. Their distances are assumed to be
that of their parent galaxies which are taken from the literature, and
range from 28-140 Mpc, with a corresponding spatial scale of 136-678
pc arcsec$^{-1}$.

The ISFOs in our spectral sample with published metallicities have
values of one third solar or higher: NGC 5291 N and NGC 5291 S with
log(O/H) $+ 12 = 8.4$ (Longmore et al. 1979; Pena et al. 1991; Duc \&
Mirabel 1998); Arp 72 with log(O/H) $+ 12 =$ 8.7 (Smith et al. 2010);
Arp 105S with log(O/H) $+ 12 =$ 8.4 and Arp 105N with log(O/H) $+ 12 =$
8.6 (Duc \& Mirabel 1994; and Arp 245 with log(O/H) $+ 12 \sim$8.65 (Duc
et al. 2000).

Ten ISFOs from this list with F$_{8\mu m} \ga $ 1 mJy were observed
with the IRS: Arp 72-S1, Arp 82-N1, Arp 84-N1, Arp 87-N1 (Higdon et
al. 2010), Ambartzumian's knot (Arp 105-S2, Higdon \& Higdon 2008,
Boquien et al. 2009), Arp 242-N3 in The Mice (Higdon et al. 2010), Arp
284-SW, SQ-A in Stephan's Quintet (Higdon \& Higdon 2008, Higdon et
al. 2010), and TDG~N and S in NGC 5291 (HHM06).

\section{OBSERVATIONS AND DATA ANALYSIS}

\subsection{Spitzer Photometry}

Infrared images for the majority of the sample were obtained as part
of our earlier programs. The IRAC 3.6 \ums, 4.5 \ums, 5.8 \um and 8
\um data for NGC 5291 were presented in HHM06; for Stephan's Quintet
in Higdon \& et al. (2008; Boquien et al. 2009; Higdon et al. 2010;
Cluver et al. 2010) and Arp 105 in Higdon \& et al. (2008; Boquien et
al. 2009; Higdon et al. 2010). The MIPS (Rieke et al. 2004)
observations are presented here for Arp 102 and NGC 5291 (also
presented in Higdon et al. 2010, Boquien et al. 2010). The remainder were presented in
Smith et al. (2007a).

The IRAC data were processed using the IRAC pipeline. To be
considered a ISFO a source must satisfy the following criteria: (1) it
must coincide with a tidal feature, either in the optical or HI, (2)
it must be point like in the 8 \um images, with at least $6$
contiguous pixels $\ge 3\sigma$ relative to the sky background, and
(3) it must be detected to $\ge 3\sigma$ in both the 3.6 \um and 8
\um bands using aperture photometry.

To estimate the likelihood that our ISFO candidates are unrelated
background sources we multiply their corresponding $8 \mu$m source
surface densities (Fazio et al. 2004a) times the surface area of the
plume or tail they are found in. For example, the four ISFO candidates
in the northern plume of Arp~65 have 8 \um magnitudes from $14.02$ to
$13.85$, with an average of $13.94$. From the 8 \um source counts in
Fazio et al. (2004a) we would expect a surface density of $0.13$
arcmin$^{-2}$ for objects in this magnitude range. The northern plume
covers $0.13$ arcmin$^{2}$, which gives a less than $2\%$ probability
that one of the four is an unrelated background source.  We derive
similar estimates for our other ISFOs, except for Arp 102, where
because of the relative faintness of the ISFOs we estimate a $20\%$
probability that one of the two candidates is unrelated. Interestingly
enough, one of these (Arp 102-N2) appears to be a background
elliptical (or possibly an old stellar cluster) based on its declining
3.6 \um through 8 \um spectral energy distribution (SED, see Table 2).

Flux densities and limits were derived using circular apertures. We
adopt the zero-point flux densities in the four IRAC bands as given in
Reach et al. (2005), i.e., $280.9 \pm 4.1$, $179.7 \pm 2.6$, $115.0
\pm 1.7$, and $64.13 \pm 0.94$ Jy at $3.6$, $4.5$, $5.8$, and 8.0
\ums, respectively.  For most sources the aperture radii are either
$3\arcsec$ or $5\arcsec$, with background measurements derived from
annuli with inner/outer radii of $3\arcsec-7\arcsec$ or
$5\arcsec-10\arcsec$ respectively. A small subset of sources (e.g.,
Arp 105-N1) were more extended (though still core dominated) and
required larger sized apertures. Aperture corrections were applied
using the recommended values in the Spitzer Observers Manual
(SOM)\footnote{\url{http://ssc.spitzer.caltech.edu/documents/som/}}.
Measured positions are listed in Table 1, while flux densities and $1
\sigma$ uncertainties for 67 ISFOs along with the object and sky
aperture sizes used are listed in Table 2. For non-detections we
calculate upper-limits ($3 \sigma$) defined as $3 \sigma \sqrt{N}$,
where $\sigma$ and $N$ are the $rms$ and number of pixels in the
source aperture respectively. These are also listed in Table 2. It is
worth noting that in Stephan's Quintet there are six objects detected
in $^{12}$CO(J$=$1-0) by Petitpas \& Taylor (2005) that are not
detected at 8 \ums.

Preliminary data processing and calibration were performed on the MIPS
24 \um data using the standard pipeline, producing a set of basic
calibrated data products. The data were further processed using the
MOPEX software package (version 18.2.2; Makovoz \& Marleau 2005).
This consisted of deleting the defective first frames in each scan,
recalculating flat-fields and background levels, and re-mosaicing each
field. The diffraction limited resolution at 24 \um is $5.7\arcsec$,
and a few targets are sufficiently bright to show faint Airy
diffraction rings (see Table 2). The $1 \sigma$ sensitivities vary
from one target field to another, but are typically $\sim 0.1$ MJy
sr$^{-1}$. We adopt a zero-point flux density of 7.14 Jy at 24 \ums,
though it should be noted that this value is uncertain at the $\pm
5\%$ level (see Engelbracht et al. 2007 for a discussion of this and
other calibration issues). 24 \um flux densities (or $3 \sigma$
limits) were derived for each ISFO candidate using circular apertures
of radii $6\arcsec$ or $13\arcsec$, depending on the source brightness
and degree of isolation. Local sky emission was measured within annuli
with inner/outer radii of $20\arcsec$ and $32\arcsec$. These are also
listed in Table 2. Point-source aperture corrections have been applied
using the recommended values in the MIPS Users Guide. $3 \sigma$
upper-limits are calculated for non-detections with $3\sigma \sqrt{N}$
as with IRAC. As a check of our photometry we derived flux densities
using the point-source fitting algorithm APEX in the MOPEX photometry
package and found very good agreement (i.e., within the quoted $1
\sigma$ uncertainties). In this work we will quote the flux densities
obtained using circular aperture photometry to be consistent with the
IRAC data, for which point-source fitting is not recommended.

\subsection{Spitzer Spectroscopy}

In this paper we present Spitzer spectra of ten ISFOs (sees Tables
3-5). The IRS low resolution spectra (IRS-LORES) were obtained with
the Short-Low module (IRS-SL), which operates between 5.2 and 7.7 \um
(IRS-SL2) and 7.4 and 14.5 $\mu$m (IRS-SL1) and the IRS Long-Low
module (IRS-LL), which operates between 14 and 21.3 \um (IRS-LL2) and
19.5 and 36.0 \um (IRS-LL1).  IRS-SL and IRS-LL have a resolving
power, R $= \frac{\lambda}{\Delta\lambda}$ $\sim$60-130. The high
resolution data (IRS-HIRES) were obtained using the IRS Short-High
module (IRS-SH), which encompasses the range 9.9 - 19.6 \ums, and the
IRS Long-High module (IRS-LH), which spans 18.7 - 37.2 \ums.
IRS-HIRES has a resolving power, $R = \frac{\lambda}{\Delta\lambda}
\sim 600$.

Observations were made in the IRS Staring Mode AOR with a high
accuracy blue peak-up using a star from the Two Micron All Sky Survey
catalog (Cutri et al. 2003).  We obtained a series of 4
$\times$ 60 s exposures in each of the two IRS-SL bands, and 4
$\times$ 120 s exposures in the two IRS-LL bands. The IRS HIRES data
was obtained with 3 $\times$ 120 s and 14 $\times$ 60 s for IRS SH
and IRS LH respectively. The Staring Mode AOR splits the integration
time between two nominal nod positions on each slit.  Only Arp 87-N1
was observed in all the IRS apertures.  Some of the data obtained as
part of the SAINTS program is not presented here as it was either of
too low signal to noise to be useful, or the slit orientation resulted
in data where emission from the ISFO was blended with the parent
galaxy. IRS-SL spectra were obtained for Arp 72-S1, Arp 82-N1, Arp
84-N1 Arp 87-N1 (Higdon et al. 2010), ``Ambartsumian's knot'' (Arp
105-S2, Higdon \& Higdon 2008; Boquien et al. 2009; Higdon \& Higdon
2010), Arp242-N3 in the Mice (Higdon et al.  2010), Arp284-SW, SQ-A
in Stephan's Quintet (Higdon \& Higdon 2008; Higdon et al. 2010), and
TDG~N and TDG~S in NGC 5291 (HHM06). There are IRS-LL spectra for Arp
72-S1 and Arp 82-N1, and no HIRES observations of Arp 105-S2 and Arp
242-N3.

The spectral data were processed as far as the un-flat-fielded two
dimensional image using the standard IRS S18.5 pipeline (see the SOM
for further observing mode and pipeline details). Due to the crowding
of the Spitzer focal plane the individual IRS apertures are not
aligned, e.g., SH is orientated $\sim$ 85$^{\circ}$ with respect to
LH. Depending on the orientation on the sky the slit may contain
emission from more than one ISFO and$/$or the parent galaxy. The
spectra were extracted and sky subtracted manually using the
Spectroscopy Modeling Analysis and Reduction Tool (SMART, Higdon et
al. 2004) software to ensure that the extracted spectra were dominated
by light from a single ISFO and that the sky spectrum was
representative of the background emission.
 
The IRS-LORES sky data is acquired serendipitously during
the standard observation. For example, SL1 and SL2 are one long slit,
so when the source position is nodded between two positions in SL1,
data is also acquired simultaneously in SL2 which is pointing off
source. Combining the SL1 and SL2 observations offers the potential
for a SL1-sky spectrum selected in either the region of the slit
adjacent to the source during the SL1 observation (nod-data) or from a
region in one of the two off-source SL1 spectra acquired during the
SL2 on-source nod-observation. To maximize the signal-to-noise all
data was inspected manually using SMART. The data from each nod for
each slit were collapsed in the cross-dispersion direction to
determine manually which data should be co-added to define the
sky. The sky data so defined are then coadded and subtracted from the
two dimensional coadded source data.  The sky-subbed IRS-LORES
two-dimensional data was then collapsed in the cross-dispersion
direction to determine the position of the ISFO in the slit and to
check for blending with other objects. A column was extracted, whose
width in the cross-dispersion direction scales with the instrument
point spread function (see the SOM, Higdon et al. 2004, and HHM06 for
further details). A full aperture extraction was applied to the HIRES
data which included dedicated off-source sky observations.

The extracted ISFO spectra were flat-fielded and flux-calibrated by
extracting and sky subtracting un-flat-fielded observations of the
calibration stars HR 7341 (IRS-SL/LL) and $\xi$ Dra (IRS-SH/LH), which
are subsequently divided by the corresponding template (Decin et
al. 2004, Cohen et al. 2003) to generate a one-dimensional relative
spectral response function (RSRF).  The RSRF was then applied to the
ISFO spectra to produce flux-calibrated spectra. The final step of
stitching spectral orders together was accomplished in the following
manner: the data from each IRS module is sky subtracted. Using the
photometry routines in SMART the SL spectra are
scaled to match the IRAC 8 \um photometry. If MIPS 24 \um photometry
is available the LL and LH spectra are scaled to match the value
reported in Table 2. If there is no 24 \um photometry the spectra are
stitched to the 8 \ums-scaled SL spectra using the overlapping
continuum regions. For example, for the observation of the ISFO in Arp
72 the LL slit was orientated along the tidal tail, the SL data was
scaled by 1.69, LL by 0.76, SH by 4.0 and LH by 0.63. As a check on
this calibration method we looked at the flux ratio of the brightest
lines observed in Arp 72-S1. The ratio of the high resolution line
fluxes to low resolution, was R(\neiiil) = 0.89, R(\siiil) = 1.03 and
R(\sivl = 1.00). The aperture scale factors are listed in Table 3.

Figure 2 shows the IRS low resolution spectra. The data are plotted in
each object's rest-frame. The broad emission features from PAHs at
6.2, 7.7, 8.6, 11.3, and 12.6 \um are clearly present.  Some of the
spectra show emission from \sivl as well as 0-0 S(3) 9.67 \um and 0-0
S(1) 17.04 \um lines from molecular hydrogen. Emission from \neiil ~is
blended with the PAH feature, but the line is distinct in the
IRS-HIRES observations. Figure 3 shows the line profiles from the
IRS-HIRES observations. The line strengths are measured using SMART
and are listed in Table 4. The strengths of the PAH features are
measured using PAHFIT (Smith et al. 2007b) and are listed in Table 5.
For Arp 72-S1 and Arp 82-N1 we have scaled LL spectra, which are
included in the PAHFIT. The results do not change when only the SL
data is fitted. For consistency between the data sets for the ISFOs
reported here we scale the NGC 5291-S spectrum to match the IRAC 8 \um
photometry (the NGC 5291 N spectrum matches the photometry), and fit
the PAHs in both NGC 5291 N and NGC 5291 S spectra using PAHFIT.
Apart from the 11.3 \um to 7.7 \um PAH ratio for NGC 5291-S, the PAH
ratios are within 20\% of those reported in HHM06.

Extinction due to silicate absorption is best constrained for Arp
72-S1 and Arp 82-N1 where the low resolution spectral coverage
includes the two silicate features, the Si-O stretching mode at 9.7
\um and O-Si-O bending mode at 18 \ums. The fits show no evidence for
absorption. The majority of our sample are limited to the scaled SL
observations where it is difficult to discriminate between silicate
absorption at 9.7 \um and PAH emission on either side of the silicate
feature.  For the analysis in this paper we assume all targets have
minimal line-of-sight silicate extinction (A$_{\rm v} \le$ 3
mag). This is also consistent with the results for the Spitzer
Infrared Nearby Galaxies Survey (SINGS, Kennicutt et al. 2003) where 5
- 38 \um spectra were available. In that sample only 8 of the 59
galaxies had measureable extinction at 9.7 \um (Smith et
al. 2007b).

\section{RESULTS AND DISCUSSION}

\subsection{Characterizing The PAHs}

The 5-38 \um SL spectra are dominated by broad emission features.
These features are generally attributed to the spontaneous emission of
an infrared photon from PAHs that have been vibrationally excited
following a single optical/UV photon absorption. The resulting
emission spectrum depends on the heat capacity (size) of the PAH, and
is largely independent of the starlight intensity. For example, the
models by Li \& Draine (2001) assume the carbonaceous grains have
PAH-like properties when the grains have $\le$10$^3$ carbon atoms,
corresponding to a size of $\sim$ 10\AA, and graphitic properties when
much larger ($\ge$ 100 \AA).  Emission from the larger grains will
peak at much longer wavelengths than observed with the IRS. The PAH
spectrum may also depend on the overall structure of the PAH, i.e.,
linear vs. concentrated (Bakes et al. 2001).

PAHs have a low ionization potential (6-7 eV) and can be easily
ionized through photoelectric emission (PAH$^+$) and electron capture
(PAH$^-$) (Allamandola et al. 1985). The grain charging is dependent
on the gas temperature, electron density, the ultraviolet radiation
field, and the cross-sections for PAH electron capture and
photoelectric emission (Draine \& Li 2007, Weingartner \& Draine
2001).

The 11.3 \um feature is a vibrational C-H out of plane bending mode
from ``solo'' CH groups implying large PAHs. The 6.2 and 7.7 \um
features are from a C-C stretching mode. Unlike the 11.3 \um PAH, the
7.7 \um feature has a much larger absorption cross-section in PAH$^+$
ions than neutral PAHs (see Figure 3 in Draine \& Li 2007) i.e., the
relative band strength of the PAH$_{11.3\mu m} /$PAH$_{7.7 \mu m}$
ratio (hereafter, R$_{11/7}$) is large for neutral PAHs and decreases
by an order of magnitude for ionized PAHs (Draine \& Li 2001, see
Figure 16 and references therein). In an empirical study of Galactic
star forming regions Galliano et al. (2008) concluded that the main
driver of R$_{11/7}$ is the PAH ionization fraction. These differences
in the relative PAH band strengths are driven by the relative
ionization to recombination rates, which is a function of the ratio of
intensity of the stellar radiation field to the electron density.
Other factors that affect the R$_{11/7}$ ratio to a lesser
extent are the PAH size, the extinction, and the hardness of the ISRF
(Draine \& Li 2001, Galliano et al. 2008).

Figure 4 displays the PAH$_{6.2\mu m} /$PAH$_{7.7 \mu m}$ ratio
(R$_{6/7}$) versus R$_{11/7}$ for 9 ISFOs.  For comparison we selected
sources in the literature which have a signal-to-noise ratio $\ge$3
for each PAH band. The PAH flux was measured using PAHFIT (Smith et
al. 2007b). We include results from SINGS, which is a comprehensive
set of observations of the inner 1-10 kpc$^2$ of 75 nearby normal
galaxies (Kennicutt et al. 2003). Each galaxy is identified as having
either a HII-like or weak AGN-like nucleus.  Figure 4 also shows 24
HII-like and 24 AGN-like SINGS galaxies (Smith et al. 2007b), 6 BCDs
(Hunt et al. 2010), an additional 6 BCDs, 16 starburst galaxies, 27
sources within three local giant HII regions: NGC 3603 in the Milky
Way, 30 Dor. in the LMC, and N 66 in the SMC (Lebouteiller et al. 2011
and references therein).

Two thirds of the ISFO sample have an R$_{6/7}$ similar to the SINGS
and starburst galaxies. The SINGS-AGN sources tend to have a smaller
ratio.  Theoretical models show that R$_{6/7}$ increases as the PAHs
get smaller (Draine \& Li 2001). In general terms the PAH size
distribution is weighted toward a population with large PAHs in
sources with a weak AGN, medium sized in the SINGS-HII and starburst
galaxies, and smallest in the BCDs and in the three local giant HII
regions (NGC 3603, 30 Dor. and N 66).  In Arp 105-S2, Arp 82-N1 and
Arp 284-SW1 the relative strengths of the PAH 6.2 and 7.7 \um emission
are similar to what is observed in BCDs and local giant HII regions
corresponding to the models with bright emission from small PAHs,
with the number of carbon atoms, N$_C$ $\le$ 50, as opposed to a model
with bright emission from larger PAHs with N$_C$$\sim$ 100, which is
appropriate for the other ISFOs and the SINGS-HII and starburst
galaxies (see Figure 16 in Draine \& Li 2001).

The ISFO R$_{11/7}$ ratios are consistent with the range seen in the
SINGS AGN and HII sources and starburst galaxies, and some of the NGC
3603 regions in the Milky Way.  The ISFO R$_{11/7}$ is lower than what
is observed in the regions in 30 Dor. and N 66 and many of the BCDs.
Again comparing the data to the Draine \& Li (2001) models the ISFOs
are consistent with a mixed charge distribution with a large fraction
of the PAHs being neutral ($\ge$ 50\%) in order to produce the
observed bright 11.3 \um PAH emission. R$_{11/7}$ for the local giant
HII regions is consistent with a mostly neutral PAH population. The
ISFO R$_{11/7}$ and R$_{6/7}$ (except R$_{6/7}$ for three ISFOs) values
are comparable to the global properties of spirals and starburst
galaxies rather than BCDs and sources in three local giant HII
regions.

\subsection{ISRF Hardness and Young Stellar Population}

Both the neon and the sulfur line flux ratios can be used to constrain
the properties of the underlying starburst. The \siv$/$\siii~ and
\neiii$/$\neii~ line flux ratios give a measure of the hardness of the
ISRF, which depends on the effective temperature of the ionizing stars
and the ionization parameter. Lowering the metallicity produces hotter
main sequence stars for a given mass, and the radiation is harder due
to reduced line blanketing and blocking.  Verma et al. (2003) used the
\siv$/$\siii~ and \neiii$/$\neii~ line flux ratios as an excitation
diagnostic for a sample of starburst and BCD galaxies.  Their data is
reproduced in Figure 5 along with our results for the ISFOs, a sample
of BCDs from Hunt et al. (2010), clumps in the ring galaxy Arp 143
(Beiraro et al. 2009) and a sample of 27 sources in three local giant
HII regions in NGC 3603, 30 Dor., and N 66 (Lebouteiller et
al. 2011). The ISFOs form a group between the starburst galaxies in
the bottom left quadrant of the plot, and the BCDs in the upper right
quadrant. The line flux ratios indicate moderate excitation consistent
with our results for NGC 5291 N and NGC 5291 S (HHM06), and with the
interacting galaxy pairs NGC 4038/NGC 4039 and NGC 3690B/C (Verma et
al. 2003), the BCDs Haro 3 and Mrk 996 (Hunt et al. 2010) and regions
in NGC 3603 and 30 Dor.  Arp 84-N1 has a lower neon ratio consistent
with the starburst galaxies.

Hunt et al. (2010) observed an increase in the \neiii$/$\neii ~flux
ratio with decreasing oxygen abundance. The BCDs tend to have larger
neon ratios and the starburst sample tend to have lower values. 
The sources in the giant HII regions mostly have neon line flux ratios
greater than one. Both Madden et al. (2006) and Lebouteillier et al
(2011) argue that PAHs are destroyed in harsh radiation fields with
\neiii$/$\neii $\ga$ 2$-$3. The ISFOs have neon line flux ratios similar
to those observed in the regions in NGC 3603 (solar metallicity). The
ISFO neon and sulfur line flux ratios are lower than the observed
ratios in the sources in N 66 (0.2 Z$_{\odot}$) and the majority of
sources in 30 Dor. (0.6 Z$_{\odot}$). The individual sources in NGC
3603, 30 Dor. and N 66 exhibit a spread in the line flux ratios for a
given abundance as they sample different parts of the giant HII
regions.  Only three ISFOs from our sample of nine with neon line flux
data have abundance estimates (see Section 2) and the average neon
line flux ratio is 1.3. If one-third to solar metallicity or higher is
typical for the ISFOs, then the ISFOs fill the gap in the distribution
of the global neon line flux as a function of metallicity, forming a
group between the starburst and BCD galaxies.

An upper limit to the age of the most recent episode of star formation
can be estimated by comparing the observed \neiii$/$\neii ~line flux ratio
to values generated for a range of population synthesis models (e.g.,
Thornley et al. 2000, Madden et al. 2006). Assuming the metallicity is
$\sim$ 0.2-1 Z$_{\odot}$ and that the star formation occurred in a
single burst gives an upper limit of 6 Myr, i.e., recent star
formation in the ISFOs.

In Figure 6 we plot R$_{11/7}$ as a function of the \neiii$/$\neii
~line flux ratio. Hunt et al. (2010) observed an increase in
R$_{11/7}$ in BCDs with neon ratios greater than one. The 7.7 \um PAH
emission arises from a fairly broad range of grain size (Schutte et
al. 1993). Hunt et al (2010) propose that a hard intense ISRF may
destroy the small grains which contribute to the 7.7 \um PAH emission
whilst not impacting on the intensity of the 11.3 \um emission.  If
the neon line flux ratio is a good proxy for the hardness of the ISRF
in the ultraviolet, the observed scatter in R$_{11/7}$ for objects
with \neiii$/$\neii $>1$ suggest that the hardness of the ISRF, as
measured by the infrared line flux ratios, is not the main driver of
R$_{11/7}$. For example, there is no significant increase in
R$_{11/7}$ in the six ISFOs with \neiii$/$\neii ~$>1$ and in one case
for a BCD when \neiii$/$\neii ~$>$ 10. Similarly, moderately high
R$_{11/7}$ values are observed in two SINGS-AGN galaxies and three
sources in the local Giant HII regions which have \neiii$/$\neii
~$<1$. Small PAHs need not be destroyed. The 7.7 \um emission is
stronger in ionized PAHs as compared to neutral PAHs, if the PAHs tend
to be neutral in the three local giant HII regions and in BCDs, then
the difference in the PAH ion fraction in these sources, as compared
to ISFOs and SINGs galaxies, could be driving the R$_{11/7}$.

Figure 7 shows the PAH 8.6 \ums/7.7 \um ratio (hereafter R$_{8/7}$) as
a function of the \neiii$/$\neii ~line flux ratio. The PAH 8.6 \um
emission is from large PAHs with a minimum of 100 carbon atoms
(Bauschlicher 2008) whereas the 7.7 \um PAH emission is from a mixture
of small and large PAHs. Both features have larger capture
cross-sections for PAH$^+$ ions than for neutral PAHs.  There is a
large scatter in R$_{8/7}$ for each source type, and especially when
\neiii$/$\neii ~$\ga$ 1.  $<$R$_{8/7}$$>$ = 0.20 $\pm$ 0.05 for 8
ISFOs which is consistent with the ratio of $<$R$_{8/7}$$>$ = 0.17
$\pm$ 0.03 for 45 SINGS galaxies. $<$R$_{8/7}$$>$ = 0.45 $\pm$ 0.33
for 19 sources in the three local giant HII regions ($<$R$_{8/7}$$>$ =
0.36 $\pm$ 0.12 for 13 sources in 30 Dor.), and $<$R$_{8/7}$$>$ = 0.31
$\pm$ 0.13 for 8 BCDs (this does not include Mrk 1315 with R$_{8/7}$=
1.8).

The ionization balance depends on the gas temperature, electron
density and ultraviolet radiation field (Weingartner \& Draine
2001). The PAH ion fraction can drive the relative strengths of the
PAH bands. As the PAH ion fraction increase and/or the fraction of
large PAHs increases the PAHs can absorb longer wavelength photons and
the observed emission is less dependent on the hardness of the
radiation field. The relative PAH band strengths in the ISFOs are
consistent with a mixed population of grain sizes that contain a much
larger fraction of PAH ions compared to PAHs in BCDs and sources in
local giant HII regions.

\subsection{Warm Molecular Gas}

Warm molecular gas is detected in 7 of 8 (88\%) of the ISFOs observed
with IRS-HIRES (see Table 4).  To derive the mass of warm molecular
hydrogen we assume that the emission is optically thin. The critical
densities of the low J levels are relatively low (n$_{\rm cr}< 10^3$
cm$^{-3}$) and we assume that the populations are in LTE. Adopting an
ortho to para ratio of 3 we construct excitation diagrams (see HHM06
for further details). The excitation temperature is the reciprocal of
the slope of the linear fit to the natural logarithm of the number of
molecules divided by the statistical weight in the upper level of each
transition versus the upper level of each transition. The mass is
derived using the calculated excitation temperature and the highest
signal-to-noise line, the IRS-HIRES 0-0 S(1) 17.03 \um line. Both the
stitching together of the individual spectra (Section 3.2) and the use
of a single temperature component model add uncertainty to the derived
molecular mass.

For Arp 72-S1 we constructed an excitation diagram using the IRS-HIRES
S1 line along with the IRS-LORES 0-0 S(0) and 0-0 S(2) lines. For a
single component model the excitation temperature is T$_{ex}$ = 210
$\pm$ 9 K with $\sim 10^6$ \msun of warm \mhs. For Arp 82-N1
we combined the IRS-HIRES detections of the 0-0 S(0) and 0-0 S(1)
lines with the IRS-LORES measurements of the 0-0 S(2) and 0-0 S(3)
lines. The excitation temperature is T$_{ex}$ = 214 $\pm$ 6 K with $\sim 10^6$ \msun of \mhs. 

In Arp87-N1, Arp284-SW and SQ-A we only detect the 0-0 S(1)
line. Adopting a value of T$_{ex}$ = 200 K we estimate that there is
$\sim 10^6$ \msun of warm \mh in Arp 87-N1 and Arp 284-SW1 and $\sim
10^7$ \msun of \mh in SQ-A.  These results are similar to those
derived for NGC 5291~N and NGC 5291~S (HHM06, note a higher excitation
temperature of $\sim$400 K was estimated with a fit limited to the 0-0
S(1) and 0-0 S(2) lines. Lowering the excitation temperature gives a
higher gas mass).

ALMA and CARMA observations are needed to constrain the cold molecular
gas mass and derive the warm gas fraction.  Based on the limited
observations of ISFOs (HHM06 reported that the warm gas mass in NGC
5291-N and NGC 5291-S is less than 1\% of the cold gas mass inferred
from $^{12}$CO (1-0) observations of Braine et al.  2001) we expect it
to be $<1$\%, which is similar to the warm gas fraction in a sample
of 59 ULIRGs (Higdon et al 2006a). These two very different classes of
objects have warm \mh excitation temperatures and masses consistent
with emission from gas in PDRs.

\subsection{Characterizing the Mid-infrared Spectral Energy Distribution}

\subsubsection{Spitzer Broadband Colors}

In Figure 8 we show four Spitzer two-color diagrams for our sample of
67 ISFOs. For comparison the SINGS sample (Dale et al 2005), a sample
of star forming dwarf galaxies collated from the literature (Smith \&
Hancock 2009 and references therein), a sample of clumps identified in
the disks of interacting galaxies collated from the literature (Arp
24, Arp 82, Arp 244, Arp 284 and Arp 285; Lapham et al. 2013 and
references therein) and some template colors of HII regions, planetary
nebulae (PNe), and supernovae remnants (SNRs) in M 33 (Verley et
al. 2007) are included. It should be noted that the individual sources
used to form each of the M~33 templates exhibit a wide range of
Spitzer colors (for example, see the middle panel of Figure 12 in
Verley et al. 2007).

In general, the ISFOs have Spitzer colors similar to star forming
clumps in the disks of interacting galaxies. However, the ISFOs are
redder on average in \iracb - \iracd colors than the integrated colors
of the SINGS galaxies or star forming dwarfs. This is likely a
consequence of bright PAH emission in the 8 \um band, as observed in
the ISFO spectra displayed in Figure 2.

This difference is apparent in panel (A) of Figure 8, where the ISFOs
lie in two zones. A red population with \iracb - \iracd $>$ 3
encompasses 70\% of the ISFOs (47/67).  Half of these (one third of
the whole sample) have \iracb -\iracd $>$ 3.7. These are redder than
the global colors of the majority of the comparison galaxies, most
likely due to enhanced non-stellar emission, particularly PAHs in the
8 \um band. The remaining third, which include the ISFOs in Arp 242,
are blue in this color, similar to the median value of the SINGS
galaxies. These may be more quiescent star forming regions, as they
tend to be bluer in the other Spitzer colors as well.

ISFOs have a \iraca - \iracb color similar to that of star forming
clumps in the disks of interacting galaxies, as well as the global
colors of spirals and dwarfs (panels A and B of Figure 8).  A \iraca -
\iracb color close to zero is expected if the emission in both bands
is predominantly starlight. 

As shown in panels (B)-(D) in Figure 8, the ISFOs have \iraca -
\mipsa colors similar to those of the interacting disk star forming
knots, SINGS-HII, dwarf galaxies and half of the SINGS-AGN galaxies.
Some of the dwarfs, star forming clumps in interacting galaxies, and
one ISFO (Arp 82-N1) are redder in this color than the SINGS galaxies,
which could be caused by intense star formation boosting the 24 \um
dust emission relative to the starlight.

The colors of ISFOs and the star forming clumps in interacting galaxy
disks are clearly separated from dwarfs when the \iraca - \mipsa color
is plotted as a function of either the \iraca - \iracd (panel (D) in
Figure 8) or \iracd - \mipsa colors (panel (C) in Figure 8), but not
the \iraca - \iracb color (panel (B) in Figure 8).  For a given \iraca -
\mipsa color the ISFOs tend to be redder in \iraca - \iracd than the
dwarfs. Likewise for a given \iraca - \mipsa color the ISFOs are bluer
in \iracd - \mipsa than the dwarfs. These color offsets are consistent
with our spectroscopic results that show the ISFOs to have bright PAH
emission. In contrast, results in the literature show that low
metallicity dwarfs have intense UV fields and weak PAH features in the
8 \um band, while the 24 \um emission arises primarily from larger
dust grains. A search for the reddest ISFOs and star forming clumps in
the disks of interacting galaxies can be made using the criteria
\iracb - \iracd $>$ 3.7.

\subsubsection{Fraction of Dust Luminosity From PDRs}

The ISFOs were selected based on their bright 8 \um emission. Some
example 3.6 - 24 \um SEDs are shown in Figure 9. The majority of the
ISFOs have a distinctive `notched-shaped' SED, i.e., the flux falls
from 3.6 to 4.5 \ums, and then steeply rises from 4.5 to 8.0 \ums.
This spectral shape is characteristic of star forming regions
illuminated by an intense stellar field, and was observed in the ISFOs
in NGC 5291 (HHM06). Some of the SEDs have a flux that decreases from
3.6 \um through 5.8 \um before rising. In one case, Arp 102-N2, the
SED slopes downward from 3.6 \um through 24 \ums, which is
characteristic of evolved stellar populations, i.e., an elliptical
galaxy. This may be a ISFO devoid of gas and dust or more likely, an
object not associated with Arp 102.

Draine et al. (2007) fit physical dust models (see Draine \& Li 2007)
to 65 SINGS galaxies using IRAC, MIPS and in some cases sub-mm
photometry. The IR/sub-mm emission from the dust depends not only on
the amount of dust, but also on the location. Draine et al. (2007)
conclude that a large fraction of the dust mass in the SINGS galaxies
is in the diffuse ISM. This diffuse ISM can be characterized with an
ISRF with an approximately constant intensity, Umin = $\alpha$U, where
$\alpha$ is a scale factor $\le$ 10, U= 0.88 G$_{o,dens}$, where
G$_{o,dens}$ is the ratio of the 6-13.6 eV energy density relative to
the value of 5.29 $\times 10^{-14}$ erg cm$^{-3}$ measured by Habing
(1968) for the local ISRF.

Most of the dust mass in galaxies is relatively cool and radiates
longward of 60 \ums. A small fraction of the total dust mass resides
in regions illuminated by a more intense radiation field (U $>$ Umin),
for example PAHs in PDRs. In one in seven SINGS galaxies the emission
from these intensely illuminated regions (likely PDRs) contribute a
significant fraction ($\ga 30$\%) of the total power emitted by the
dust grains (Draine et al. 2007).

Our dataset is limited to photometry at wavelengths $< 30$ \um and is
insufficient to be used to fit models, and to derive the dust mass and
properties in detail. However we can use the Draine \& Li (2007)
models, and the Draine et al. (2007) results for the SINGS galaxies,
to estimate what fraction of the dust luminosity is likely to come
from intensely illuminated (U $> 10^2$) regions (e.g., PDRs
f$_{Intense}$), and where the bulk of the IR emission originates in
the ISFOs.

Emission from dust (including PAHs) in the ISM heated by
a diffuse ISRF, has $\lambda$F$_{\lambda}(8
\mu m)$ $ > \lambda$F$_{\lambda} (24 \mu m)$ (Draine \& Li 2007).  In
regions illuminated by intense starlight the emission at 24 \um is
from both single-photon heating of PAHs and multi-photon heating of
larger grains. Emission from these more intensely illuminated regions,
i.e., PDRs, has $\lambda$F$_{\lambda}(8 \mu m)$ $ \le
\lambda$F$_{\lambda}(24 \mu m)$.

Following the recipe in Draine \& Li (2007) we will assume that the
flux measured in the IRAC 3.6 \um band apertures is stellar in origin,
and can be approximated by emission from a 5000 K blackbody. We
extrapolate this stellar contribution to 8 \um and 24 \um in order to
calculate the non-stellar flux in these bands using F$_{\nu(NS)}$(8
\ums) = F$_{\nu}$(8 \ums) - 0.260 F$_{\nu}$(3.6 \ums) and
F$_{\nu(NS)}$(24 \ums) = F$_{\nu}$(24 \ums) - 0.0326 F$_{\nu}$(3.6
\ums). In the absence of extinction and the presence of red giants and
AGB stars this simple blackbody subtraction will overestimate the
non-stellar (dust) emission. In the presence of bright 3.3 \um PAH
emission the stellar emission will be over-estimated. The 3.3 \um PAH
emission in the 3.6 \um band is sensitive to the abundance of the
smallest PAHs and the PAH charge state (Draine \& Li 2007).

The non-stellar emission in the 8 \um band is dominated by single
photon heating of PAHs, and it is proportional to total starlight
power absorbed by the dust (i.e., $\nu F_{\nu}(71 \mu m) + \nu
F_{\nu}(160 \mu m)$ flux; Draine \& Li 2007). Using the photometry for
23 SINGS-HII spiral galaxies (Dale et al. 2007) we find $<\nu
F_{\nu}(71 \mu m) + \nu F_{\nu}(160 \mu m)> = (5.00 \pm 1.5) \times
<\nu F_{\nu NS}(8 \mu m)>$.

Draine \& Li (2007) show that the fraction of the total emission
supplied from high intensity (likely PDR) regions with U $> 10^2$ is related
to the observed non-stellar power at 24 \um and 8 \ums.

\begin{equation}
P_{24} - 0.14P_{8} = \frac{<\nu F_{\nu NS}(24 \mu m) >- 0.14 <\nu F_{\nu NS}(8 \mu m)>}
{<\nu F_{\nu} (71 \mu m)> + <\nu F_{\nu}(160 \mu m)>}
\end{equation}

The numerator in Equation (1) is a measure of the emission from 
large grains that will only radiate at 24 \um when exposed to
intense starlight. The 24 \um emission from  single-photon heating is
subtracted using the weighted 8 \um flux, as the 8 \um emission is
dominated by single-photon heating of PAHs. f$_{Intense}$ for the SINGS
galaxies is given by the following relation (see Figure 24, Draine et
al. 2007):

\begin{equation}
f(Intense; L_d; U>10^2) \sim 1.05 (P_{24} - 0.14P_{8} - 0.035)^{0.75}
\end{equation}

 The ISFOs and SINGS-HII galaxies have similar \iracd - \mipsa colors,
 and we assume the ISFOs follow this relation. The ISFOs were not
 observed with MIPS in the 71 \um and 160 \um bands. However by
 substituting $(5.00 \pm 1.5) \times <\nu F_{\nu NS}(8 \mu m)>$ for
 the denominator in Equation (1) we can estimate f$_{Intense}$ using
 Equation (2). The largest uncertainty is the assumption that the PAH
 abundance in the ISFOs is similar to that observed in the SINGS
 sample. We do not know of any objects with an enhanced PAH abundance
 relative to SINGS-HII, but a PAH deficit is observed in AGN
 environments and low metallicity systems.  The ISFOs in our spectral
 sample with published metallicities have values of one third solar or
 higher, this is above the regime where dwarfs are observed with a PAH
 deficit (e.g., Madden et al. 2006), and the ISFOs have bright PAH
 emission. If the other ISFOs in the photometric sample have a lower
 abundance and a corresponding PAH deficit, we will underestimate the
 total starlight power absorbed by the dust using this method, and
 overestimate f$_{Intense}$. However, we do not expect there to be a
 PAH deficit, the ISFO spectra have bright 8\um features, and the
 larger photometric sample have a `notched-shaped' SED characteristic
 of star forming regions with bright PAH emission.

Some example ISFO SEDs are given in Figure 9. We have overlaid the
SED for NGC 3190, which is characteristic of a region dominated by
emission from grains in the diffuse ISM, and Mrk 33 which is
characteristic of emission from grains in PDRs. We have 24 \um data
for 57 sources.  The emission between 8 \um and 24 \um in the ISFOs is
a mix of these two templates.

For the 41 ISFOs with 24 \um detections we can estimate $f_{Intense} $
directly, and for an additional 16 ISFOs the 24 \um limit is
consistent with a $\le$ 10\% $f_{Intense}$.  Two thirds (37/57) of the
ISFOs are dominated by emission from the diffuse ISM with a $\la $10\%
contribution from PDRs based on their 8 - 24 \um emission.
%Ten ISFOs have 15\% $\la f_{Intense} \la 25$\%. 
One in six ISFOs have a significant PDR component, with $f_{Intense} \sim
30$\% in Arp 72-S2 through Arp 72-S5, Arp82-S2, Arp 82-S4, NGC
5291-13, NGC 5291-26 (TDG-N), NGC5291-28, and $f_{Intense} \sim 60$\% in
Arp 82-N1. This is comparable to SINGS, where one in seven galaxies
have $f_{Intense} \ga 30$\% (Draine et al. 2007).

ISFOs with $f_{Intense} \ge 30$\% tend to have a red \iracb - \iracd
color. However the reddest ISFOs (\iracb - \iracd $>$ 3.7) are not
correlated with $f_{Intense}$. The PAH emission in ISFOs is predominantly
from a diffuse ISM.

\section{Summary}

We have investigated a sample of 67 ISFOs from 14 systems, consisting
of 13 interacting galaxy pairs and Stephan's Quintet. The ISFOs range
from classical TDGs at the tips of tidal tails, groups of sources that
together define ``beads-on-a-string'', plus a luminous ``hinge clump''
at the base of a tidal feature. In the introduction we posed a number
of questions concerning the mid-infrared properties of ISFOs. Here we
summarize our response to those questions.

Spitzer IRS observations of 10 ISFOs in Arp 72, Arp 82, Arp 84, Arp
87, Arp 105, Arp 242, Arp 284, NGC 5291 and Stephan's Quintet indicate
that:

1) Two thirds of the ISFO sample have an R$_{6/7}$ similar to the
SINGS and starburst galaxies, which corresponds to the models with
bright emission from large PAHs (N$_C$ $\sim$ 100). In Arp 105-S2,
Arp 82-N1 and Arp 284-SW1 R$_{6/7}$ is similar to what is observed in
BCDs and local giant HII regions, corresponding to the models with
bright emission from small PAHs (N$_C$ $\le$ 50).

2) PAH models with the observed ISFO R$_{11/7}$ indicate a mix of neutral
and charged PAHs with $\ge$50\% of the PAHs being neutral. The
$<$R$_{11/7}>$ for the ISFOs is consistent with SINGS AGN and HII
sources and starburst galaxies. Sources in the three local giant HII
regions (NGC 3603, 30 Dor. and N 66) and BCDs tend to have a much
larger fraction of neutral PAHs.

3) The \neiii$/$\neii ~and \siv$/$\siii ~line flux ratios are used as
an excitation diagnostic and as a proxy for the slope of the ISRF in
the UV. The ionized gas in the ISFOs indicates moderate levels of
excitation with an ISRF that is harder than the majority of the SINGS
and starburst galaxies but softer than BCDs and local giant HII
regions. The neon line flux ratios are consistent with population
synthesis models for recent star formation, i.e., a burst of star
formation $\la$ 6 million years ago. The wide range in the observed
R$_{11/7}$ for different galaxies for a given value of neon line flux
ratio implies that R$_{11/7}$ is driven by a change in the PAH ion
fraction.  $<$R$_{8/7}$$>$ = 0.20 $\pm$ 0.05 for 8 ISFOs, and is
consistent with the $<$R$_{8/7}$$>$ observed in SINGS galaxies. The
ionization fraction can drive the relative strengths of the PAH
bands. As the PAH ion fraction increases and/or the PAH grain size
increases the PAHs can absorb longer wavelength photons and the
observed emission is less dependent on the hardness of the radiation
field.

4) Emission from the low-J rotational lines from warm molecular
hydrogen is detected in 88\% (7/8) of the ISFOs corresponding to 
$\sim 10^6$ \msun of warm \mhs.

Analysis of the IRAC and MIPS photometry for 67 ISFOs confirms that
the ISFOs have a `notched-shaped' SED, which is characteristic of star
forming regions. 

5) The ISFOs separate into two groups in \iracb - \iracd
color. The blue group (\iracb - \iracd $< 3$) has colors similar to
SINGS galaxies. The red group is redder on average than normal
spirals, dwarf irregulars and BCD galaxies. This is caused by bright
emission at 8 \ums, most likely from PAHs. The observed color offsets
between dwarfs and ISFOs when the \iraca - \mipsa color is plotted as
a function of either the \iraca - \iracd, or \iracd - \mipsa color,
are consistent with our spectroscopic results that show the majority
of the ISFOs have a ISM with bright PAHs, illuminated by diffuse
stellar light, in contrast to low metallicity dwarfs.  ISFOs have
colors similar to either star forming knots in the disks of
interacting galaxies (red \iracb - \iracd color) or global properties
of SINGS-HII galaxies (blue \iraca - \iracd color), but not the global
properties of dwarf galaxies.

6) In two thirds (37/57) of the ISFOs the infrared power is dominated
by emission from grains in the diffuse ISM illuminated by a ISRF with
G$_o$ $\le$ 10. One in six ISFOs have a significant PDR component,
with $f_{Intense} \sim 30$\% in Arp 72-S2 through Arp 72-S5, Arp82-S2,
Arp 82-S4, NGC 5291-13, NGC 5291-26 (TDG-N), NGC5291-28, and
$f_{Intense} \sim 60$\% in Arp 82-N1.

This paper compared the mid-infrared properties of ISFOs to other
galaxy types and sources in local giant HII regions. The ISFOs are
observed to have bright PAH emission with a significant
fraction of PAH ions, that is located in a diffuse ISM. This is
similar to what is observed in SINGS-HII galaxies. In contrast the
relative PAH band strength in BCDs indicates a population of mainly
neutral PAHs, consistent with the emission from sources in local
giant HII regions, which are exposed to more intense stellar radiation
fields.

\acknowledgements

This work is based [in part] on observations made with the Spitzer
Space Telescope, which is operated by the Jet Propulsion Laboratory,
California Institute of Technology under NASA contract 1407. Support
for this work was provided by NASA through Contract Number 1257184
issued by JPL/Caltech.  This research has made use of the excellent
NASA/IPAC Extragalactic Database (NED) which is operated by the Jet
Propulsion Laboratory, California Institute of Technology, under
contract with the National Aeronautics and Space Administration;
Partial funding for this work was provided by Spitzer/NASA grants RSA
No.s 1346930 (Higdon \& Higdon), 1353814 (Smith \& Hancock). We thank
Bruce Draine and the referee for constructive comments, which have
improved this paper.

\references

\reference{} Allamandola, L. J., Tielens, A. G. G. M. \& Barker, J. R. 1985, \apjl, 290, L25

\reference{} Arp, H., 1966, \apjs, 14, 1

\reference{} Bakes, E. L. O., Tielens, A. G. G. M., Bauschlicher, C. et al. 2001, \apj, 560, 261

%from hunt paper 2010 
\reference{} Bauschlicher, C. W., Peeters, E. \&  Allamandola, L. 2008, \apj, 678, 316

\reference{} Beiraro, P., Appleton, P. N., Brandl, B. R. et al. 2009, \apj, 693, 1650

\reference{} Boquien, M. et al. 2009, \aj, 137, 4561

\reference{} Boquien, M. et al. 2010, \apj, 140, 2124

\reference{} Bournaud, F \& Duc, P.-A. 2006, A\&A, 456, 481

\reference{} Braine, J. et al. 2001, A\&A, 378, 51

\reference{} Cluver, M. E., Appleton, P. N., Boulanger, F. et al. 2010, \apj, 710, 248

\reference{} Cohen, M., Megeath, T.G., Hammersley, P.L. et al.  2003, \aj, 125, 2645

\reference{} Cutri, R. et al. 2003, The IRSA 2MASS All-Sky Point Source Catalog, NASA/IPAC Infrared Science Archive

\reference{} Dale, D. A.,Bendo, G. J., Engelbracht, C. W. et al. 2005 \apj, 633,857 

\reference{} Decin, L., Morris, P. W., Appleton, P. N. et al. 2004, ApJS, 154, 408

\reference{} de Mello, D. F., Urrutia-Viscarra, F., Mendes de Oliveira, C et al. 2012, MNRAS, 426, 2441

\reference{} Draine, B. T., Dale, D. A., Bendo, G., et al.\ 2007, ApJ, 663, 866

\reference{} Draine, B. T. \& Li, A. 2001, ApJ, 551, 807

\reference{} Draine, B. T. \& Li, A.  2007, \apj 657, 810

\reference{} Duc, P.-A. \& Mirabel, I. 1994, \aap, 289, 83

\reference{} Duc, P.-A. \& Mirabel, I. 1998, \aap, 333, 813

\reference{} Duc, P.-A. \& Paudel, S., McDermid, R., et al. 2014, \mnras, Accepted

\reference{} Duc, P.-A. et al. 2014, MNRAS, 440, 1458

\reference{} Elmegreen, B. G. \& Efremov, Y. N. 1996, \apj, 466, 802

\reference{} Elmegreen, B. G., Kaufman, M., \& Thomasson, M. 1993, \apj, 412, 90

\reference{} Engelbracht, C. W., Gordon, K. D., Rieke, G. H., et al.  2005, \apj, 628, 29

\reference{} Engelbracht, C. W. et al.  2007, PASP, 119, 994

\reference{} Fazio, G., Ashby, M., Barmby, P., et al. 2004a, \apj, 154, 39

\reference{} Fazio, G. G., Hora, J. L., Allen, L. E. et al., 2004b, ApJS, 154, 10

\reference{} Galliano, F., Madden, S. C., Jones, A. P. et al. 2003, A\&A, 407, 159

\reference{} Galliano, F. et al. 2008, \apj, 679, 310

\reference{} Gordon, K. et al. 2008, \apj, 682, 336

\reference{} Habing, H. J. 1968, BAN, 19, 421

\reference{} Hancock, M., Smith, B. J., Struck, C. et al., 2007, AJ, 133, 676

\reference{} Hibbard, J., Guhathakurta, P, van Gorkom, J., \&
Schweizer, F. 1994, \aj, 107, 67 

\reference{} Hibbard, J. E. \& Mihos, J. C. 1995, \aj, 110, 140

\reference{} Hibbard, J. E. \& van Gorkom, J. H. 1996, \aj, 111, 655

%ulirg
\reference{} Higdon, S. J. U., Armus, L., Higdon, J. L. et al. 2006a, \apj, 648, 323

\reference{} Higdon, S. J. U., Devost, D., Higdon, J. L. et al. 2004, PASP, 116, 975

\reference{} Higdon, S. J. U., \& Higdon, J. L. 2008, in IAU Symp. 244
Dark Galaxies and Lost Baryons,ed. J. I. Davies \& M. J. Disney
(Cambridge, Cambridge Univ. Press), 356

\reference{} Higdon, S. J. U., Higdon, J. L., \& Marshall, J. HHM06 2006b, \apj, 640, 768

\reference{} Higdon, S. J. U., Higdon, J. L., Smith, B. J., Hancock, M., Struck, C. 2010, in ASP Conf. Ser. 423, Galaxy Wars: Stellar Populations and Star Formation in Interacting Galaxies , ed. Smith, B. J., Bastian, N. Higdon, S. J. U. \& Higdon, J. L. (San Francisco, CA:ASP),  271

\reference{} Houck, J. R., Charmandaris, V, Brandl, B et al., 2004a ApJS, 154, 211

%IRS paper
\reference{} Houck, J. R., Roellig, T., Van Cleve, J. et al. 2004b ApJS, 154, 18

\reference{} Hunsberger, S. D., Charlton, J., \& Zaritsky, D. 1996, \apj, 462, 50

\reference{} Hunt, L. K., Thuan, T., Izotov, Y. \& Sauvage, M.  2010, \apj, 712, 164

\reference{} Kaviraj, S., Darg, D., Lintott,C., Schawinski, K. \& Silk, J. 2012, MNRAS, 419, 70

\reference{} Kennicutt, R. C., Armus, L., Bendo, G. et al. 2003, \pasp, 115, 928

\reference{} Lapham, R., Smith, B. \& Struck, C. 2013, \aj, 145, 130

\reference{} Lebouteiller, V. et al. 2011, \apj, 728, 45

\reference{} Li, A \& Draine, B. T. 2001, \apj, 554, 778

\reference{}  Longmore, A. J., Hawarden, T. G., Cannon, R. D., et al. 1979 MNRAS, 188, 285 (L79)

\reference{} Madden, S. C. 2000, NewAR, 44, 249

\reference{} Madden, S. C.,Galliano, F., Jones, A. P. \& Sauvage,
M. 2006, A\&A, 446, 877

\reference{} Makovoz, D., \& Marleau, F. R. 2005, \pasp, 117, 1113

\reference{} Mirabel, I. F., Lutz, D., \& Maza, J. 1991, \aap, 243, 367

\reference{} Morris, S. L. \& van den Bergh, S. 1994, \apj, 427, 696

\reference{} Pena, M., Ruiz, M.. T. \& Maza, J. 1991, A\&A, 251, 417

\reference{} Petitpas, G. R. \& Taylor, C. L. 2005, \apj, 633, 138 %SQ co 

\reference{} Plante, S \& Sauvage, M. 2002, \aj, 124, 1995

\reference{} Reach, W. et al. 2005, PASP, 117, 978

\reference{} Rieke, G. H., Young, E. T., Engelbracht, C. W. et al. 2004, \apjs, 154, 25

\reference{} Rosenberg, J. L., Ashby, M. L. N., Salzer, J. J. \&
Huang, J. -S. 2006, \apj, 636, 742

\reference{} Rosenberg, J. L., et al. 2008, \apj, 674, 814

\reference{} Sanders, D. B. \&  Mirabel, I. F. 1996, ARAA, 34, 749 

% pah size 
\reference{} Schutte, W. A., Tielens, A. G. G. M., \& Allamandola 1993 \apj, 415, 397

\reference{} Schweizer, F. 1978, in Structure and Properties of Nearby Galaxies,
ed. E. M. Berkhuijsen \& R. Wielebinski (Dordrecht: Reidel), 279.

\reference{} Schombert, J. M., Wallin, J. F. \& Struck-Marcell, C. 1990,
\apj, 99, 497

% Galex Spiral bridges and tails
\reference{} Smith, B. J., Giroux, M. L., \& Struck, C., et al. 2010, AJ, 139, 1212

%Dwarf sample
\reference{} Smith, B. J. \& Hancock, M. 2009, \aj, 138, 130 

%orbit crowidin
\reference{} Smith, B. J. \& Struck, C. 2012, MNRAS, 422, 2444

%SBT paper
\reference{} Smith, B. J., Struck, C. , Hancock, M., et al. 2007a, \aj, 133, 791

%arp 285 beads on string
\reference{} Smith, B. J., Struck, C. , Hancock, M., et al. 2008, \aj, 135, 2406

% PAH
\reference{}  Smith, J. D. T., Draine, B. T., Dale, D. A.  et al. 2007b, ApJ, 656, 770

%arp 284 paper
\reference{} Struck, C. \& Smith, B. J. 2003, \apj, 589, 157

\reference{} Struck,C. \& Smith, B. 2012, MNRAS, 422, 2444

\reference{} Thornley, M. D., Forster Schreiber, M., Lutz, et
al. 2000, ApJ 539 641

\reference{} Thuan, T. X., Sauvage, M. \& Madden, S. 1999, \apj, 516, 783

\reference{} Toomre, A. \& Toomre, J. 1972, \apj, 178, 623

\reference{} Van der Hulst, J. M.,  1979, \aap, 71, 131

\reference{} Verley, S., Hunt, L. K., Corbelli, E., \& Giovanardi, C. 2007 A\&A, 476, 1161

\reference{} Verma, A. Lutz, D., Sturm, E., et al. 2003, A\&A, 403, 829

\reference{} Weilbacher, P. M., Duc, P.-A., \& Fritze-v. Alvensleben, U.  2003, A\&A, 397, 545

\reference{} Weingartner, J. C., \& Draine, B. T. 2001, \apj, 563, 842
 
\reference{} Wu, Y., Charmandaris, V., Hunt, L. K. et al. 2007, \apj, 662, 952

\reference{} Zwicky, F. 1956, Ergebnisse der Exakten Naturwissenchaften,
              29, 344

\clearpage

%%%%%%%%%%%%%%%%%%%%%

%%%%%%%%%% Table 1 
\begin{deluxetable}{lllr} 
\tabletypesize{\footnotesize}
%\tabletypesize{\scriptsize}
%\rotate
\tablecolumns{4} 
\tablewidth{0pc}
\tablecaption{Spitzer 8 \um ISFOs}
\tablehead{
\colhead{Object}&
\colhead{RA (J2000)}&
\colhead{Dec (J2000)} & 
\colhead{D} \\
\colhead{}&
\colhead{h:m:s}&
\colhead{$^{\circ}$:':''}&
\colhead{Mpc}
}
\startdata 

Arp 65-N1&00:21:47.74&+22:24:55.9&75.4\\
Arp 65-N2&00:21:49.43&+22:24:41.2&75.4\\
Arp 65-N3&00:21:50.18&+22:24:34.5&75.4\\
Arp 65-N4&00:21:50.74&+22:24:26.3&75.4\\
Arp 65-S1&00:21:51.53&+22:23:38.4&75.4\\
Arp 65-S2&00:21:53.48&+22:23:27.9& 75.4\\
Arp 65-S3&00:21:53.14&+22:23:27.9& 75.4\\
Arp 72-S1&15:46:58.24&+17:52:32.7& 46.5\\
Arp 72-S2&15:46:57.02&+17:52:22.1& 46.5\\
Arp 72-S3&15:46:57.47&+17:52:27.2& 46.5\\
Arp 72-S4&15:46:57.02&+17:52:22.9& 46.5\\
Arp 72-S5&15:46:56.77&+17:52:19.4& 46.5\\
Arp 82-N1&08:11:13.93&+25:13:09.6& 57.7\\
Arp 82-S1&08:11:12.56&+25:11:40.9& 57.7\\
Arp 82-S2&08:11:12.97&+25:11:25.7& 57.7\\
Arp 82-S3&08:11:13.06&+25:11:21.2& 57.7\\
Arp 82-S4&08:11:14.02&+25:11:13.8& 57.7\\
Arp 84-N1&13:58:33.52&+37:27:42.3& 48.9\\
Arp 87-N1&11:40:45.11&+22:25:58.8&100.2\\
Arp 102-N1&17:19:17.20&+49:04:45.2&101.8\\
Arp 102-N2&17:19:17.14&+49:04:38.7&101.8\\
Arp 104-S1&13:32:01.69&+62:41:32.9& 45.7\\
Arp 105-N1&11:11:12.68&+28:45:55.0&123.6\\
Arp 105-N2&11:11:12.59&+28:45:37.6&123.6\\
Arp 105-S1&11:11:13.37&+28:41:24.6&123.6\\
Arp 105-S2&11:11:13.46&+28:41:16.3&123.6\\
Arp 107-E1&10:52:14.78&+30:04:06.5&140.1\\
Arp 107-SW1&10:52:12.82&+30:03:49.6&140.1\\
Arp 107-N1&10:52:12.63&+30:04:21.5&140.1\\
Arp 242-N1&12:46:10.45&+30:45:31.0& 93.1\\
Arp 242-N2&12:46:10.45&+30:45:22.6& 93.1\\
Arp 242-N3&12:46:10.46&+30:45:12.1& 93.1\\
Arp 242-N4&12:46:10.55&+30:45:04.0& 93.1\\
Arp 242-N5&12:46:10.46&+30:44:53.2& 93.1\\
Arp 242-N6&12:46:10.42&+30:44:34.0& 93.1\\
Arp 242-N7&12:46:10.37&+40:44:26.2& 93.1\\
Arp 242-N8&12:46:10.33&+30:44:20.8& 93.1\\
Arp 284-E1&23:36:18.53&+02:09:26.5& 39.5\\
Arp 284-SW1&23:36:13.37&+02:09:03.7& 39.5\\
Arp 284-SW2&23:36:12.96&+02:09:01.7& 39.5\\
Arp 285-N1&09:24:18.49&+49:15:18.6& 27.9\\
SQ-A&22:35:58.91&+33:58:49.5& 90.8\\
SQ-B&22:36:10.38&+33:57:20.0& 90.8\\
NGC 5291-13&12:46:10.45&-30:45:31.0& 62.0\\
NGC 5291-15&12:46:10.45&-30:45:31.0& 62.0\\
NGC 5291-16&12:46:10.45&-30:45:31.0& 62.0\\
NGC 5291-17&12:46:10.45&-30:45:31.0& 62.0\\
NGC 5291-18&12:46:10.45&-30:45:31.0& 62.0\\
NGC 5291-19&12:46:10.45&-30:45:31.0& 62.0\\
NGC 5291-22&12:46:10.45&-30:45:31.0& 62.0\\
NGC 5291-25&12:46:10.45&-30:45:31.0& 62.0\\
NGC 5291-26&12:46:10.45&-30:45:31.0& 62.0\\
NGC 5291-28&12:46:10.45&-30:45:31.0& 62.0\\
NGC 5291-30&12:46:10.45&-30:45:31.0& 62.0\\
NGC 5291-31&12:46:10.45&-30:45:31.0& 62.0\\
NGC 5291-32&12:46:10.45&-30:45:31.0& 62.0\\
NGC 5291-33&12:46:10.45&-30:45:31.0& 62.0\\
NGC 5291-36&12:46:10.45&-30:45:31.0& 62.0\\
NGC 5291-39&12:46:10.45&-30:45:31.0& 62.0\\
NGC 5291-41&12:46:10.45&-30:45:31.0& 62.0\\
NGC 5291-42&12:46:10.45&-30:45:31.0& 62.0\\
NGC 5291-43&12:46:10.45&-30:45:31.0& 62.0\\
NGC 5291-45&12:46:10.45&-30:45:31.0& 62.0\\
NGC 5291-47&12:46:10.45&-30:45:31.0& 62.0\\
NGC 5291-49&12:46:10.45&-30:45:31.0& 62.0\\
NGC 5291-50&12:46:10.45&-30:45:31.0& 62.0\\
NGC 5291-51&12:46:10.45&-30:45:31.0& 62.0\\
\enddata
\end{deluxetable}

%\clearpage

%%%%%%%%%%%%%% new table 2 &&&&&&&&&&&&&&&

%%%%%%%%%%%%% Table 2
\begin{deluxetable}{lrrrrrr} 
\tabletypesize{\scriptsize}
\rotate
\tablecolumns{7} 
\tablewidth{0pc}
\tablecaption{Spitzer Photometry}
\tablehead{
\colhead{Object}&  
\colhead{F$_{3.6 \mu m}$}&  
\colhead{F$_{4.5 \mu m}$} &   
\colhead{F$_{5.8 \mu m}$}&  
\colhead{F$_{8 \mu m}$}&   
\colhead{F$_{24 \mu m}$} &  
\colhead{Aper\tablenotemark{a}}\\
\colhead{}&  
\colhead{$\mu$Jy}&  
\colhead{$\mu$Jy}&  
\colhead{$\mu$Jy}&  
\colhead{$\mu$Jy}&  
\colhead{$\mu$Jy}&  
\colhead{}
}
\startdata

Arp 65-N1&   29.3 $\pm$ 4.5&  17.6 $\pm$ 3.6&  46.3 $\pm$ 6.8&  181.4 $\pm$ 11.8& $\le$113.7& 3,2 \\
Arp 65-N2 &  36.2 $\pm$ 5.0&  24.4 $\pm$ 4.2&  67.9 $\pm$ 7.3&  157.9 $\pm$ 11.2& $\le$132.5& 2,2 \\
Arp 65-N3 &  29.4 $\pm$ 4.9&  17.6 $\pm$ 3.5&  68.0 $\pm$ 7.3&  154.8 $\pm$ 11.1& $\le$169.0& 2,2 \\
Arp 65-N4 &  45.1 $\pm$ 5.6&  26.7 $\pm$ 4.3&  62.4 $\pm$ 7.1&  176.0 $\pm$ 11.8& $\le$184.1& 2,2 \\
Arp 65-S1 & 157.2 $\pm$ 5.2& 108.7 $\pm$ 4.5& 244.4 $\pm$ 6.7&  615.5 $\pm$ 11.9&  848.5 $\pm$ 64.6& 3,2 \\
Arp 65-S2\tablenotemark{b}&   34.5 $\pm$   2.9&   15.4 $\pm$   2.8&    73.5 $\pm$   6.6&   283.9 $\pm$   6.4&    346.2 $\pm$   41.2&  2,2\\
Arp 65-S3\tablenotemark{b}&   39.4 $\pm$   3.0&   28.8 $\pm$   2.8&    95.1 $\pm$   8.1&   276.8 $\pm$   6.4&    346.2 $\pm$   41.2&  2,2\\
Arp 72-S1  &690.2$\pm$ 7.3&    520.1 $\pm$  6.6 &  1927.7 $\pm$ 13.3 &   5425.9 $\pm$  14.0 & 13148.0  $\pm$ 248.0 &  3,2 \\ 
Arp 72-S2 & 234.3 $\pm$ 13.3& 157.5 $\pm$ 11.3& 558.8 $\pm$ 20.1&  1649.6 $\pm$ 35.6&  5149.6 $\pm$ 156.3& 2,2\\
Arp 72-S3 & 189.5 $\pm$ 11.4& 135.5 $\pm$ 9.7&  435.3 $\pm$ 17.6&  1192.5 $\pm$ 30.4&  4101.4 $\pm$ 166.2& 2,2\\
Arp 72-S4 & 149.7 $\pm$ 10.1& 101.0 $\pm$ 8.9&  326.9 $\pm$ 15.5&   861.3 $\pm$ 26.7&  2938.4 $\pm$ 142.7& 2,2\\
Arp 72-S5 & 125.3 $\pm$ 9.3&   92.2 $\pm$ 7.9&  279.3 $\pm$ 14.4&   712.1 $\pm$ 23.7&  2502.8 $\pm$ 130.1& 2,2\\  
Arp 82-N1 & 259.3 $\pm$   5.0&  225.8 $\pm$   4.3&   973.4 $\pm$   9.8&  2906.1 $\pm$  16.1&  20926.1 $\pm$  260.0&  2,3\\
Arp 82-S1 & 146.9 $\pm$   4.3&  103.3 $\pm$   4.8&   428.5 $\pm$  11.0&  1271.7 $\pm$  14.7&   3388.6 $\pm$  131.0&  2,1\\
Arp 82-S2  & 144.1 $\pm$ 9.4&  86.0 $\pm$ 7.6&  298.3 $\pm$ 18.9&  912.1 $\pm$ 23.8&  3079.1 $\pm$ 120.5& 2,2\\
Arp 82-S3  & 93.8 $\pm$ 7.6&  49.0 $\pm$ 6.0&  175.3 $\pm$ 12.3&  540.9 $\pm$ 20.4&  1557.1 $\pm$ 91.9& 2,2\\
Arp 82-S4  & 60.3 $\pm$ 6.1&  34.5 $\pm$ 5.2&  143.8 $\pm$ 11.5&  453.2 $\pm$ 18.8&  1575.6 $\pm$ 92.3& 2,2\\
Arp 84-N1\tablenotemark{c}&  371.3 $\pm$   4.4&  251.5 $\pm$   4.9&   850.9 $\pm$  19.2&  2626.3 $\pm$  15.3&     \nodata     &  3,-\\
Arp 87-N1\tablenotemark{c}&  139.9 $\pm$   6.2&   96.7 $\pm$   4.7&   226.2 $\pm$  13.1&   854.4 $\pm$  18.1&      \nodata     &  3,-\\
Arp 102-N1&   63.7 $\pm$   13.5&   57.8 $\pm$   12.9&    $\le$27.4 &   176.9 $\pm$  24.0&    218.8 $\pm$ 27.7&  1,1  \\
Arp 102-N2&  347.9 $\pm$   30.9&  202.2 $\pm$   23.7&   111.4 $\pm$  18.7&    47.7 $\pm$  13.5&    $\le$ 83.5 &  4,1  \\
Arp 104-S1&  239.3 $\pm$   6.6&  184.6 $\pm$   4.7&   109.0 $\pm$  12.2&   391.9 $\pm$  12.8&    622.7 $\pm$   56.0&  3,1  \\
Arp 105-N1&  201.8 $\pm$  10.4&  144.2 $\pm$   8.7&   331.4 $\pm$  20.3&   617.7 $\pm$  23.7&     \nodata &  6,-  \\
Arp 105-N2&   19.0 $\pm$   4.7&   23.4 $\pm$   4.1&    68.9 $\pm$  10.3&   235.3 $\pm$  12.0&     \nodata &  3,-  \\
Arp 105-S1&   47.8 $\pm$   5.8&   30.9 $\pm$   4.2&    92.1 $\pm$  10.4&   277.1 $\pm$  13.1&      \nodata &  1,-  \\
Arp 105-S2&  336.4 $\pm$  10.9&  271.0 $\pm$  10.4&   677.7 $\pm$  20.6&  1970.7 $\pm$  24.9&    \nodata &  4,-  \\
Arp 107-E1&  349.9 $\pm$   7.9&  203.8 $\pm$   7.4&   648.2 $\pm$  18.2&  2586.2 $\pm$  18.4&   2184.9 $\pm$   36.5&  5,1  \\
Arp 107-SW1&  124.4 $\pm$   5.2&   83.2 $\pm$   4.1&   340.3 $\pm$  11.6&  1219.3 $\pm$  16.2&    889.2 $\pm$   35.0&  3,1  \\
Arp 107-N1&   85.4 $\pm$   5.8&   49.7 $\pm$   5.5&    63.3 $\pm$  10.5&   177.6 $\pm$  13.1&    105.5 $\pm$   37.0&  3,1 \\
Arp 242-N1&   90.0 $\pm$   8.6&   52.9 $\pm$   5.9&   101.6 $\pm$   9.8&   233.3 $\pm$  14.2&    263.5 $\pm$   46.1&  2,1  \\
Arp 242-N2&  267.1 $\pm$  14.5&  154.2 $\pm$  10.7&   165.7 $\pm$  19.5&   403.6 $\pm$  13.7&    $\le$ 120.0 &  3,1 \\
Arp 242-N3&  428.6 $\pm$   4.2&  267.3 $\pm$   4.2&   407.8 $\pm$  13.2&   946.7 $\pm$  14.7&    536.3 $\pm$   44.1&  3,1  \\
Arp 242-N4&  163.5 $\pm$   2.7&   85.4 $\pm$   2.6&   107.3 $\pm$   9.2&   286.5 $\pm$  11.0&    $\le$ 120.0 &  2,1  \\
Arp 242-N5&  411.0 $\pm$   4.3&  254.2 $\pm$   4.5&   243.5 $\pm$  14.3&   558.2 $\pm$  12.8&    369.2 $\pm$   54.6&  3,1  \\
Arp 242-N6&  503.5 $\pm$   4.7&  315.7 $\pm$   4.2&   548.3 $\pm$  14.9&  1277.1 $\pm$  19.0&   1555.1 $\pm$   82.0&  3,1  \\
Arp 242-N7&  239.7 $\pm$   3.2&  149.8 $\pm$   2.9&   252.8 $\pm$  10.1&   567.5 $\pm$  11.6&   $\le$ 120.0 &  2,1 \\
Arp 242-N8&  236.8 $\pm$   4.6&  144.7 $\pm$   3.8&   232.3 $\pm$  11.4&   587.3 $\pm$  10.1&   $\le$ 120.0 &  2,1  \\
Arp 284-E1\tablenotemark{c}&   61.1 $\pm$   5.8&   47.6 $\pm$   7.1&   219.9 $\pm$  15.2&   721.8 $\pm$  16.5&     \nodata&  3,- \\
Arp 284-SW1\tablenotemark{c}&  524.6 $\pm$   8.1&  432.4 $\pm$   6.6&  1217.7 $\pm$  16.6&  3177.9 $\pm$  18.8&    \nodata&  2,-  \\
Arp 284-SW2\tablenotemark{c}&  282.5 $\pm$   6.6&  205.7 $\pm$   6.9&   917.6 $\pm$  20.0&  2513.5 $\pm$  16.6&     \nodata&  2,-  \\
Arp 285-N1&   23.0 $\pm$   3.5&   17.5 $\pm$   3.6&    64.4 $\pm$  12.1&   353.9 $\pm$  18.7&      \nodata&  2,- \\
SQ-A&  458.5 $\pm$  15.8&  328.5 $\pm$  12.1&  1271.8 $\pm$  29.9&  3695.2 $\pm$  43.9&   9308.7 $\pm$   23.2&  4,3  \\
SQ-B&  260.2 $\pm$   9.6&  186.8 $\pm$   8.0&   831.6 $\pm$  21.1&  2553.4 $\pm$  39.9&   5717.3 $\pm$   25.3&  4,3  \\
NGC 5291-13&  140.0 $\pm$   3.5&  107.0 $\pm$   3.4&   170.0 $\pm$  15.2&   419.0 $\pm$  10.2&   1978.1 $\pm$   55.9&  3,3  \\
NGC 5291-15&   39.9 $\pm$   3.5&   29.1 $\pm$   3.4&    86.6 $\pm$  15.2&   252.0 $\pm$  10.2&    429.0 $\pm$   36.5&  3,3\\
NGC 5291-16&  145.0 $\pm$   3.5&  122.0 $\pm$   3.4&    92.3 $\pm$  15.1&   231.0 $\pm$  10.2&    396.7 $\pm$   47.8&  3,3 \\
NGC 5291-17&   89.9 $\pm$   3.5&   57.1 $\pm$   3.4&   259.0 $\pm$  15.2&   663.0 $\pm$  10.2&    656.7 $\pm$   49.4&  3,3\\
NGC 5291-18&   45.0 $\pm$   3.5&   24.8 $\pm$   3.4&   130.0 $\pm$  15.2&   352.0 $\pm$  10.3&    505.1 $\pm$   36.7&  3,3  \\
NGC 5291-19&   30.7 $\pm$   3.5&    9.6 $\pm$   3.4&    $\le$  36.3 &   188.0 $\pm$  10.2&    $\le$146.1         &  3,3 \\
NGC 5291-22&   49.5 $\pm$   2.6&   32.1 $\pm$   2.9&    40.7 $\pm$  10.8&   106.0 $\pm$   9.2&    173.7 $\pm$   33.8&  3,3\\
NGC 5291-25&   69.3 $\pm$   2.6&   38.0 $\pm$   2.9&   402.0 $\pm$  10.8&   486.0 $\pm$   9.2&    778.7 $\pm$   48.4&  3,3\\
NGC 5291-26&  513.0 $\pm$   6.5&  389.3 $\pm$   6.2&  1844.5 $\pm$  28.2&  4804.0 $\pm$  18.8&  14809.2 $\pm$   76.1&  3,3\\
NGC 5291-28&   19.8 $\pm$   2.6&    8.1 $\pm$   2.9&    18.9 $\pm$  10.9&    71.1 $\pm$   9.3&    306.5 $\pm$   32.1&  3,3 \\
NGC 5291-30&  373.0 $\pm$   3.5&  256.0 $\pm$   3.4&   286.0 $\pm$  15.2&  1350.0 $\pm$  10.3&    943.2 $\pm$   35.3&  3,3  \\
NGC 5291-31&    8.9 $\pm$   2.6&   10.4 $\pm$   2.9&    52.3 $\pm$  10.9&   158.9 $\pm$   9.2&    $\le$135.2 &  3,3 \\
NGC 5291-32&  118.0 $\pm$   3.5&   72.5 $\pm$   3.4&   300.0 $\pm$  15.2&   853.0 $\pm$  10.2&    778.0 $\pm$   37.9&  3,3\\
NGC 5291-33&  295.0 $\pm$   6.5&  160.9 $\pm$   6.1&   617.8 $\pm$  28.2&  1930.0 $\pm$  18.8&   2233.6 $\pm$   37.0&  3,3\\
NGC 5291-36&   27.4 $\pm$   2.6&   17.8 $\pm$   2.9&    28.4 $\pm$  10.9&   174.0 $\pm$   9.2&    $\le$130.0 &  3,3 \\
NGC 5291-39&   17.1 $\pm$   2.6&   15.0 $\pm$   2.9&    27.3 $\pm$  10.8&   113.0 $\pm$   9.3&    $\le$107.0 &  3,3 \\
NGC 5291-41&   25.0 $\pm$   2.6&   10.8 $\pm$   2.9&    40.0 $\pm$  10.9&   137.0 $\pm$   9.2&    196.9 $\pm$   30.9&  3,3\\
NGC 5291-42&   51.0 $\pm$   3.5&   32.7 $\pm$   3.4&   105.0 $\pm$  15.2&   258.0 $\pm$  10.2&    212.7 $\pm$   34.5&  3,3\\
NGC 5291-43&   62.8 $\pm$   2.6&   40.5 $\pm$   2.9&    62.1 $\pm$  10.9&   180.0 $\pm$   9.2&    217.7 $\pm$   34.5&  3,3\\
NGC 5291-45&   76.0 $\pm$   2.6&   62.0 $\pm$   2.9&    25.1 $\pm$  10.8&   198.0 $\pm$   9.2&    $\le$110.7 &  3,3\\
NGC 5291-47&   38.6 $\pm$   2.6&   17.3 $\pm$   2.9&    97.6 $\pm$  10.8&   284.0 $\pm$   9.2&    379.4 $\pm$   35.5&  3,3\\
NGC 5291-49&   27.5 $\pm$   2.6&   15.1 $\pm$   2.9&    50.1 $\pm$  10.8&   158.0 $\pm$   9.2&    $\le$121.1 &  3,3\\
NGC 5291-50&   27.7 $\pm$   2.6&   21.0 $\pm$   2.9&    57.2 $\pm$  10.8&   162.0 $\pm$   9.3&    166.0 $\pm$   37.4&  3,3\\
NGC 5291-51&  134.0 $\pm$   2.6&   88.9 $\pm$   2.9&   142.0 $\pm$  10.8&   313.0 $\pm$   9.2&    256.0 $\pm$   37.0&  3,3\\
\enddata

\tablenotetext{a} {Aperture and sky annulus radii code. The first integer refers to IRAC observations and the second integer 
refers to MIPS. A dash indicates that no data was available. The IRAC code is: 1-($3\arcsec$,$3-7\arcsec$), 
2-($3\arcsec$,$10-20\arcsec$), 3-($5\arcsec$,$10-20\arcsec$), 4-($6\arcsec$,$10-20\arcsec$), 5-($8\arcsec$,$10-20\arcsec$), 
6-($10\arcsec$,$10-20\arcsec$), 7-($13\arcsec$,$20-32\arcsec$). The aperture code for MIPS is: 1-($6\arcsec$,$20-32\arcsec$), 
2-($6.6\arcsec$,$20-32$), 3-($13\arcsec$,$20-32\arcsec$).}
\tablenotetext{b} {Two sources are visible in IRAC but only one in MIPS. We divide the $24 \mu$m flux equally between them}.
\tablenotetext{c} {ISFOs are buried by bright emission (diffraction patterns) from main galaxy at $24 \mu$m.} 

\end{deluxetable}

%%%%%%%%%%%%%%%%%%%% table 3
\begin{deluxetable}{lcccccccccc} 
%\rotate
%\footnotesize
%\scriptsize
\tabletypesize{\footnotesize}
\tabletypesize{\scriptsize}
\tablecolumns{9} 
\tablewidth{0pc}
\tablecaption{Aperture Scale Factors\tablenotemark{a}}
\tablehead{
\colhead{Module}&
\colhead{Arp 72}&
\colhead{Arp 82}&
\colhead{Arp 84}&
\colhead{Arp 87}&
\colhead{Arp 105}&
\colhead{Arp 242}&
\colhead{Arp 284}&
\colhead{NGC 5291}&
\colhead{NGC 5291}&
\colhead{SQA}\\
\colhead{}&
\colhead{S1}&
\colhead{N1}&
\colhead{N1\tablenotemark{b}}&
\colhead{N1\tablenotemark{b}}&
\colhead{S2}&
\colhead{N3}&
\colhead{SW1\tablenotemark{b}}&
\colhead{N\tablenotemark{c}}&
\colhead{S\tablenotemark{c}}&
\colhead{}\\

}

\startdata
SL&1.69&0.87&1.41&1.0&1.19&1.43&0.90&1.04&1.72&4.25\\
LL&0.76&0.95&0.64&0.24&\nodata&\nodata&\nodata&\nodata&\nodata&\nodata\\
SH&4.0&0.94&2.81&0.34&\nodata&\nodata&0.76&\nodata&\nodata&7.41\\
LH&0.63&0.78&0.49&0.23&\nodata&\nodata&0.40&\nodata&\nodata&\nodata\\
\enddata

\tablenotetext{a} {Apertures listed in Table 2}
\tablenotetext{b} {No MIPS 24\um photometry}
\tablenotetext{c} {No sky data available for HIRES see HHM06}

\end{deluxetable}

%%%%%%%%%%%% Table 4
%\begin{onecolumn}

\begin{deluxetable}{lrrrrrrrrr} 
%\rotate
%\footnotesize
%\scriptsize
\tabletypesize{\footnotesize}
\tabletypesize{\scriptsize}
\tablecolumns{9} 
\tablewidth{0pc}
\tablecaption{Line Flux}
\tablehead{
\colhead{Line\tablenotemark{a}}&
\colhead{Arp 72 S1\tablenotemark{b}}&
\colhead{Arp 82 N1\tablenotemark{c}}&
\colhead{Arp 84 N1}&
\colhead{Arp 87 N1}&
\colhead{Arp 284 SW1}&
\colhead{NGC 5291N\tablenotemark{d}}&
\colhead{NGC 5291S\tablenotemark{d}}&
\colhead{SQA}\\
}

\startdata
$[$SIV$]$& 11.61$\pm$  0.94&  8.80$\pm$  0.46&  $\le$4.50&\nodata &  5.51$\pm$  0.49&  7.40$\pm$  0.15&  1.05$\pm$  0.22&  7.66$\pm$  2.16\\
 10.51 \um& -0.16& -0.15&\nodata&\nodata& -0.05& -0.02& -0.00& -0.05\\
$[$NeII$]$& 13.34$\pm$  5.71& 15.35$\pm$  0.73&  5.29$\pm$  0.34&  0.99$\pm$  0.05& 15.97$\pm$  0.54& 10.65$\pm$  0.06&  4.05$\pm$  0.31& 26.24$\pm$  2.20\\
12.81 \um& -0.09& -0.17& -0.13& -0.02& -0.02& -0.02& -0.01& -0.22\\
$[$NeIII$]$& 21.74$\pm$  4.99& 30.30$\pm$  0.49&  1.34$\pm$  0.25&  0.51$\pm$  0.17& 31.14$\pm$  0.44& 25.31$\pm$  1.03&  5.47$\pm$  0.11& 35.37$\pm$  2.60\\
15.55 \um& -0.31& -0.46& -0.07& -0.10& -0.56& -0.06& -0.02& -0.60\\
$[$SIII$]$& 15.45$\pm$  1.66& 17.90$\pm$  0.52&  2.20$\pm$  0.44&  0.71$\pm$  0.18& 19.44$\pm$  0.76& 15.45$\pm$  0.58&  4.66$\pm$  0.17& 23.26$\pm$  3.18\\
18.71 \um& -0.21& -0.23& -0.20& -0.54& -0.25& -0.04& -0.01& -0.35\\
$[$SIII$]$  &1.55$\pm$0.17&3.64$\pm$0.32& \nodata&\nodata &4.78$\pm$0.08 &\nodata & \nodata&\nodata \\
33.48 \um &-0.21&-0.23&\nodata&\nodata&-0.24 &\nodata&\nodata&\nodata\\
$[$SiII$]$  &1.66$\pm$0.40&1.96$\pm$0.61&\nodata&\nodata&\nodata&\nodata&\nodata&\nodata\\
34.81 \um &-0.13&\nodata&\nodata&\nodata&\nodata&\nodata&\nodata&\nodata\\
H$_2$ S2&  $\le$8.4&  $\le$1.95& $\le$1.29 &  0.40$\pm$  0.13&  $\le$1.35&  1.90$\pm$  0.80&  0.90$\pm$  0.40&  $\le$3.09\\
12.28 \um&\nodata&\nodata& & -0.04&\nodata& -0.00& -0.00& -0.04\\
H$_2$ S1& 3.1$\pm$0.9 &  2.57$\pm$  0.84&$\le$2.04  &  0.80$\pm$  0.21&  3.01$\pm$  0.37&  1.10$\pm$  0.40&  1.30$\pm$  0.30& 14.07$\pm$  3.47\\
17.03 \um&\nodata& -0.02&\nodata& -0.14& -0.04& -0.00& -0.00& -0.22\\
H$_2$ S0&$\le$0.31&  7.48$\pm$  1.50&\nodata&\nodata&\nodata  & \nodata &\nodata&\nodata\\
28.22 \um&\nodata& -0.05&\nodata&\nodata&\nodata&\nodata&\nodata&\nodata\\
\enddata

\tablenotetext{a} { $[$ W cm$^{-2}]$ scaled by 10$^{22}$. Negative equivalent widths $[\mu m]$ 
denote line emission.}

\tablenotetext{b} {The molecular mass is calculated using the line
  fluxes measured in the IRS-LORES spectrum: \mh S0 $= (4.01\pm0.4)$ 10$^{22}$W cm$^{-2}$, \mh S1
  $= (3.1\pm0.9)$ 10$^{22}$W cm$^{-2}$, \mh S2$ = (8.0\pm1.7)$ 10$^{22}$W cm$^{-2}$.} 

\tablenotetext{c} {The lines are blueshifted wrt to the rest
  wavelength for Arp 82 by $\sim$ 200 km s$^{-1}$. The molecular mass
  is calculated using the HIRES fluxes for the \mh S0 and S1 lines and the
  IRS-LORES fluxes for the \mh S2 $= (2.6\pm0.3)$ 10$^{22}$W cm$^{-2}$,
  \mh S3 $= (4.1\pm1.0)$10$^{22}$W cm$^{-2}$.}
\tablenotetext{d} {data from HHM06}

\end{deluxetable}

%%%%%%%%% Table 5

\begin{deluxetable}{lrrrrrrrrrr} 
\rotate
%\footnotesize
%\scriptsize
%\tabletypesize{\footnotesize} %bigger
\tabletypesize{\scriptsize} %
\setlength{\tabcolsep}{0.02in} 
\tablecolumns{11} 
\tablewidth{0pc}
\tablecaption{PAH Flux}
\tablehead{
\colhead{Flux\tablenotemark{a}}&
\colhead{Arp 72\tablenotemark{b}}&
\colhead{Arp 82\tablenotemark{b}}&
\colhead{Arp 84}&
\colhead{Arp 87}&
\colhead{Arp 105}&
\colhead{Arp 242}&
\colhead{Arp 284}&
\colhead{SQ}&
\colhead{NGC 5291}&
\colhead{NGC 5291}\\
\colhead{EW\tablenotemark{c}}&
\colhead{S1}&
\colhead{N1}&
\colhead{N1}&
\colhead{N1}&
\colhead{S2}&
\colhead{N3}&
\colhead{SW1}&
\colhead{A}&
\colhead{N}&
\colhead{S}\\
}
\startdata

P$_{6.2}$ Flux   & 13.89 $\pm$  0.09 & 11.98 $\pm$  2.70 &  7.37 $\pm$  0.35 &  2.95 $\pm$  0.10 &  6.60 $\pm$  0.09 &  2.58 $\pm$  0.46 &  9.66 $\pm$  0.15 &  \nodata &  12.0 $\pm$ 0.6 &  5.09 $\pm$  0.30\\
P$_{6.2}$ EW  & - 2.1 & - 6.0 & - 4.2 & -3.5 & - 5.9 & - 4.1 & - 3.1 &   & - 8.3 & - 5.8\\
P$_{7.7C}$ Flux  & 48.28 $\pm$  0.92 & 18.32 $\pm$  4.35 & 22.22 $\pm$  1.53 &  9.89 $\pm$  0.74 & 14.46 $\pm$  0.35 &  7.69 $\pm$  0.20 & 23.82 $\pm$  0.97 & 36.51 $\pm$  0.25  &  42.40 $\pm$0.51   & 17.77 $\pm$  0.64\\
P$_{7.7C}$ EW & - 6.8 & - 5.7 & - 8.3 & - 12.9 & - 8.0 & -12.2 & - 4.3 & -90.6  & -12.1 & -10.0\\
P$_{8.3}$ Flux  &  4.40 $\pm$  0.53 &  3.32 $\pm$  1.63 &  3.20 $\pm$  0.34 &  1.24 $\pm$  0.25 &  1.54 $\pm$  0.09 &  0.66 $\pm$  0.13 &  2.38 $\pm$  0.20 &  5.48 $\pm$  0.11  &  2.67 $\pm$ 0.21   &  2.68 $\pm$  0.17\\
P$_{8.3}$ EW  & - 0.6 & - 0.9 & - 1.1 & - 1.8 & - 0.7 & - 1.1 & - 0.4 & - 7.3 & - 0.6 & - 1.3\\
P$_{8.6}$ Flux   &  8.51 $\pm$  0.50 &  3.59 $\pm$  0.44 &  3.40 $\pm$  0.11 &  1.02 $\pm$  0.08 &  2.88 $\pm$  0.06 &  2.17 $\pm$  0.03 &  4.59 $\pm$  0.26 & 10.54 $\pm$  0.21 & 11.45 $\pm$  0.12 &  3.38 $\pm$  0.10\\
P$_{8.6}$ EW  & - 1.3 & - 0.9 & - 1.1 & - 1.6 & - 1.2 & - 3.9 & - 0.7 & -11.6 & - 2.2 & - 1.5\\
P$_{11.3C}$  & 15.51 $\pm$  1.08 &  5.46 $\pm$  0.10 &  6.21 $\pm$  0.15 &  2.99 $\pm$  0.13 &  3.82 $\pm$  0.09 &  2.42 $\pm$  0.04 &  8.54 $\pm$  0.13 & 14.71 $\pm$  0.68 & 10.5 $\pm$  0.07 &  6.23 $\pm$  0.06\\
P$_{11.3C}$ EW  & - 3.0 & - 0.8 & - 2.1 & - 8.4 & - 0.8 & - 7.6 & - 1.1 & - 5.7 & - 1.1 & - 2.5\\
P$_{12.0}$ Flux  &  5.36 $\pm$  0.50 &  1.39 $\pm$  0.19 &  2.56 $\pm$  0.23 &  1.12 $\pm$  0.09 &  1.47 $\pm$  0.08 &  0.95 $\pm$  0.09 &  0.47 $\pm$  0.17 &  0.95 $\pm$  1.57  &  2.81 $\pm$  0.11 &  2.12 $\pm$  0.03\\
P$_{12.0}$ EW  & - 1.1 & - 0.2 & - 0.9 & - 3.7 & - 0.3 & - 3.5 & - 0.1 & - 0.3 & - 0.3 & - 0.9\\
 P$_{12.6C }$ Flux  &  4.64 $\pm$  0.48 &  2.59 $\pm$  0.10 &  3.05 $\pm$  0.37 &  1.80 $\pm$  0.10 &  3.01 $\pm$  0.12 &  2.17 $\pm$  0.14 &  4.93 $\pm$  0.10 & 11.66 $\pm$  0.53 &  4.31 $\pm$  0.06 &  2.96 $\pm$  0.14\\
P$_{12.6C}$ EW  & - 1.0 & - 0.3 & - 1.2 & - 7.0 & - 0.5 & - 8.9 & - 0.7 & - 3.6 & - 0.4 & - 1.3\\
P$_{13.6}$ Flux  &  4.81 $\pm$  0.31 &  0.12 $\pm$  0.15 &  1.44 $\pm$  0.35 &  \nodata &  0.69 $\pm$  0.27 &  0.22 $\pm$  0.09 &  1.33 $\pm$  0.13 &  5.57 $\pm$  0.20 & \nodata &  2.26 $\pm$  0.17 \\
P$_{13.6}$ EW  & - 1.0 &  0.0 & - 0.6 & & - 0.1 & - 0.8 & - 0.2 & - 1.6 & \nodata & - 1.1 \\
P$_{14.2}$ Flux  &  2.29 $\pm$  0.18 & \nodata &  0.11 $\pm$  0.21 & \nodata &  \nodata &  \nodata &  2.59 $\pm$  0.08 &\nodata & \nodata& \nodata \\
P$_{14.2}$ EW  & - 0.5 & & - 0.1 &   & & & - 0.4 &  & & \\
P$_{16.4}$ Flux  &  1.99 $\pm$  0.04 & \nodata & \nodata &  \nodata & \nodata & \nodata& \nodata & \nodata& \nodata & \nodata \\
P$_{16.4}$ EW  & - 0.3 &  & &  & & &  &  & &  \\
P$_{17C}$ Flux  &  9.69 $\pm$  0.17 &  4.63 $\pm$  2.93 & \nodata & \nodata & 97.09 $\pm$  0.00 &  \nodata & \nodata & \nodata & \nodata &\nodata \\
P$_{17C}$ EW  & - 1.5 & - 0.4 & &  & -10.3  & & & & & \\
P$_{6.2}/$P$_{7.7}$  &  0.29 $\pm$  0.01 &  0.65 $\pm$  0.21 &  0.33 $\pm$  0.03 &  0.30 $\pm$0.03 &  0.46 $\pm$  0.01 &  0.34 $\pm$  0.06 &  0.41 $\pm$  0.02 & \nodata&  0.28 $\pm$  0.01 &  0.29 $\pm$  0.02 \\
P$_{11.3}/$P$_{7.7}$  &  0.32 $\pm$  0.02 &  0.30 $\pm$  0.07 &  0.28 $\pm$  0.02 &  0.30$\pm$0.03 &  0.26 $\pm$  0.01 &  0.31 $\pm$  0.01 &  0.36 $\pm$  0.02 &  0.40 $\pm$  0.02 &  0.25 $\pm$  0.01 &  0.35 $\pm$  0.01 \\

\enddata

\tablenotetext{a} {Flux $[$W cm$^{-2}]$ is scaled by 10$^{21}$ }

\tablenotetext{b} {Fit includes LL spectrum }

\tablenotetext{c} {A negative equivalent width $[$\um$]$ designates line emission. }

\end{deluxetable}

\clearpage

%%%%FIGURES%%%%%%%%%%%%%%%%%%%%%%%%%%%%%%%%%%%%%%%%%%%%%%%%%%%%%%%%%%%%%%
%\includegraphics[angle=90,scale=.75]{fig1_lowres.ps} 
%\includegraphics[angle=0,scale=.75]{Figure2_sarah.eps} 
%\includegraphics[angle=0,scale=.75]{f2.eps}

% Figure 1 finder charts
\begin{figure}
\includegraphics[width=1.0\textwidth]{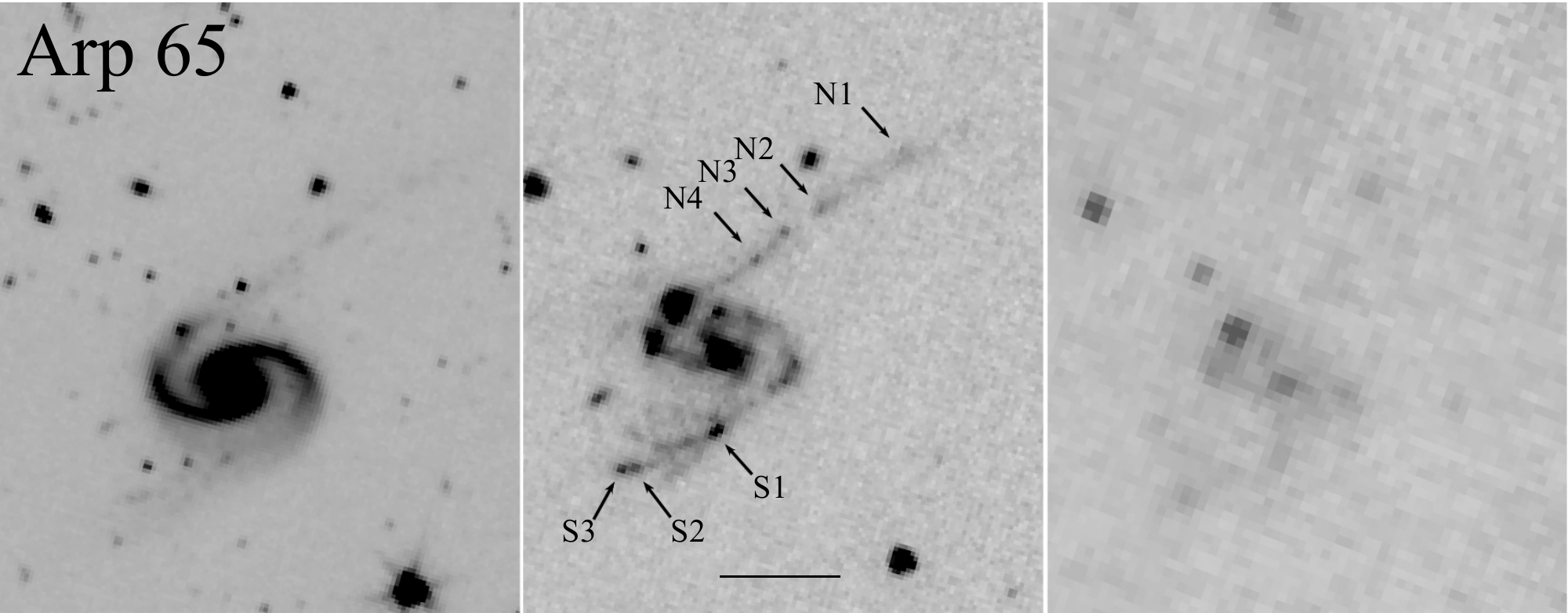}
%\epsscale{.80}
%\plotone{Figures/Fig1a-arp65findercloseup.eps}
\caption{ Images for the 14 interacting systems, except for NGC 5291
  (see HHM06) are displayed in three panels: left-right 3.6, 8.0 \& 24
  \ums. North is at the top and west is to the right.  ISFOs are
  indicated with arrows.  A linear transfer function is used in all
  images, which are displayed on the same scale. A $30''$ scalebar is
  shown in each middle panel. Arp 65 is a widely separated equal mass
  pair of galaxies. The western galaxy has two tails. The northern
  tail has beads of star formation terminated with a hinge clump at
  the base of the tail. The southern tail has an offset between old
  and young stars (Smith et al. 2007). }
% Our target is the largest candidate TDG at the tip of the southern tail
\end{figure}

\addtocounter{figure}{-1}
\begin{figure}
\includegraphics[width=1.0\textwidth]{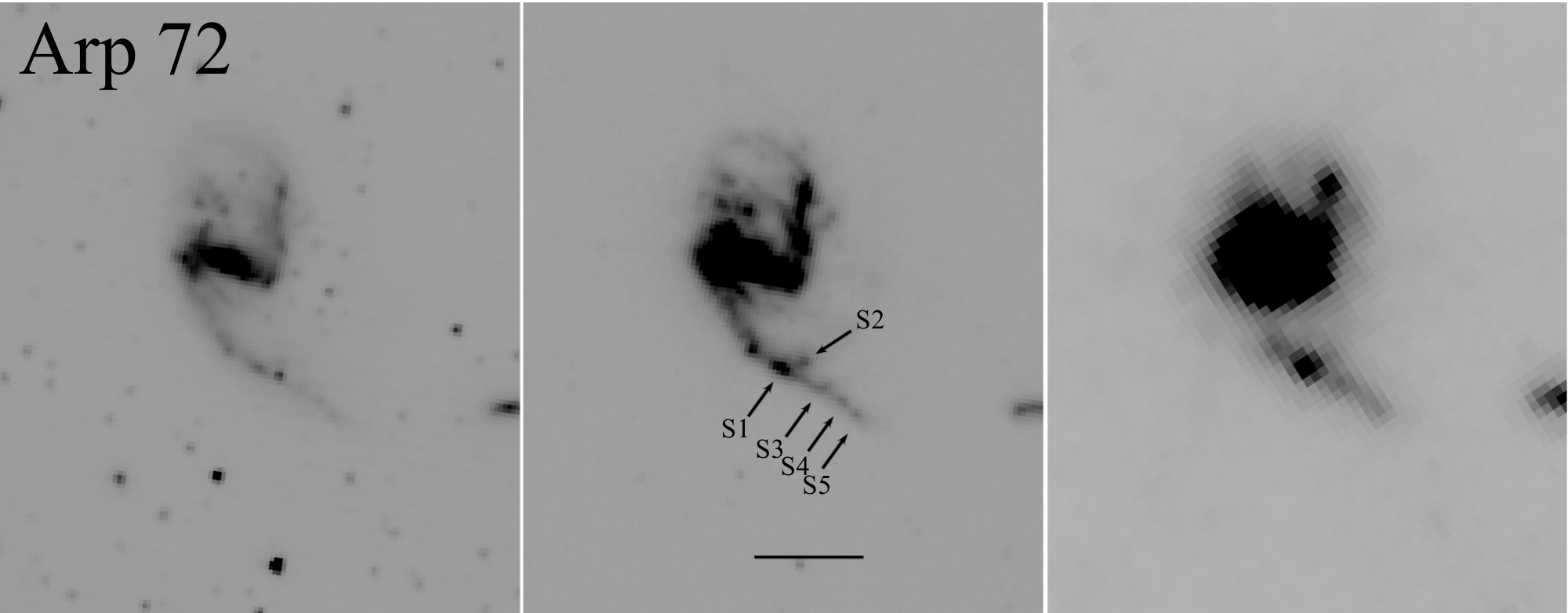}
\caption{ Arp 72 consists of the peculiar starburst (NGC 5996) with a
  strong interaction with its smaller companion NGC 5994. The eastern
  arm of NGC 5996 is prominent in the UV/visible while the western
  arm is prominent at both UV/visible wavelengths and at 8 \ums. The
  western arm forms a bridge with the companion. The brightest
  8 \um ISFO is in the bridge and is at the base of the bifurcation of
  the bridge/arm material.}
%\epsscale{.80}
%\plotone{Figures/Fig1b-arp72finder.pdf}
\end{figure}

\addtocounter{figure}{-1}
\begin{figure}
\includegraphics[width=1.0\textwidth]{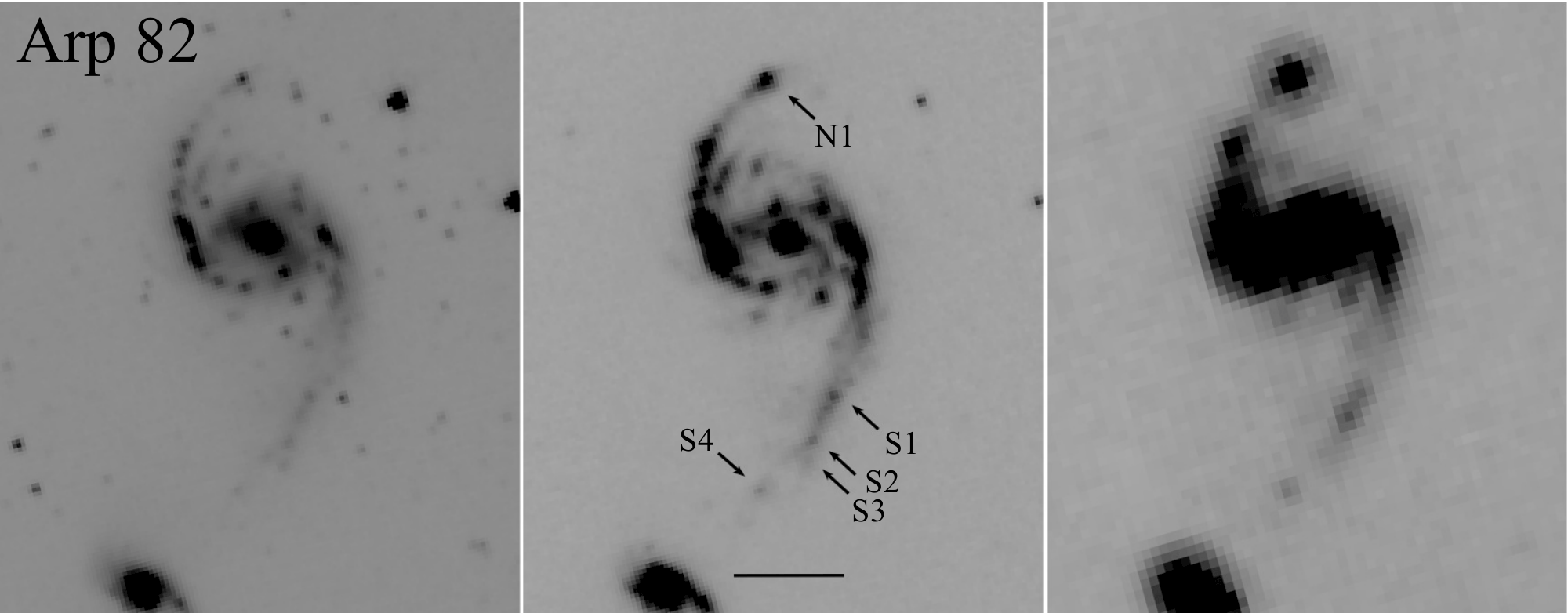}
%\epsscale{.80}
%\plotone{Figures/Fig1d-arp84finder.ps}
\caption{ cont. Arp 82 is an M51-like system. NGC 2335 is a
  knotty spiral with a small companion NGC 2536 on the extended
  southern arm. The southern arm is visible at UV/optical and infrared
  wavelengths whereas the northern arm is prominent in UV/optical
  images and absent at 8 \ums. We identify a bright hinge-clump
  Arp82-N1 at the base of the northern tail. The brightest ISFO in the
  southern tail is labeled Arp 82-S1.}
\end{figure}

\addtocounter{figure}{-1}
\begin{figure}
\includegraphics[width=1.0\textwidth]{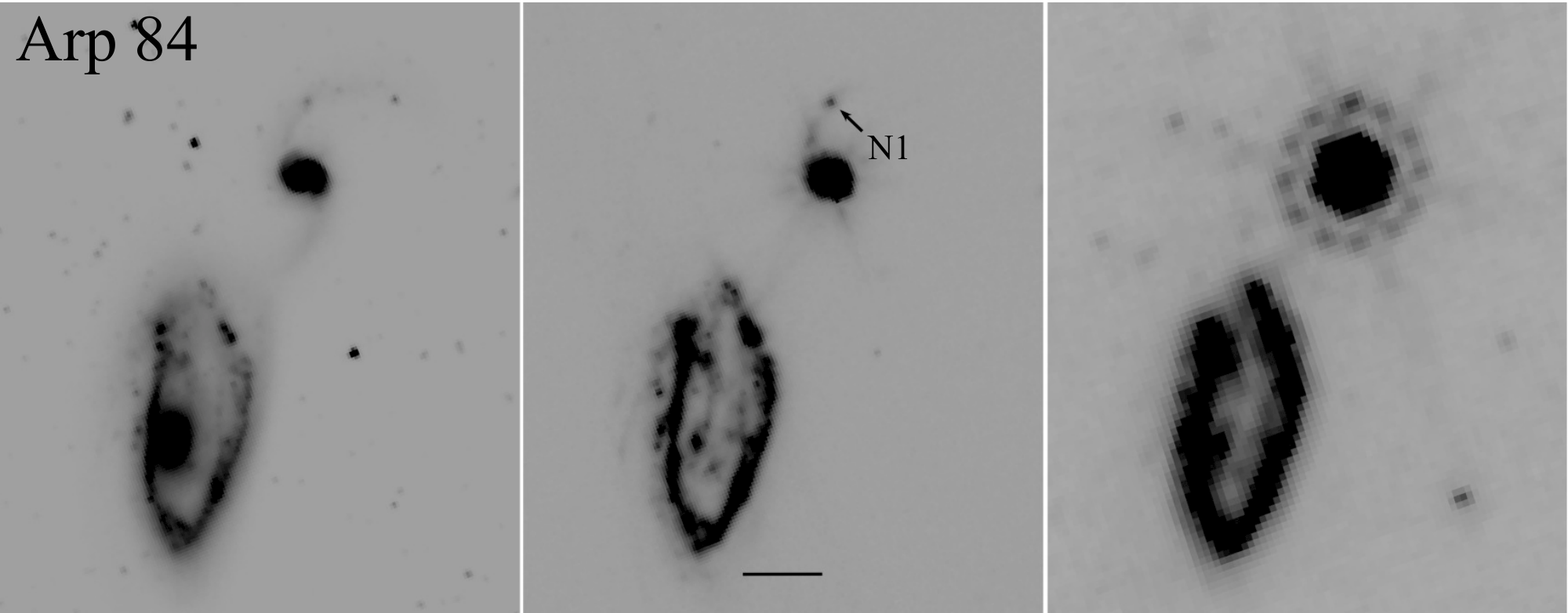}
%\epsscale{.80}
%\plotone{Figures/Fig1d-arp84finder.ps}
\caption{ cont. Arp 84. NGC 5395 is the large companion and is an
  asymmetric spiral. NGC 5394 has two open arms. The southern arm
  connects to NGC 5395. We observed the brightest 8 \um bead/ISFO in
  the northern tail. Emission from the ISFO at 24 \um is confused by
  diffraction from NGC 3808A. Smith et al. (2007a) note that there may
  be some accretion material associated with the northern tidal
  tail. }
\end{figure}

\addtocounter{figure}{-1}
\begin{figure}
\includegraphics[width=1.0\textwidth]{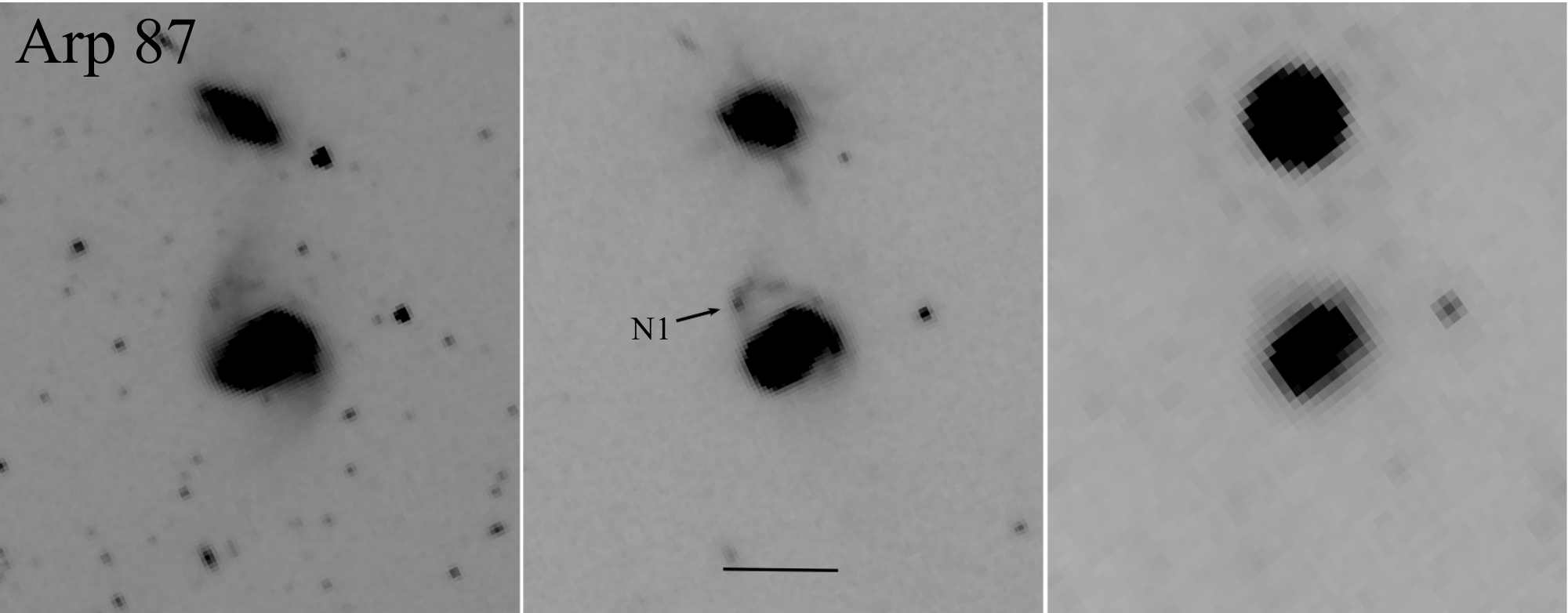}
%\epsscale{.80}
%\plotone{Figures/Fig1e-arp87finder.pdf}
\caption{cont. Arp 87 is a more equal-mass interacting pair, NGC 3808
  and NGC 3808A. NGC 3808 has a polar ring-like structure. We have
  observed the brightest 8 \um knot in the bridge star forming region,
  Arp 87-N1. Emission from the ISFO at 24 \um is confused by
  diffraction from the main galaxy. }
\end{figure}

\addtocounter{figure}{-1}
\begin{figure}
%\epsscale{.80}
%\plotone{Figures/Fig1f-arp102findercloseup.pdf}
\includegraphics[width=1.0\textwidth]{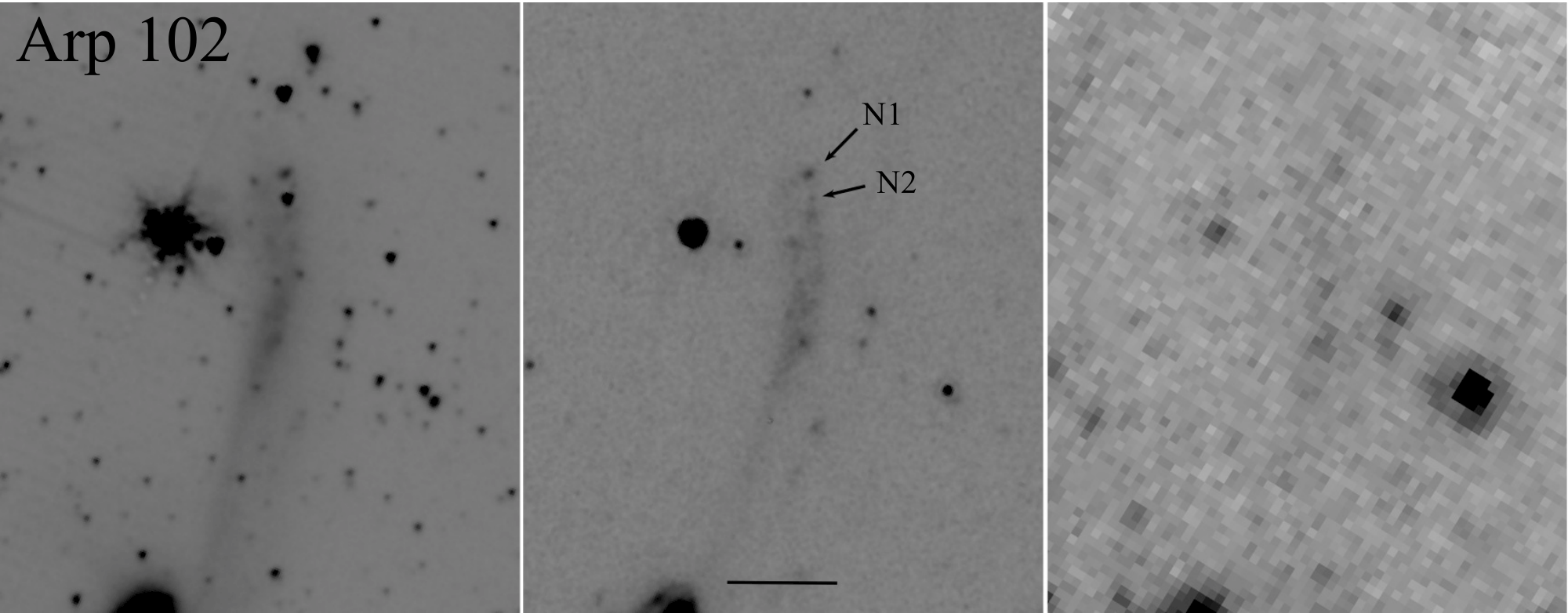}
\caption{cont. Arp 102 is an interaction between a spiral and
  elliptical galaxy. We imaged the tidal tail to the north of the
  northern spiral galaxy. Schombert et al. (1990) note that the tail
  is divided into two by an absorption band and looks like a spiral
  arm that has been straightened by the interaction between the galaxy
  pair. }
\end{figure}

%  do we see the bifurcation clearly at 3.6 or 8 \um ?

\addtocounter{figure}{-1}
\begin{figure}
\includegraphics[width=1.0\textwidth]{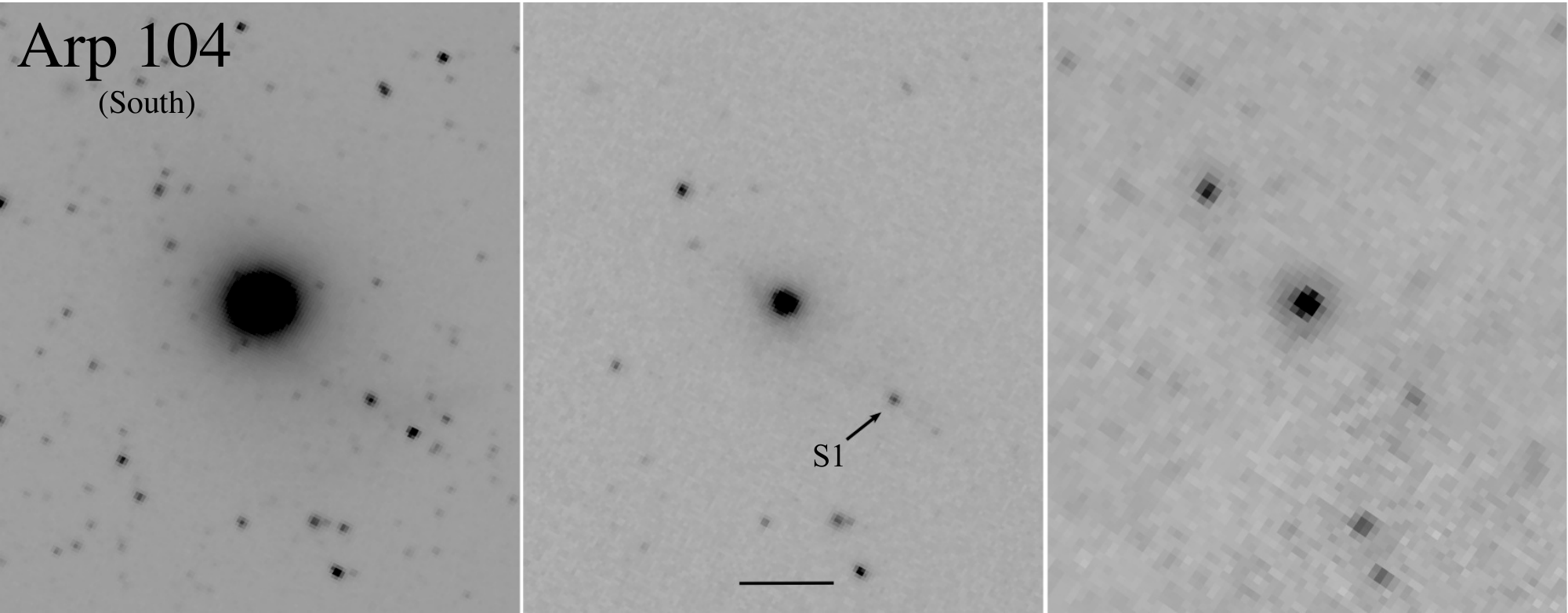}
%\epsscale{.80}
%\plotone{Figures/Fig1g-arp104findersouth.pdf}
\caption{cont. Arp 104. NGC 5216 is a peculiar Elliptical connected by
  an HI bridge to its northern companion (NGC 5218). Arp 104-S1 is at
  the tip of the tail extending away from NGC 5216. }
\end{figure}

%DO WE SEE THE BRIDGE OR TAIL AT 3.6 OR 8 UM?  NEED FINDER

\addtocounter{figure}{-1}
\begin{figure}
%\epsscale{.80}
\includegraphics[width=1.0\textwidth]{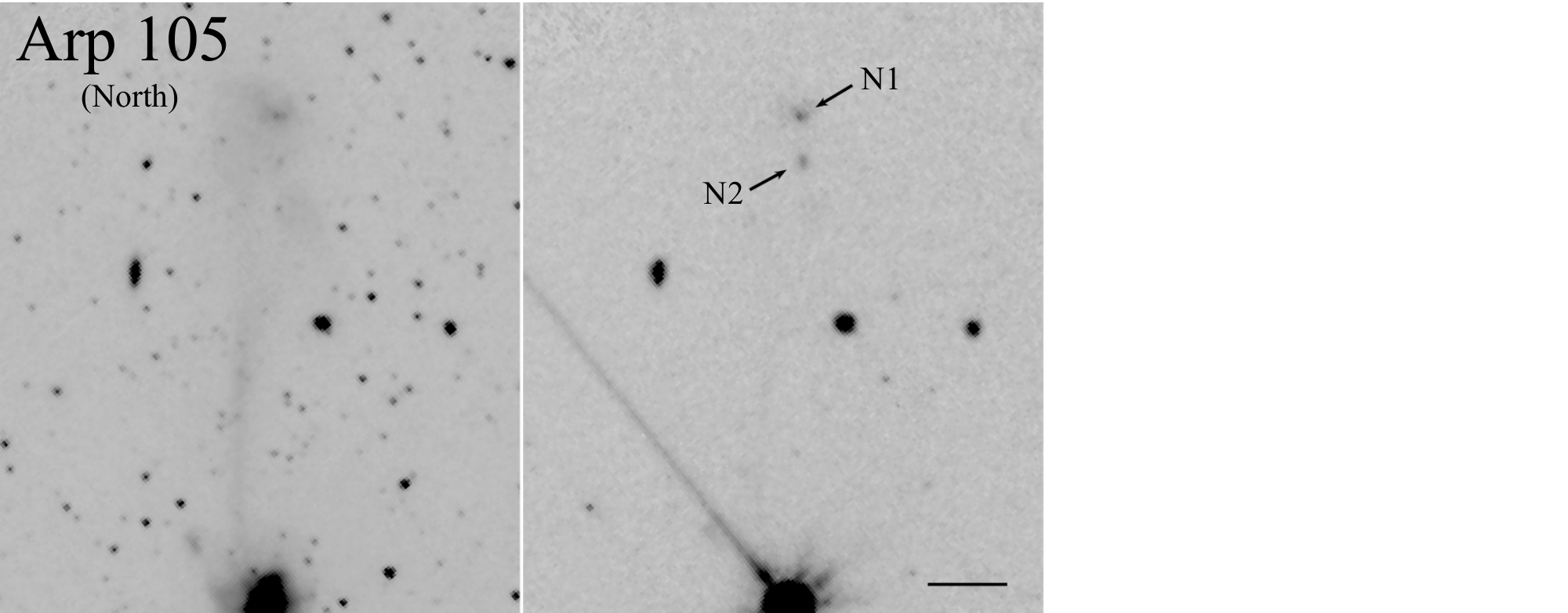}
%\plotone{Figures/Fig1h-arp105findernorth.pdf}
\caption{cont. Arp 105 North. Arp 105/The Guitar (Abell 1185) consists
  of a distorted spiral (NGC 3561A) and a very close lenticular
  companion (NGC 3561B). This figure shows the northern tail. We
  identify two ISFOs at the tip of this tail. There are no clumps
  identified along the rest of the 2 arcminute northern tail }
\end{figure}

\addtocounter{figure}{-1}
\begin{figure}
%\epsscale{.80}
\includegraphics[width=1.0\textwidth]{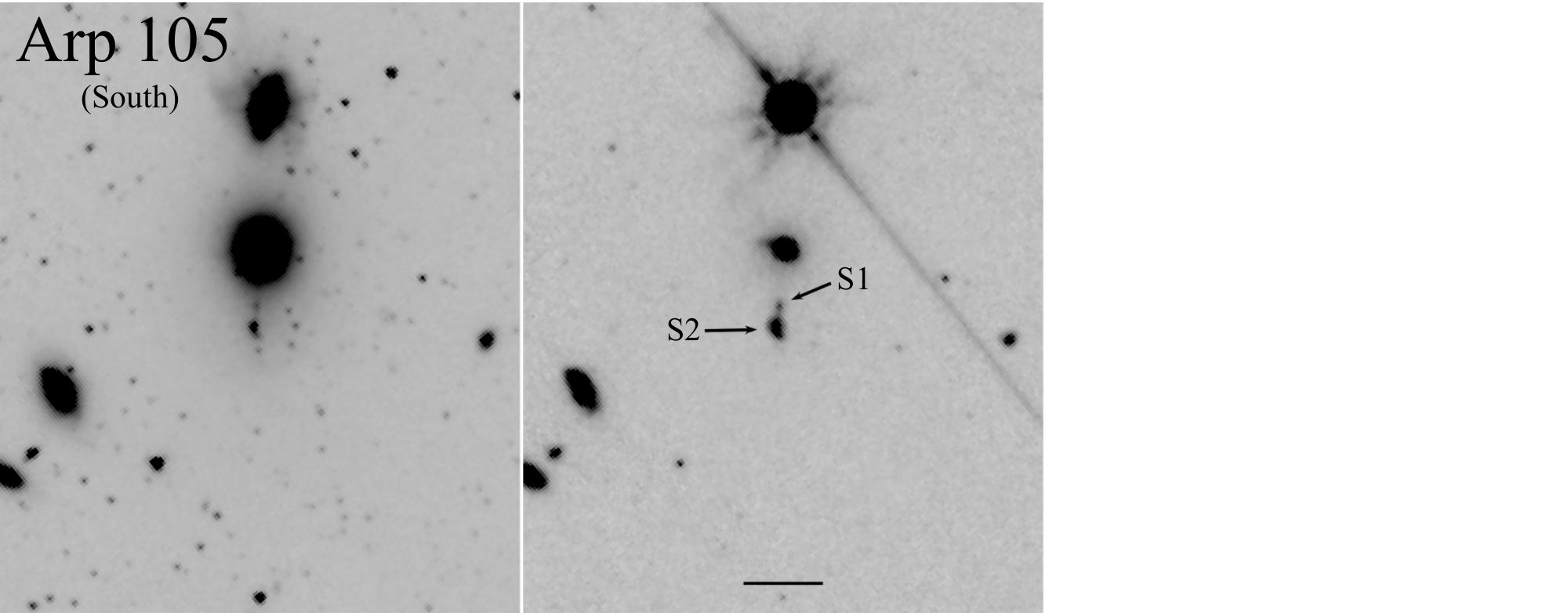}
%\plotone{Figures/Fig1i-arp105findersouth.pdf}
\caption{cont. Arp 105 South. Smith et al. (2010) considered the
  southern tail to be an `accretion tail', formed by material from the
  spiral that fell into the gravitational potential of the elliptical,
  overshot that potential, and is now forming stars. Two ISFOs are
  identified. An analysis of th IRS spectrum of ``Ambartsumian's
  knot'' (S2) at the tip of the short southern tail is presented in this
  paper.}
\end{figure}

\addtocounter{figure}{-1}
\begin{figure}
\includegraphics[width=1.0\textwidth]{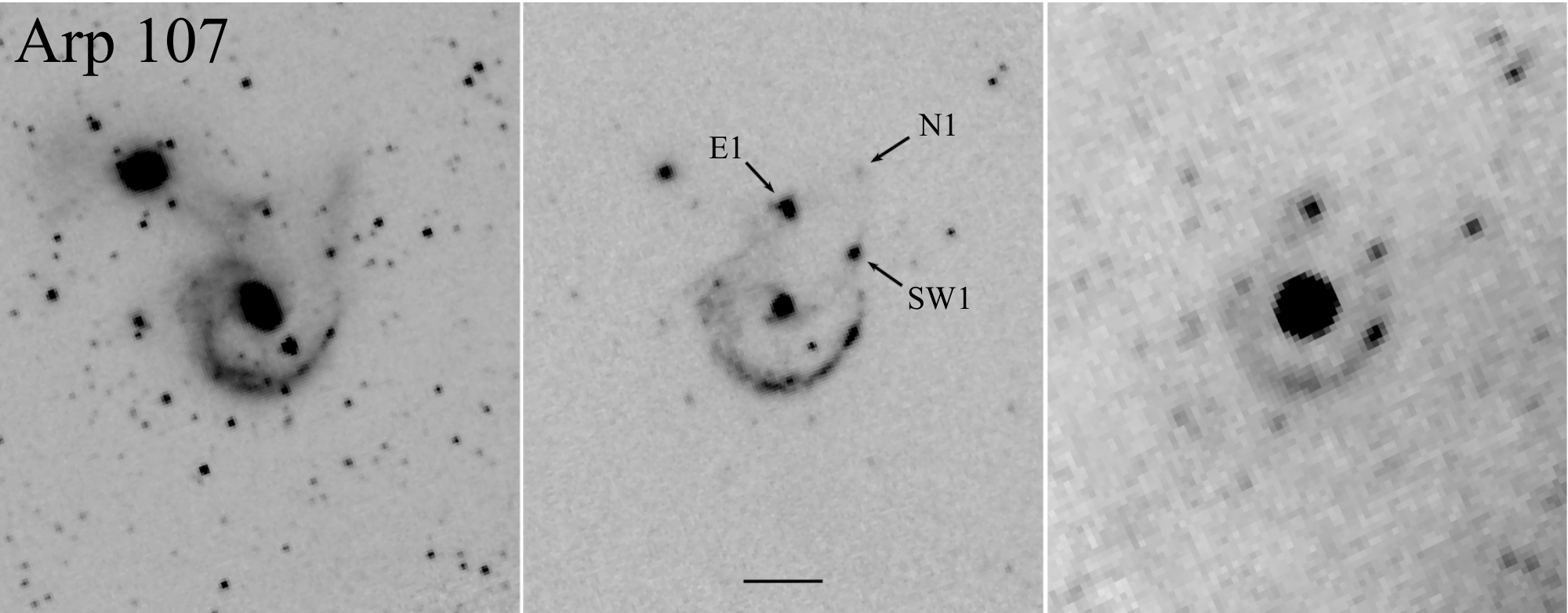}
%\epsscale{.80}
%\plotone{Figures/Fig1j-arp107finder.pdf}
\caption{cont.  Arp 107 has a prominent ring-like structure connected
  via a bridge to an elliptical-like companion. Smith et al. (2005)
  reproduced the basic morphology as a collisional ring galaxy with a
  prograde planar intruder passage. Arp 107-SW1 is at the tip of the
  southern tail associated with the elliptical-like galaxy. }
\end{figure}

\addtocounter{figure}{-1}
\begin{figure}
\includegraphics[width=1.0\textwidth]{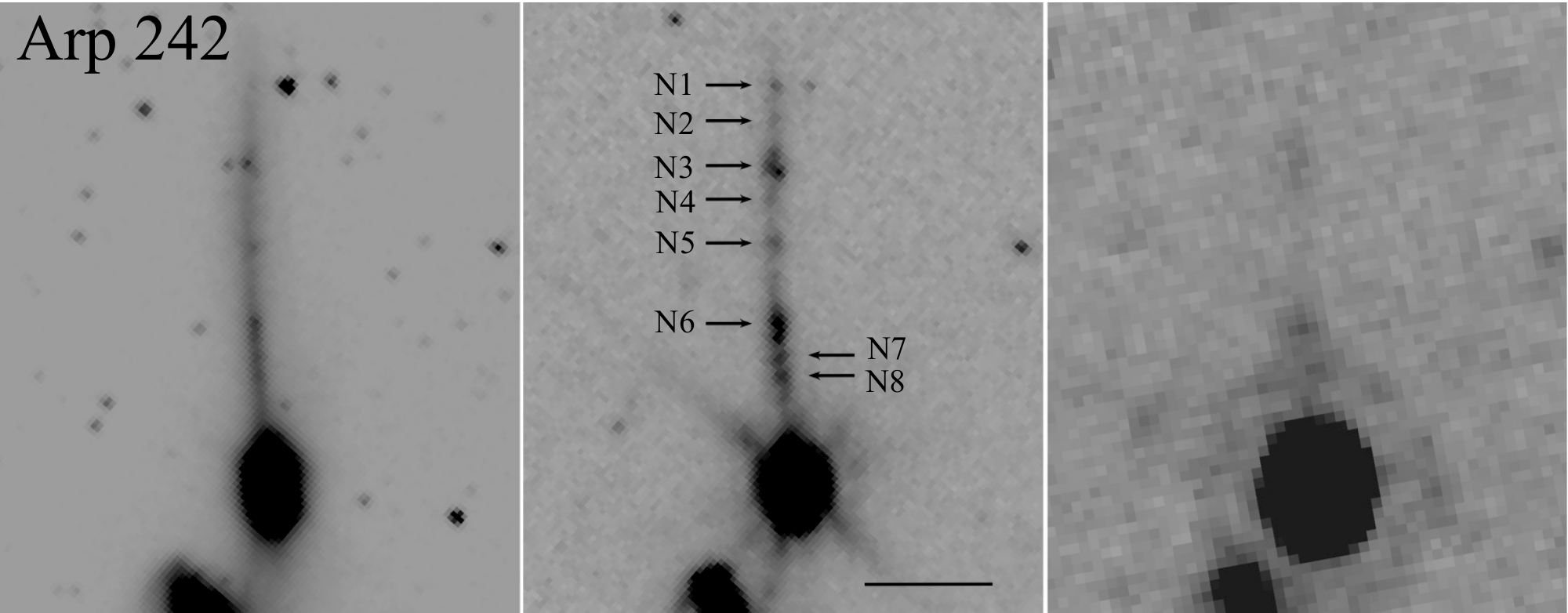}
%\epsscale{.80}
%\plotone{Figures/Fig1k-arp242finder}
\caption{cont. Arp 242$/$Mice$/$NGC 4676 is a classic example of a
  tidal interaction between two spirals. Two long tidal tails are
  visible in the original Arp image (Arp 1966). Only the northern tail
  is bright at 8 \ums. Arp 242-N3 is the brightest knot in the outer
  third of the tail. }
\end{figure}

%  Smith \& Higdon 1994  CO DATA

\addtocounter{figure}{-1}
\begin{figure}
\includegraphics[width=1.0\textwidth]{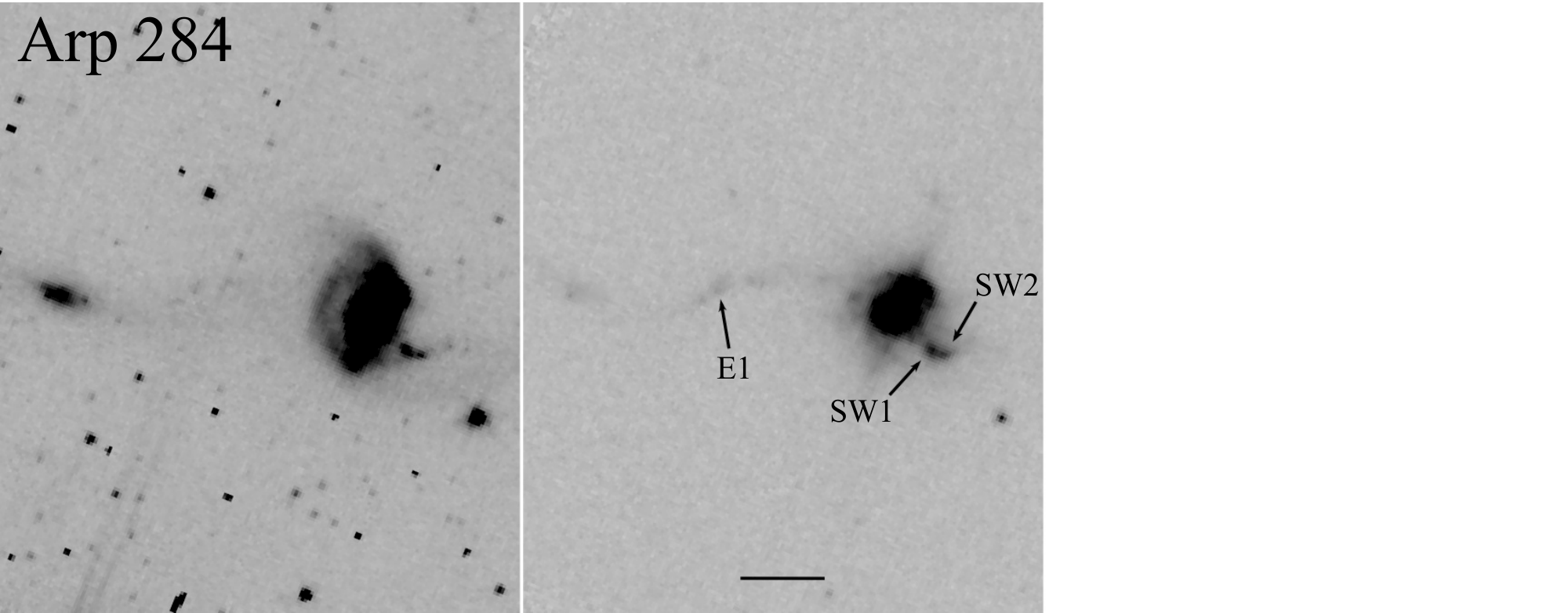}
%\epsscale{.80}
%\plotone{Figures/Fig1l-arp284finder.ps}
\caption{cont. Arp 284 consists of two interacting spirals, NGC
  7714/5. NGC 7714 is a classic starburst galaxy, with a partial ring
  with three tails and a bridge connecting it to its edge-on companion
  NGC 7715. Struck \& Smith (2003) reproduced the basic morphology
  with a pro-grade, near head-on collision, with the western tail
  being formed via accretion from the bridge and the HI loop being a
  classical tidal feature.  Arp 284-SW1 is a knot in the accretion tail
  and Arp 284-E1 is a knot in the bridge joining the two galaxies.}
\end{figure}

\addtocounter{figure}{-1}
\begin{figure}
\includegraphics[width=1.0\textwidth]{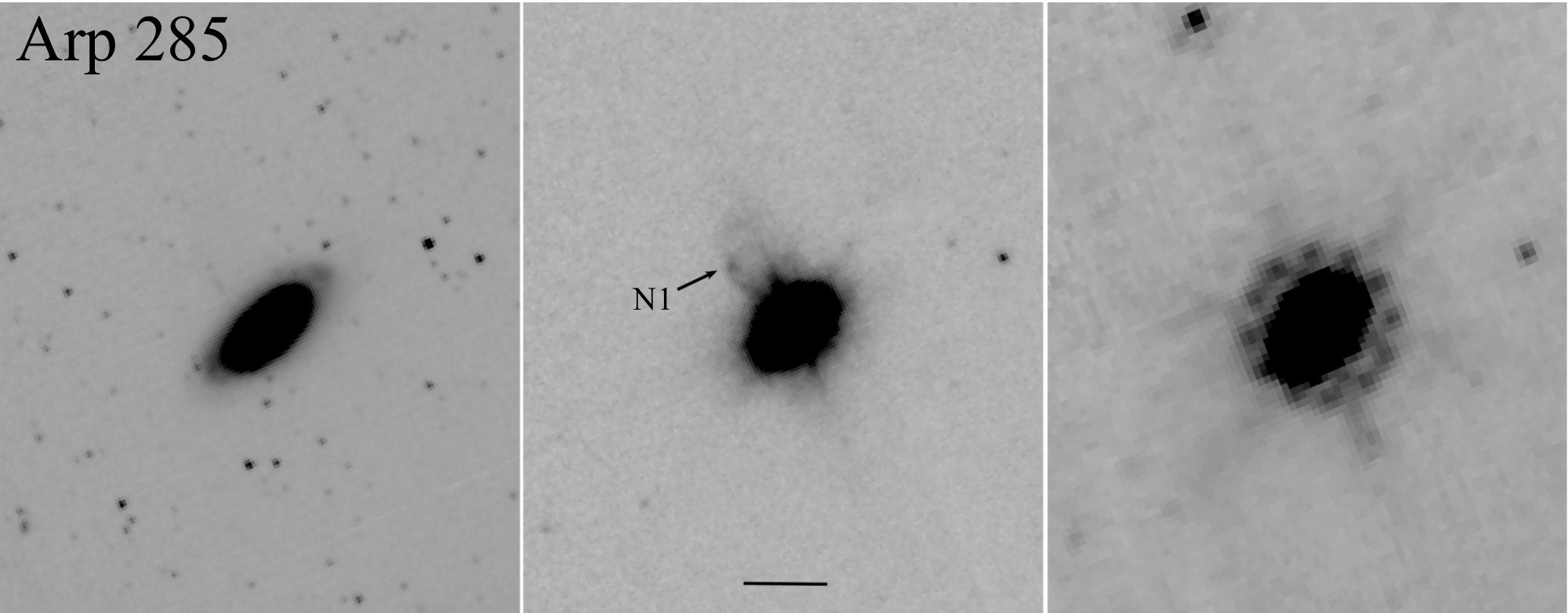}
%\epsscale{.80}
%\plotone{Figures/Fig1m-arp285finder.ps}
\caption{cont. Arp 285. The northern tail-like
structure is perpendicular to the disk of NGC 2856. Simulations
by Smith et al. (2008) suggest this feature may have been accreted
from material along the bridge joining NGC 2856 to NGC 2854. }
\end{figure}
%our image containts a beautiful spiral north of this system NGC 2857
%Arp 1 spirals wit low surf bright - 2 very long thin arms.

\addtocounter{figure}{-1}
\begin{figure}
\includegraphics[width=1.0\textwidth]{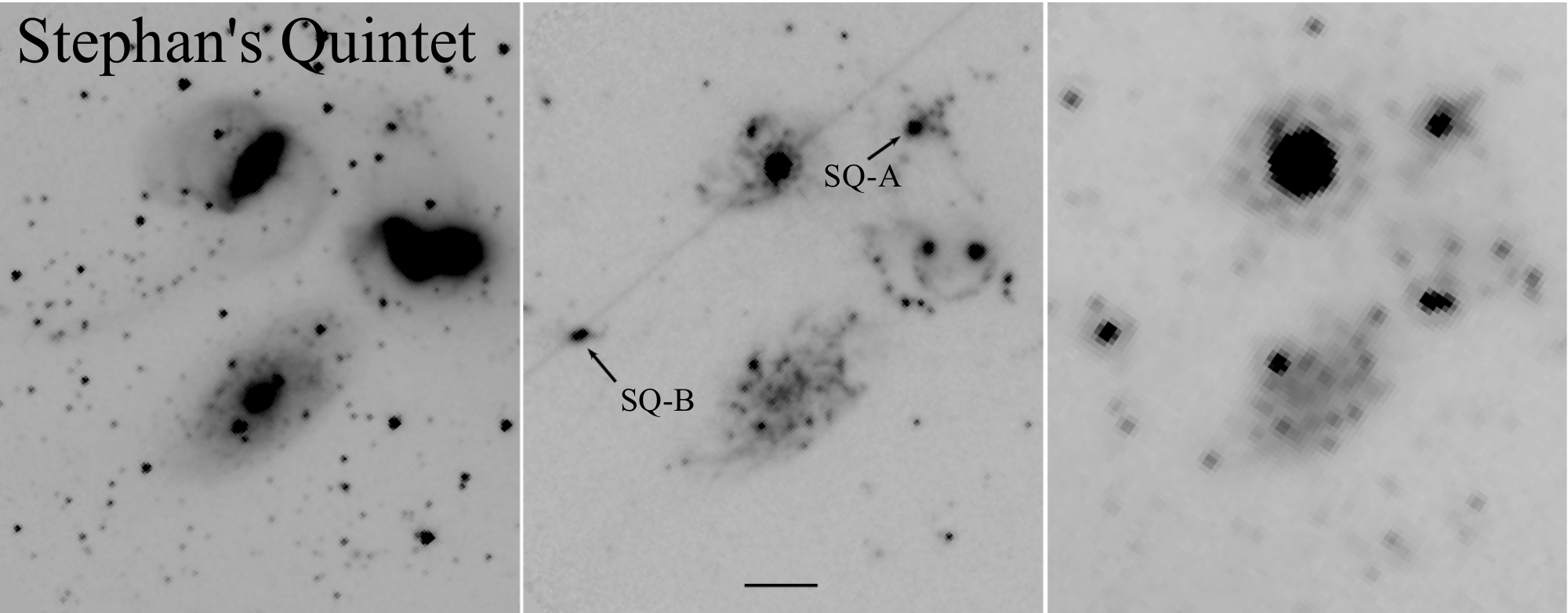}
%\epsscale{.80}
%\plotone{Figures/Fig1n-stephan5finder.eps}
\caption{cont. Stephan's Quintet (SQ, Arp 319) is a prototype compact
  group (Hickson 92).  There are five galaxies in SQ: NGC 7317(E), NGC
  7318A(E), NGC 7318B (Sbc pec), NGC 7319 (Sbc pec Sey 2) and NGC 7320
  (Sd foreground galaxy). Two 100 kpc parallel tails stretch from NGC
  7319 towards the intruder NGC 7320c.  We observed SQ-A with the IRS.
  (spectra are not included for SQ-B (Arp 1973) in the tidal tail of
  NGC7319/HCG 92C, due to a pointing issue with the data).}
\end{figure}

\clearpage

%Figure 2
\begin{figure}
\centering
\includegraphics[width=.7\textwidth]{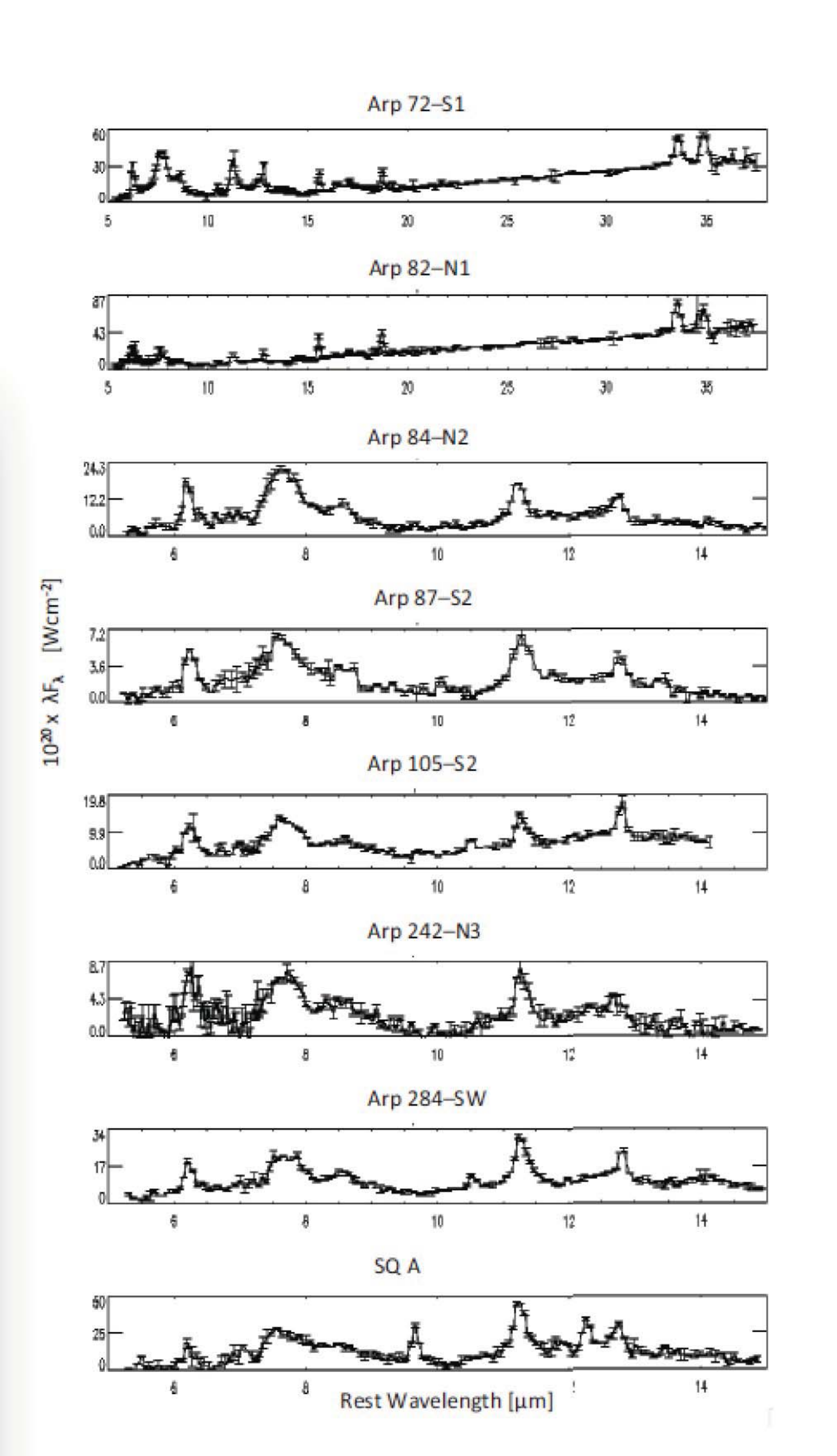}

\caption{IRS LORES Spectra. Note, the spectra of Arp 72-S1 \& Arp
  82-N1 include IRS-LL data. }
\end{figure}

%Figure 3

\begin{figure}
\includegraphics[width=.4\textwidth]{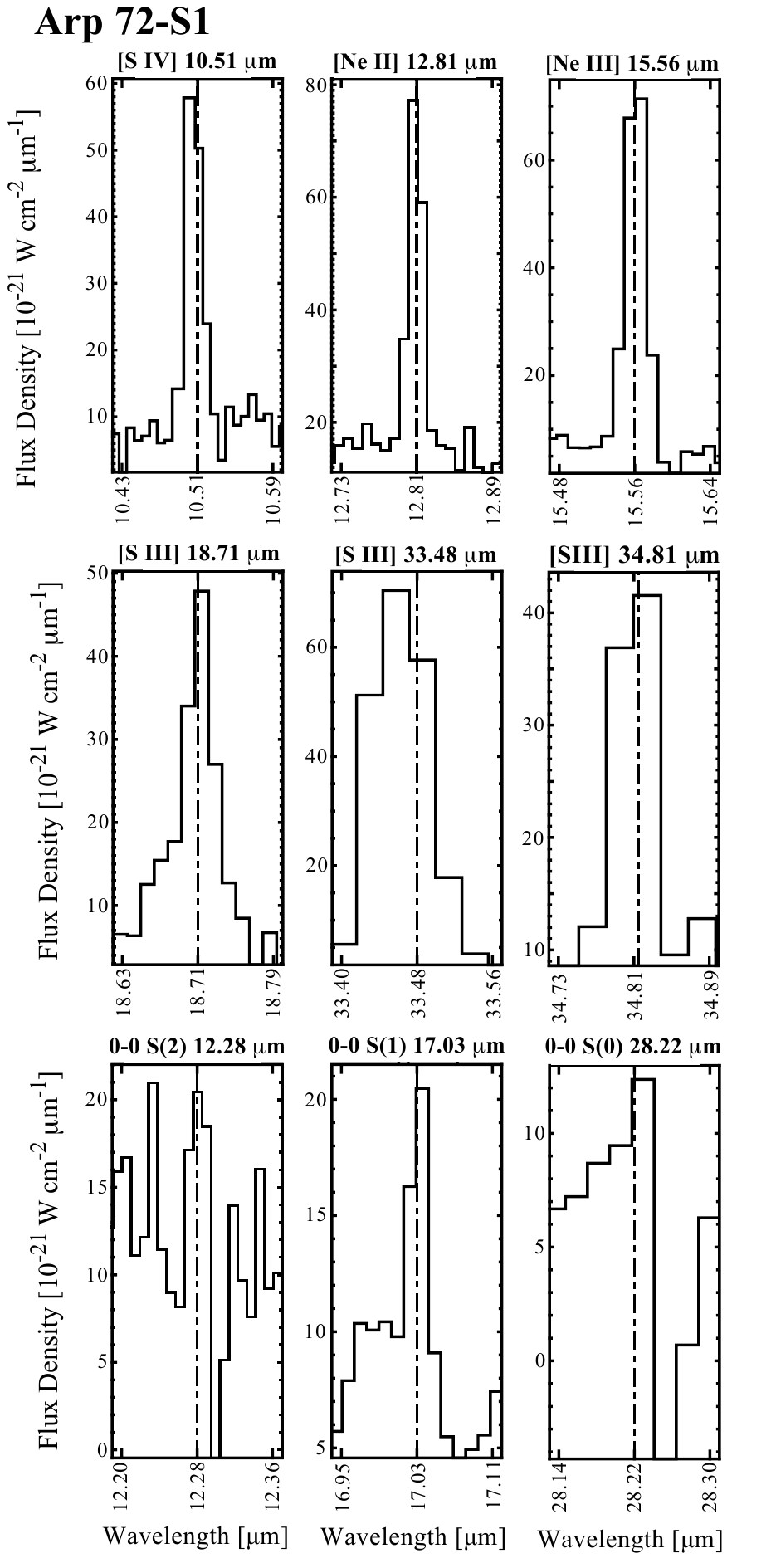}
%\epsscale{.80}
%\plotone{hardnessbw.eps}
\caption{IRS HIRES emission line profiles - Arp 72-S1 }
\end{figure}

\addtocounter{figure}{-1}
\begin{figure}
\includegraphics[width=.4\textwidth]{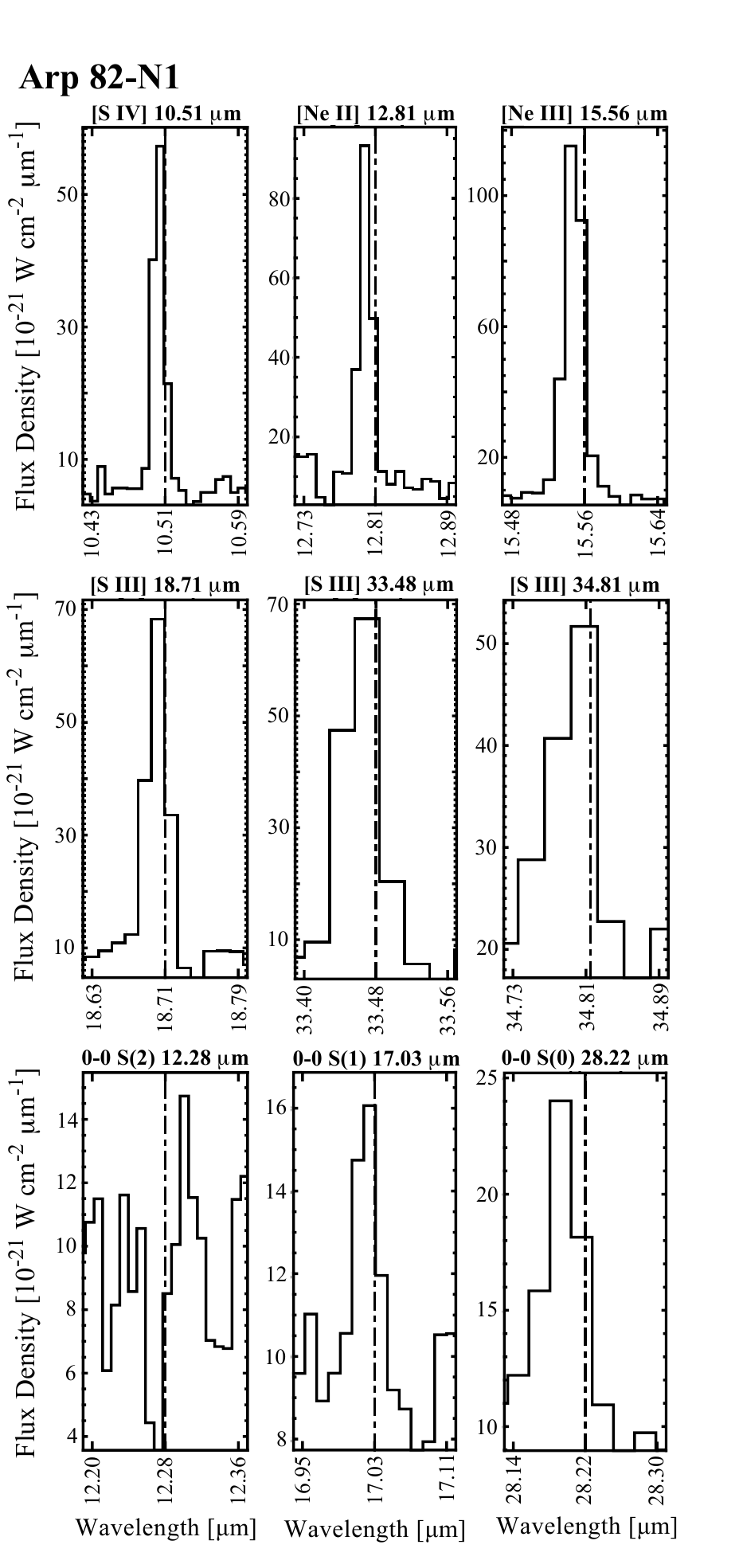}
%\epsscale{.80}
%\plotone{hardnessbw.eps}
\caption{cont. Arp 82-N1}
\end{figure}

\addtocounter{figure}{-1}
\begin{figure}
\includegraphics[width=.4\textwidth]{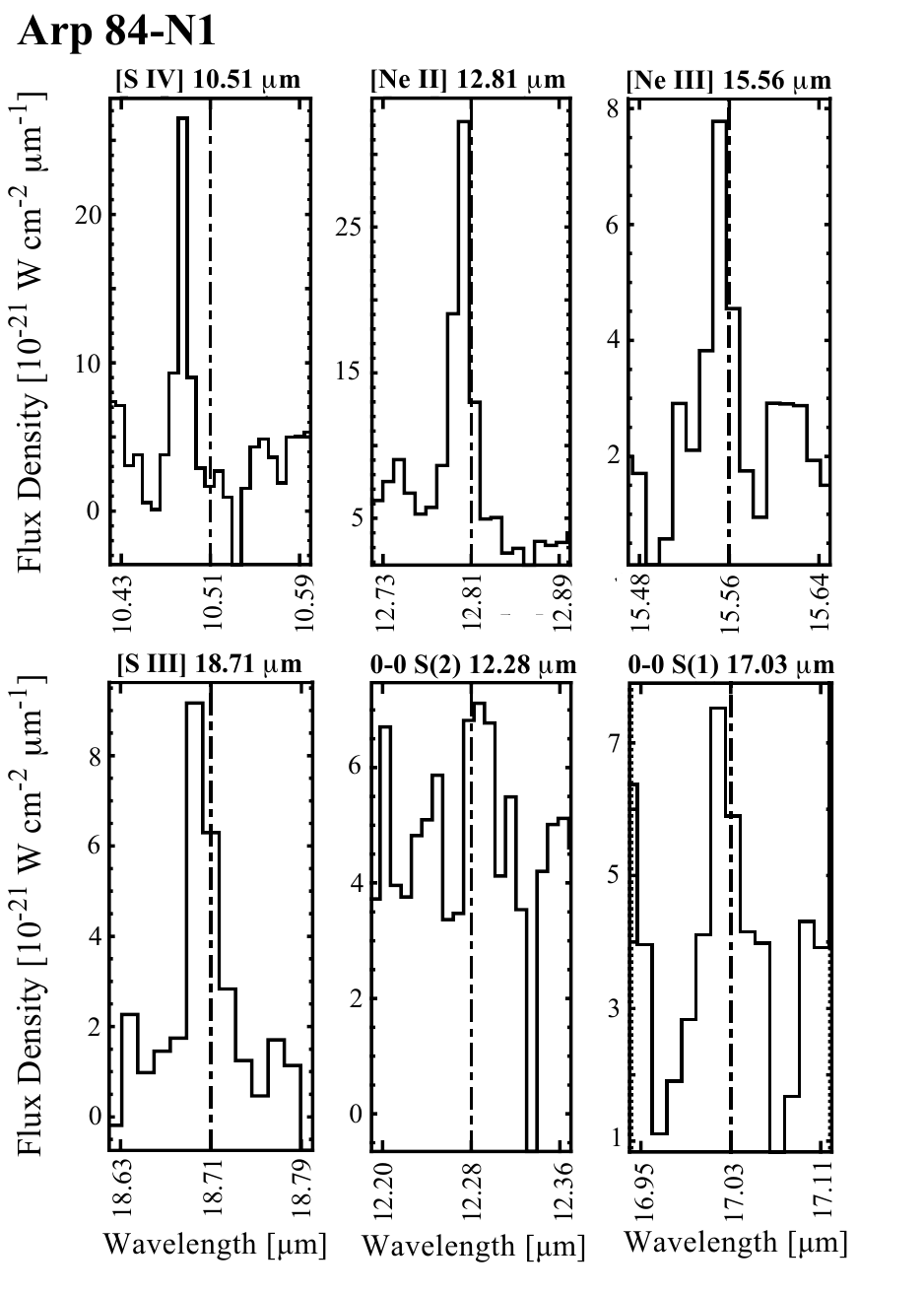}
%\epsscale{.80}
%\plotone{hardnessbw.eps}
\caption{cont. Arp 84-N1}
\end{figure}

\addtocounter{figure}{-1}
\begin{figure}
\includegraphics[width=.4\textwidth]{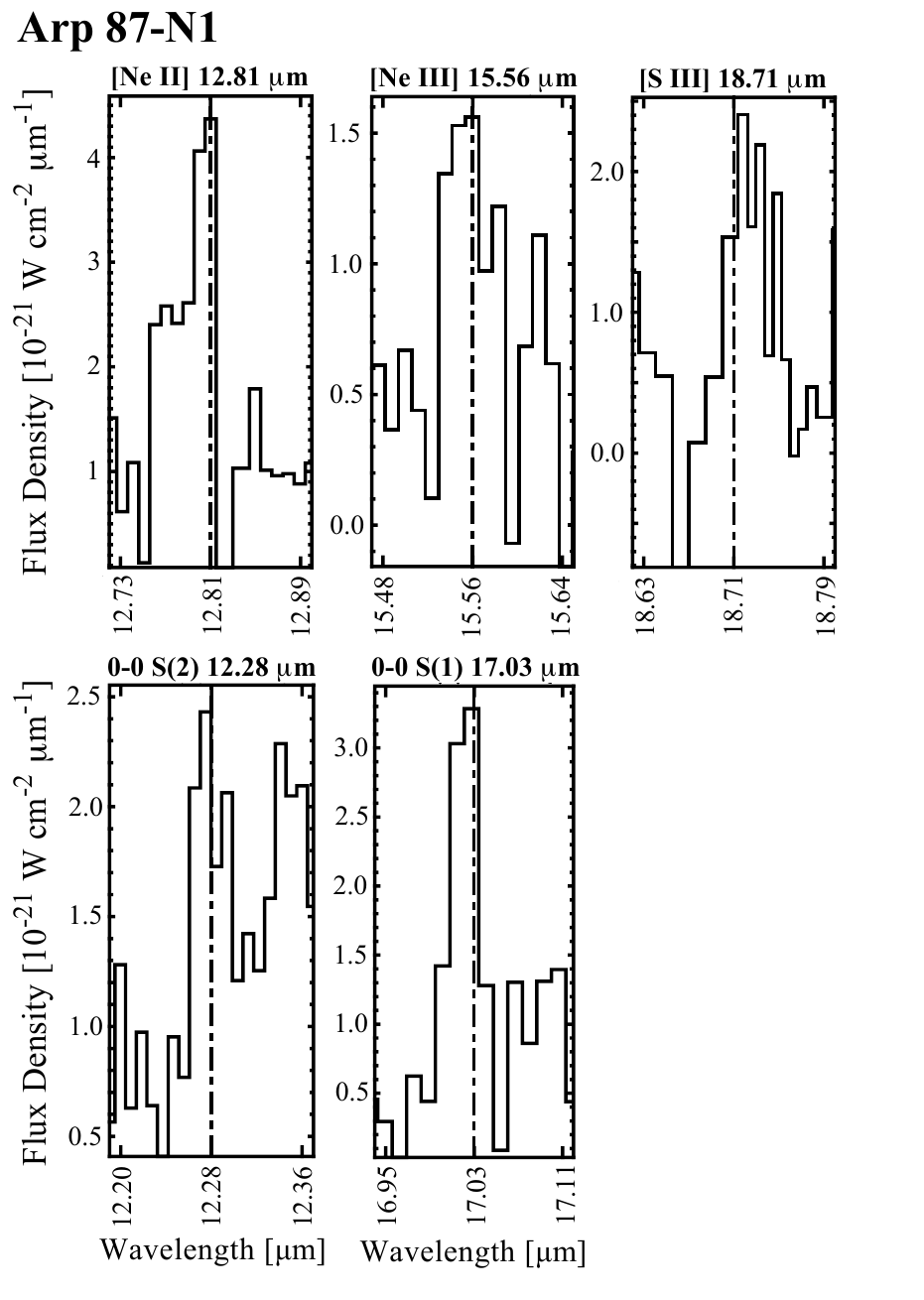}
%\epsscale{.80}
%\plotone{hardnessbw.eps}
\caption{cont. Arp 87-N1}
\end{figure}

\addtocounter{figure}{-1}
\begin{figure}
\includegraphics[width=.4\textwidth]{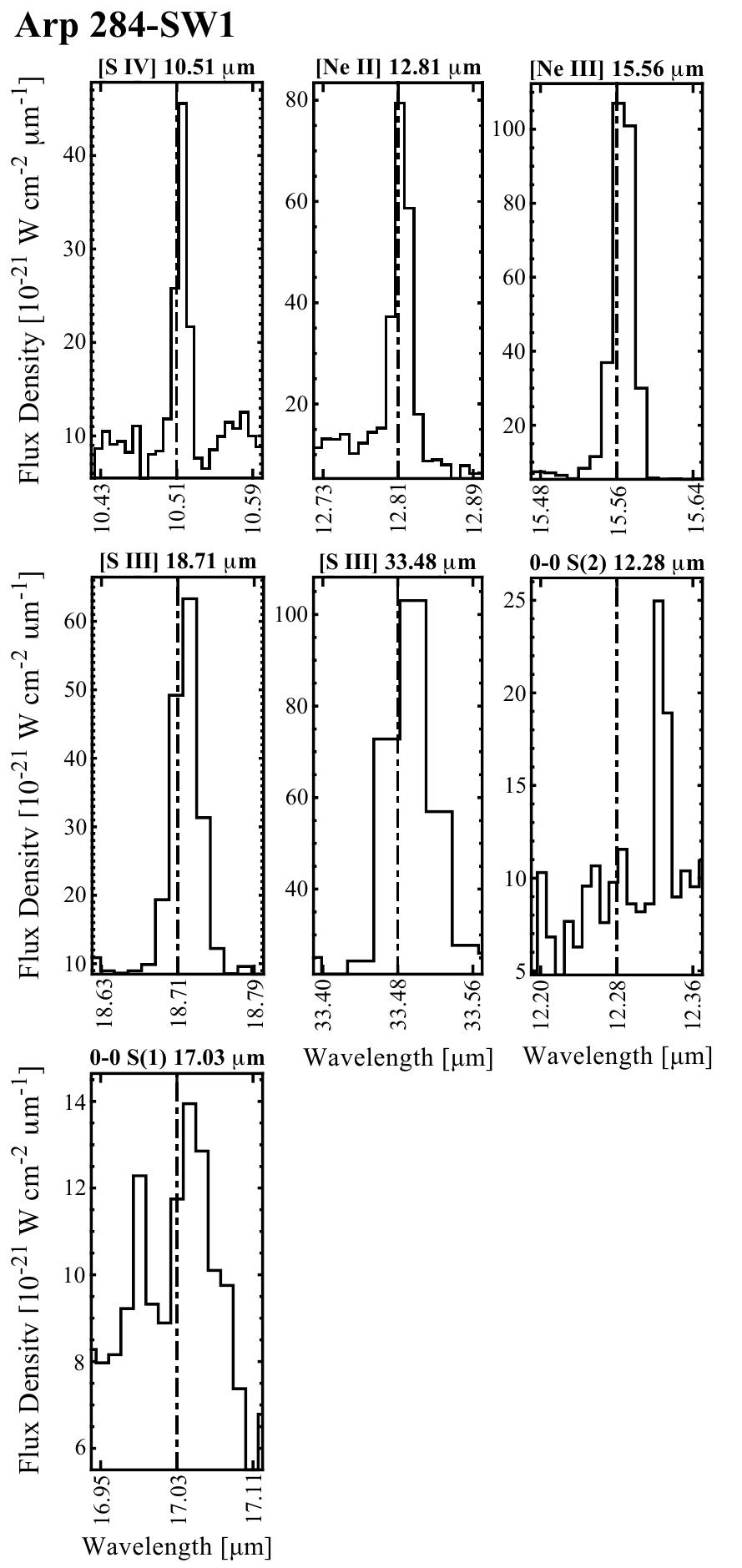}
%\epsscale{.80}
%\plotone{hardnessbw.eps}
\caption{cont. Arp 284-SW1}
\end{figure}

\addtocounter{figure}{-1}
\begin{figure}
\includegraphics[width=.4\textwidth]{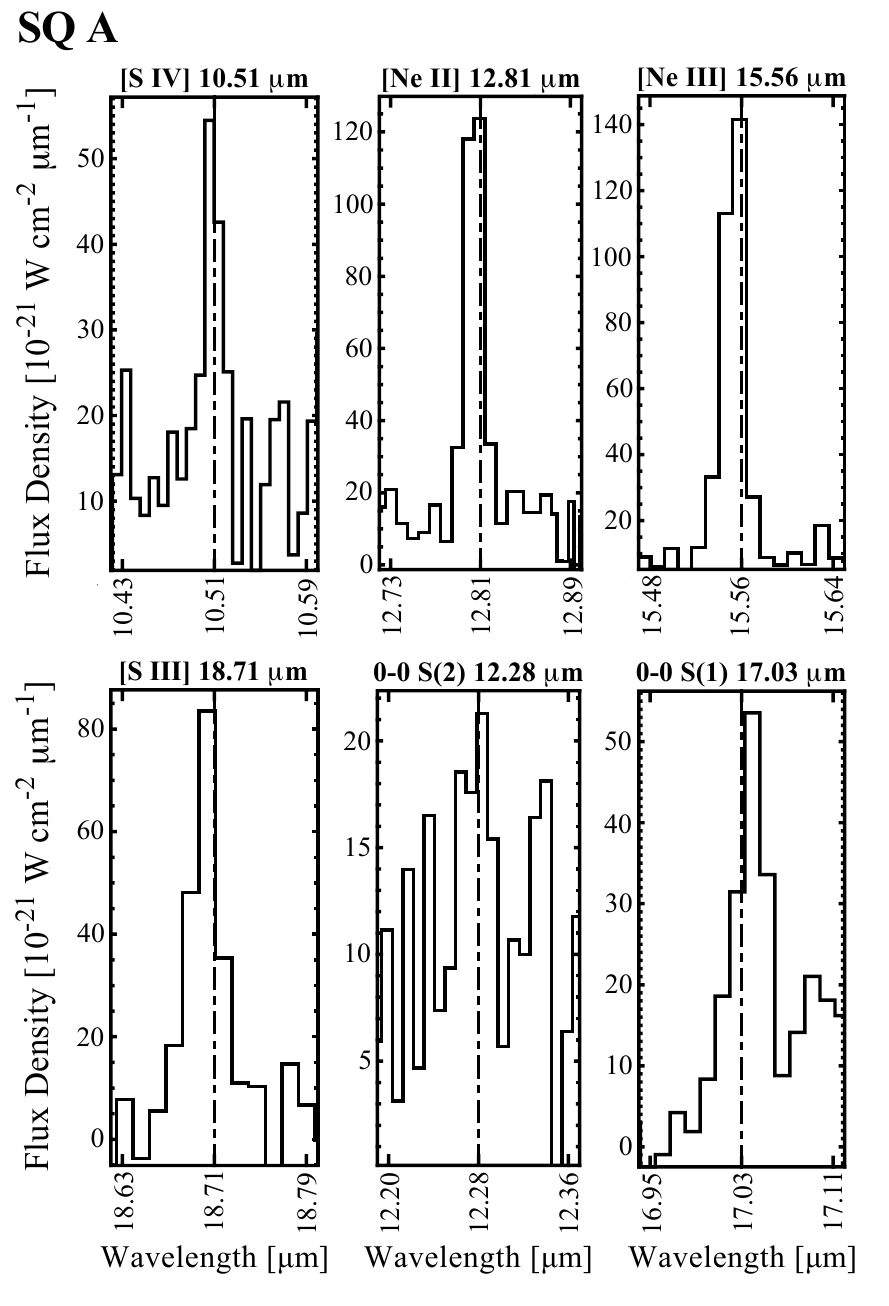}
%\epsscale{.80}
%\plotone{hardnessbw.eps}
\caption{cont. SQ A }
\end{figure}

\clearpage
% Figure 4
\begin{figure}
\includegraphics[width=.9\textwidth]{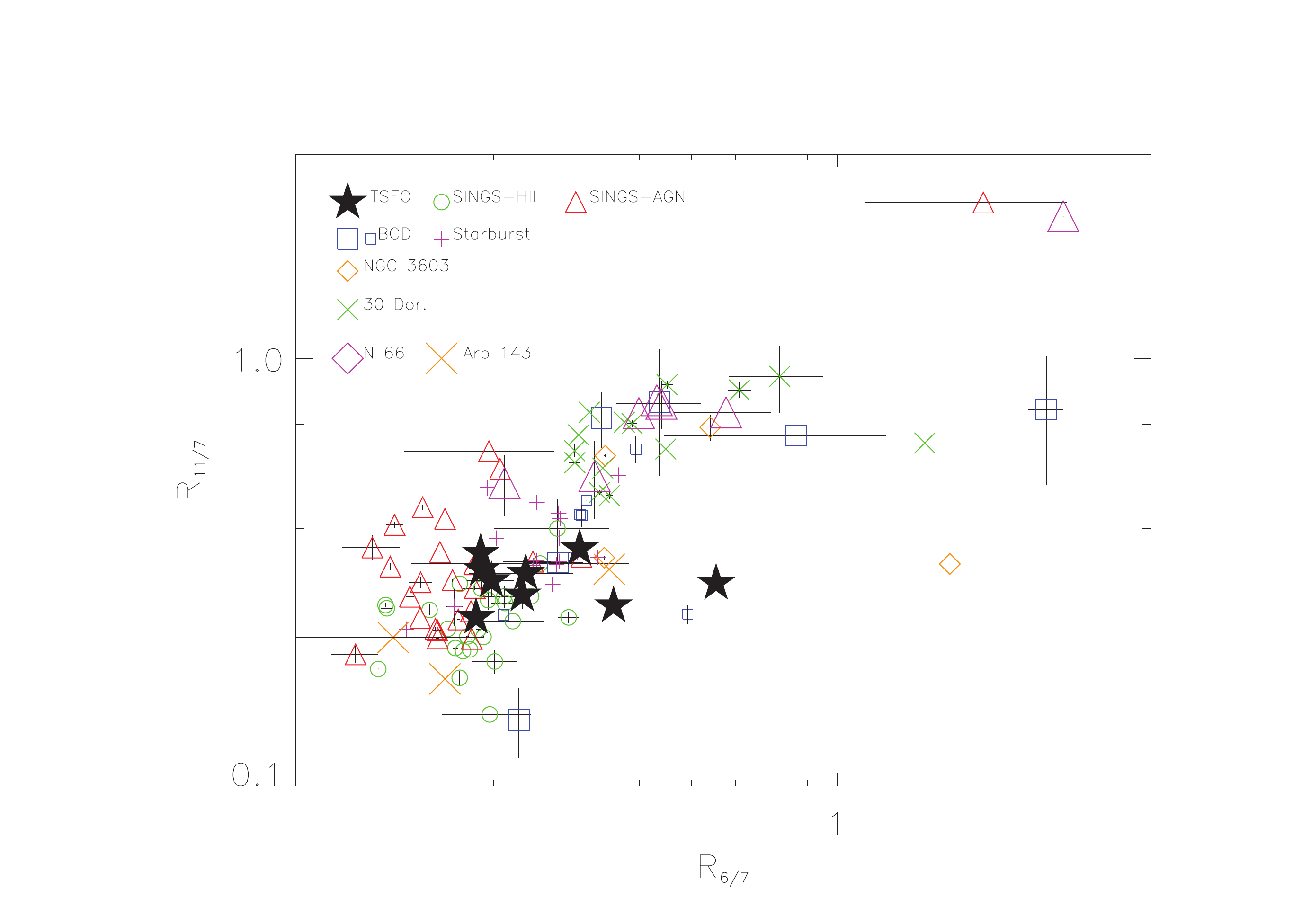}
%\centering
%\includegraphics[width=60mm]{pah-draine-bw.jpg}
%  \includegraphics[width=60mm]{myfig.png}
%  \includegraphics[height=60mm]{myfig.jpg}
%  \includegraphics[scale=0.75]{myfig.pdf}
%  \includegraphics[angle=45,width=52mm]{myfig.jpg}
%\MakePicture{pah-draine-bw.jpg}{angle=0, scale=0.5}[h]
%\MakeCaption{Figure PAH/Draine -see pah-draine-bw.jpg}
%\includegraphics[angle=-90,scale=.5]{pahdrainebw.eps} 
% MMP ISRF \chi = 1.23 ,123 -add note from Draine  PAH/Draine
\caption{Relative strengths of the PAH bands. R$_{11/7}$ increases
  with decreasing PAH ion fraction and R$_{6/7}$ increases with
  decreasing PAH size. ISFOs are shown as filled stars. The
  SINGS-AGN are shown as (red) open triangles and the SINGS-HII
  (green) open circles (Smith et al. 2007b). BCDs are shown as (blue)
  large open squares (Hunt et al. 2010). Clumps in Arp 143 are shown
  as (orange) large X's (Beiraro et al. 2009). Star forming regions in
  NGC 3603 in the Milky Way are shown as (orange) open diamonds, in 30
  Dor. in the LMC as (green) small x's and in N 66 in the SMC as
  (pink) large open triangles, BCDs (blue) small open squares,
  starburst galaxies (pink) pluses (Lebouteiller et al. 2011). A color
  version of this figure is available in the online journal.}
\end{figure}

\clearpage

% Figure 5
\begin{figure}
\includegraphics[width=1.0\textwidth]{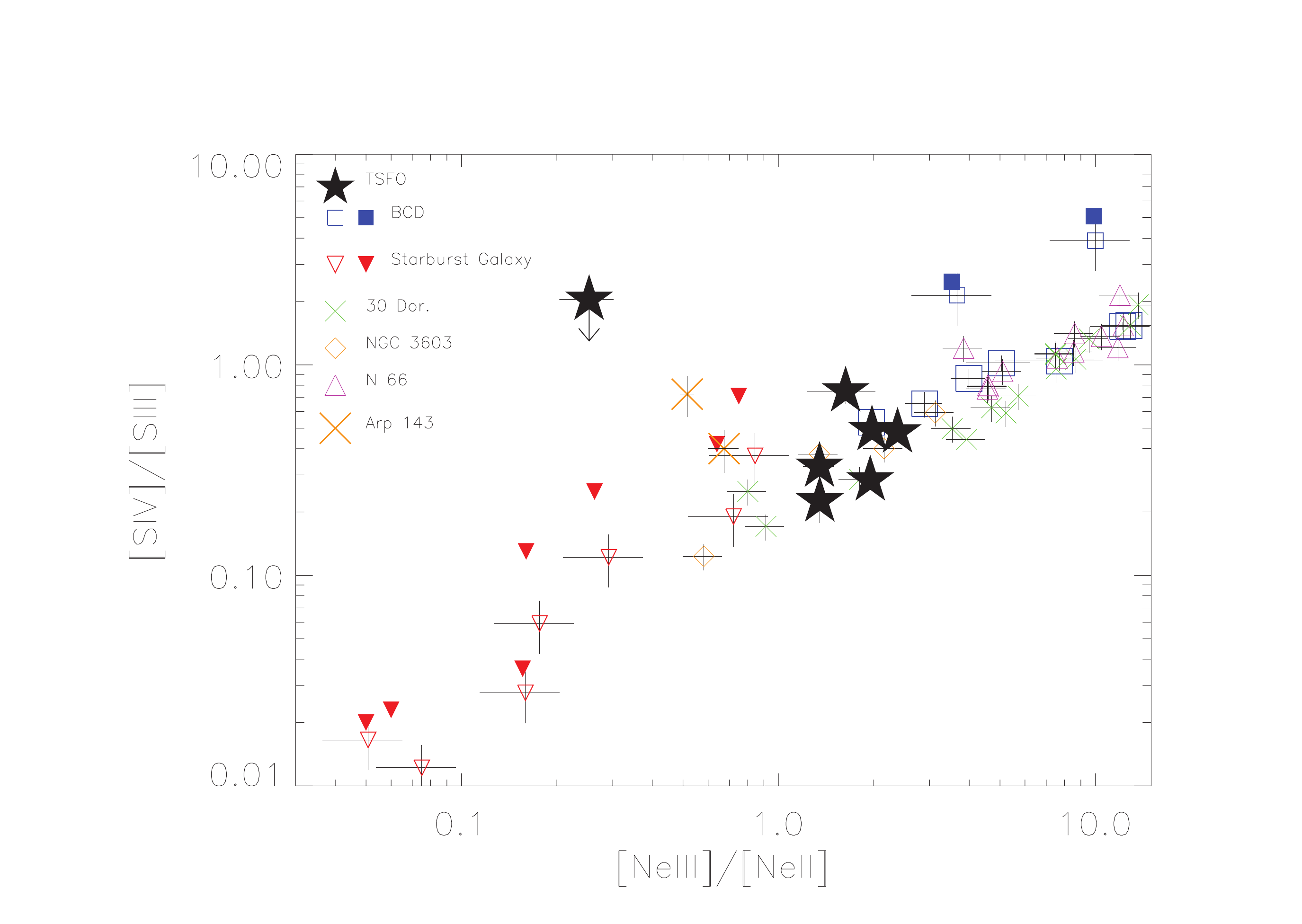}
%\epsscale{.80}
%\plotone{hardnessbw.eps}
\caption{Mid-infrared excitation diagram. ISFOs are displayed with a
  filled star symbol. Dusty starbursts are shown as (red) open
  inverted triangles, and extinction corrected data as (red) filled
  inverted triangles; BCDs are shown as (blue) small open squares and
  filled squares (extinction corrected, Verma et al. 2003. We adopted
  a 20\% uncertainty in the line fluxes). Additional BCDs are
  displayed as (blue) large open squares (Hunt et al. 2010).  Star
  forming regions in 30 Dor, NGC 3603 and N 66 are displayed using the
  same symbols as in Figure 4. The ISFOs are assumed to have minimal
  extinction (A$_V \le$ 3) and the line ratios are not corrected. A
  color version of this figure is available in the online journal.}
\end{figure}

\clearpage

% Figure 6

\begin{figure}
\includegraphics[width=1.0\textwidth]{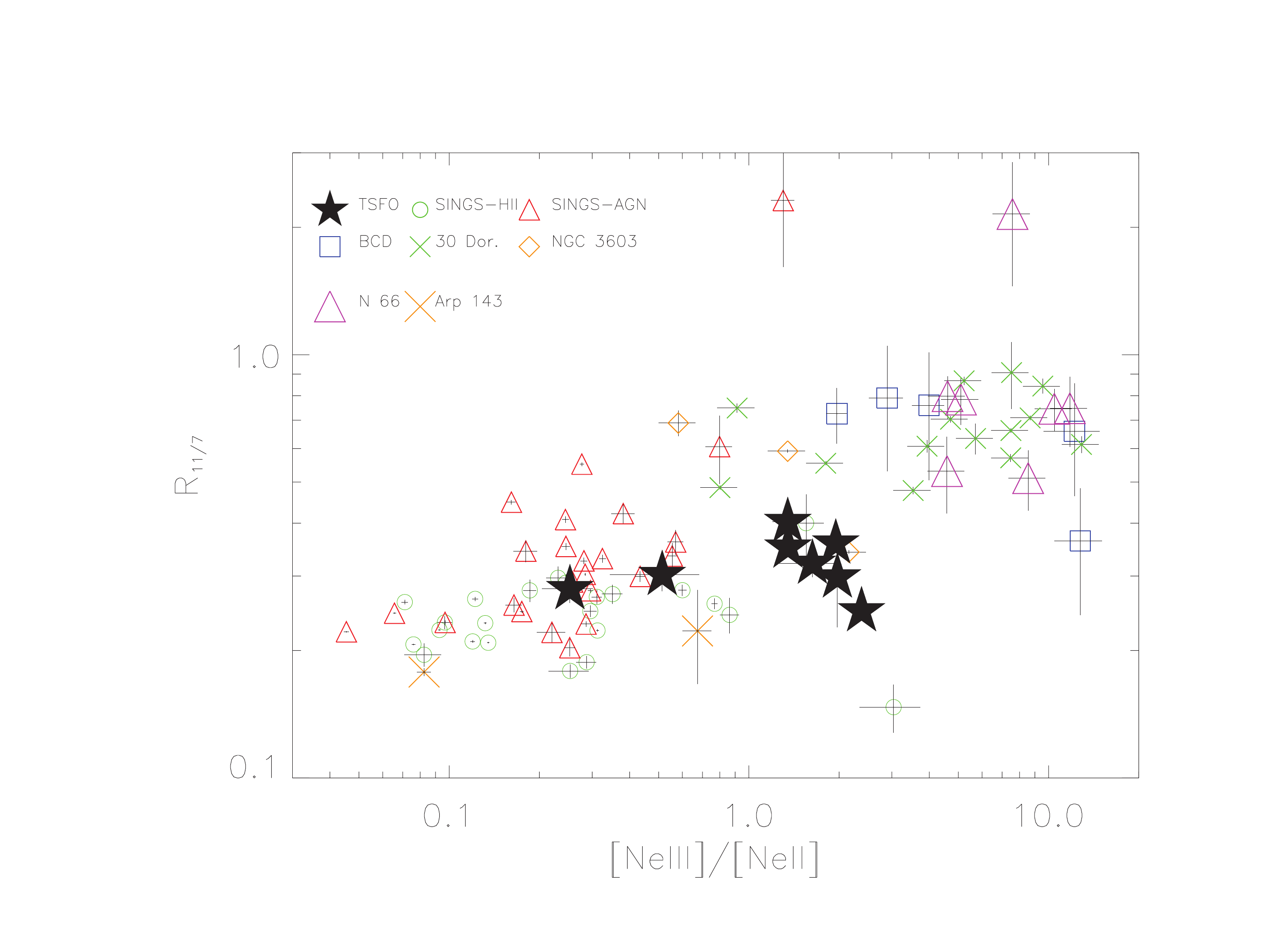}
%\epsscale{.80}
%\plotone{hardnessbw.eps}
\caption{R$_{11/7}$ as a function of the hardness of the ISRF as
  measured by the \neiii$/$\neii ~line flux ratio. Symbols are the same
  as for Figure 4. The ISFOs display no significant increase in
  R$_{11/7}$ when the neon line ratio is $>$1.  A color version of
  this figure is available in the online journal.}
\end{figure}

%Figure 7  neonr 87
\begin{figure}
\includegraphics[width=.9\textwidth]{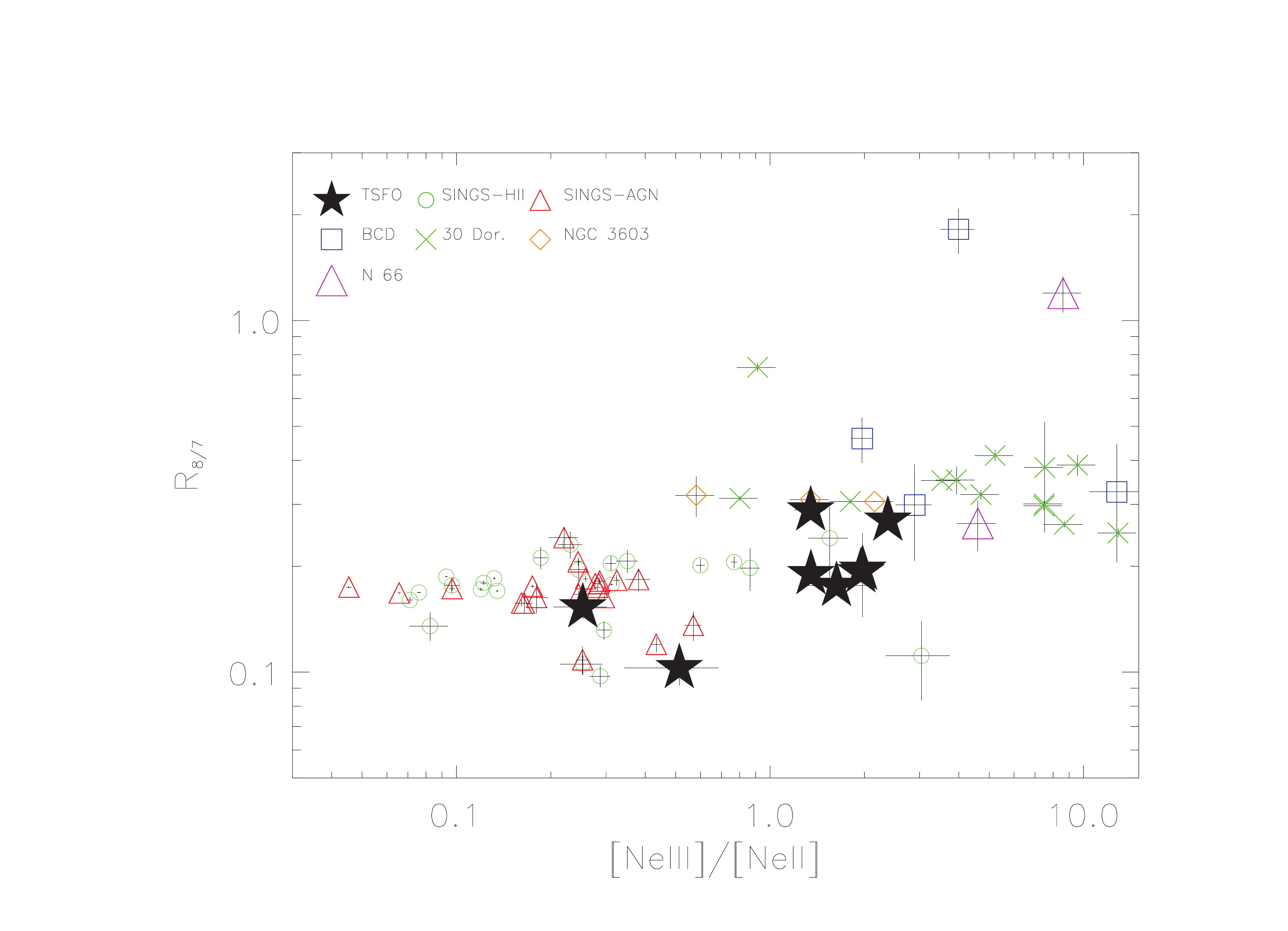}
\caption{R$_{8/7}$ as a function of the hardness of the ISRF as
  measured by the \neiii$/$\neii ~line flux ratio. Symbols are
  the same as for Figure 5. A color version of this figure is
  available in the online journal.}
%\label{fig:graphics}
\end{figure}

\clearpage
% Figure 8
\begin{figure}
\includegraphics[width=1.0\textwidth]{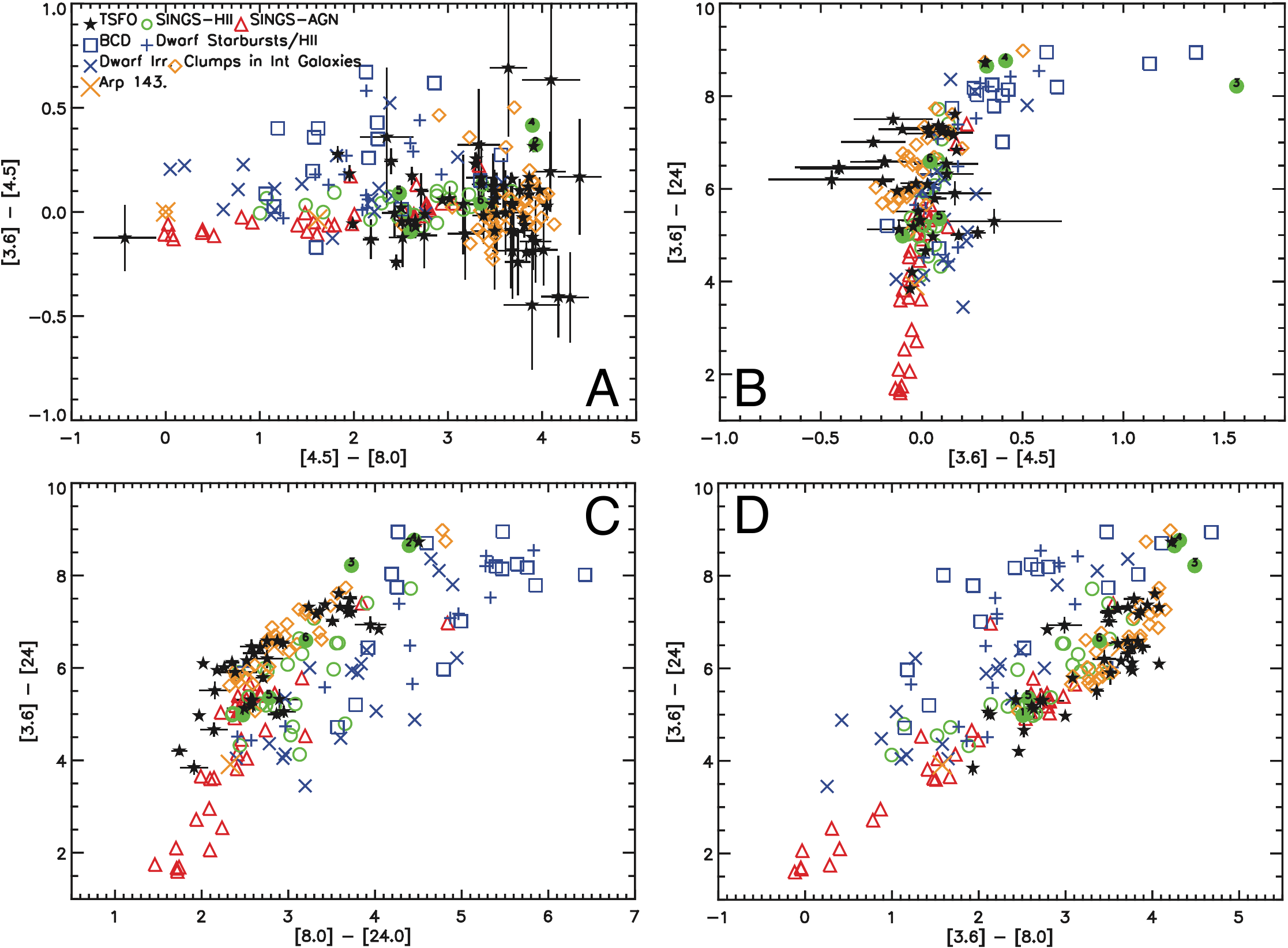}
\caption{ Spitzer Two-Color Diagrams. The ISFOs are shown as filled
  stars, while the SINGS-HII are shown as (green) open circles,
  SINGS-AGN are shown as (red) open triangles. A sample of dwarf
  galaxies (Smith \& Hancock 2009) are shown as (blue) open squares
  (BCDs), (blue) pluses (starburst/HII), and (blue) Xs (irregular). A
  selection of clumps from the inner disks of interacting systems are
  shown as (orange) open diamonds (Lapham et al. 2013 and
  references therein). Data is shown for regions in Arp 24, Arp 82,
  Arp 244, Arp 284 \& Arp 285. Data for Arp 143 is shown as large
  (orange) X's (Beiraro et al. 2009). Template colors for discrete
  sources identified in M 33 (Verley et al. 2007) are numbered (green
  solid circles): (1) PNe, (2) HII (infrared sample), (3) HII (radio
  sample), (4) HII (optical sample), (5) SNR, (6) Unknown source type. For
  clarity only the uncertainties for the ISFOs are shown. The very
  blue \iracb - \iracd ISFO in the far left of Panel (A) is most likely
  a background elliptical galaxy. A color version of this figure is
  available in the online journal. }
\end{figure}

% Figure 9
\begin{figure}
\includegraphics[width=1.0\textwidth]{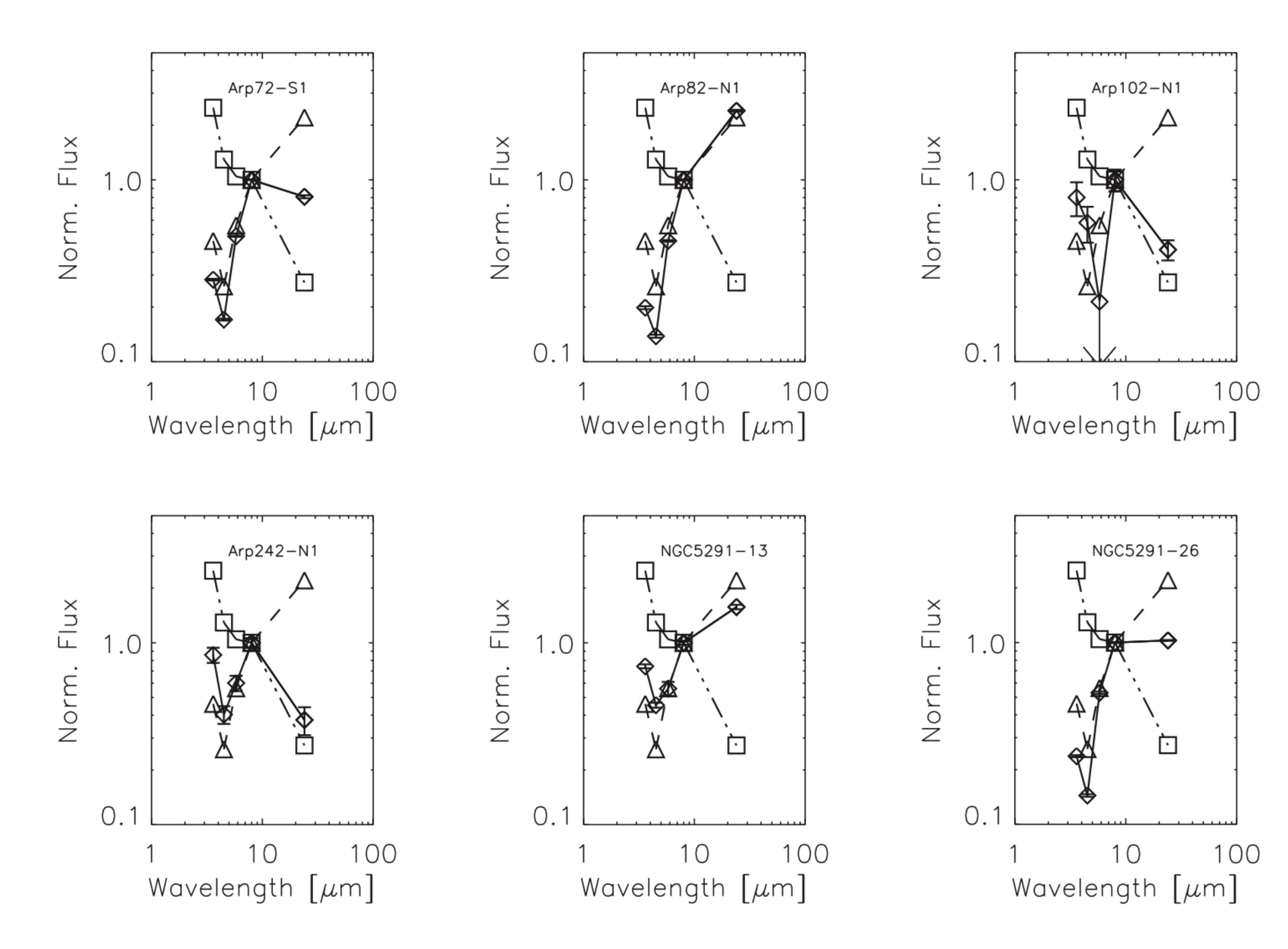}
%\epsscale{.80}
%\plotone{sed1bw.eps}
\caption{ A selection of ISFO SEDs shown as diamond symbols.  Overlaid
  are two galaxy SEDs from Draine et al. (2007), which serve as
  templates for regions dominated by emission from dust in the diffuse
  ISM (NGC 3190) shown with square symbols and from dust in PDRs (Mrk 33)
  shown with triangles. Each data set has been normalized to the
  8 \um flux ($\lambda F_{\lambda} (8 \mu m) = 1$).}
\end{figure}

\end{document}